\documentclass[oldversion]{aa}  

\usepackage{xcolor}
\usepackage{graphicx}
\usepackage{txfonts}
\usepackage{amsmath}
\begin{document}

   \title{JWST/NIRSpec and MIRI observations of an expanding, jet-driven bubble of warm H$_2$ in the radio galaxy 3C~326~N}
   \author{J.~H.~Leftley\inst{1} \and N.~P.~H.~Nesvadba\inst{1} \and G.~Bicknell\inst{2} \and R.~M.~J.~Janssen\inst{3,4} \and D.~Mukherjee\inst{5} \and R.~Petrov\inst{1} \and M.~B.~Shende\inst{5} \and H.~R.~M.~Zovaro\inst{2,6}}
   \institute{Universit\'e de la C\^ote d'Azur, Observatoire de la C\^ote d'Azur,
  CNRS, Laboratoire Lagrange, Bd de l'Observatoire, CS 34229, 06304
  Nice cedex 4, France
  \and
  Research School of Astronomy and Astrophysics, Australian National University, Canberra, ACT 2611, Australia
  \and
  Jet Propulsion Laboratory, California Institute of Technology, 4800 Oak Grove Dr., Pasadena, CA 91109, USA
  \and
  Department of Astronomy, California Institute of Technology, 1216 E California Blvd., Pasadena, CA 91125, USA 
  \and
  Inter-University Centre for Astronomy and Astrophysics, Pune-411007, India
  \and
  ARC Centre of Excellence for All Sky Astrophysics in 3 Dimensions (ASTRO 3D), Australia}

   \date{Received ; accepted }

 
  \abstract
    {The physical link between AGN activity and the suppression of star formation in their host galaxies is one of the major open questions of the AGN feedback scenario. The Spitzer space mission revealed a subset of powerful nearby radio galaxies with unusually bright line emission from warm ($T\ge 100$~K) molecular hydrogen, while typical star-formation tracers like PAHs or a dust continuum, were exceptionally faint or undetected. Here we present JWST NIRSpec and MIRI {MRS} IFU observations of one of the best studied galaxies of this class, 3C~326~N at z=0.09. We identify a total of 19 lines of the S, O, and Q series of ro-vibrational H$_2$ emission with NIRSpec at 0.11\arcsec\ spatial resolution that probe a small amount ($1.4\times 10^4$ M$_{\odot}$) of  gas at temperatures $T\sim 1000$~K. We also map the rotational mid-infrared lines of H$_2$ 0--0 S(3), S(5), and S(6) at a spatial resolution of 0.4\arcsec\ with MIRI/MRS, which probe most of the $2\times 10^9$ M$_{\odot}$ of warm H$_2$ in this galaxy. CO band heads show a stellar component consistent with a "slow-rotator", typical of a massive ($3\times10^{11}$\,M$_\odot$) galaxy, and provide us with a reliable systemic redshift of $z=0.08979\pm 0.0003$. Extended line emission shows a bipolar bubble expanding through the molecular disk at velocities of up to 380\,km\,s$^{-1}$, delineated by several bright clumps along the Northern outer rim, potentially from gas fragmentation. Throughout the disk, the H$_2$ is very broad, FWHM $\sim 100-1300$~km s$^{-1}$, and has complex, dual-component Gaussian line profiles. Extended [FeII]$\lambda$1.644 and Pa$\alpha$ follow the same morphology, however [NeIII]$\lambda$15.56 is more symmetric about the nucleus. We show that most of the gas, with the exception of [NeIII]$\lambda$15.56, is predominantly heated by shocks driven by the radio jets into the gas, both for the ro-vibrational and rotational H$_2$ lines, and that the accompanying line broadening is sufficient to suppress star formation in the molecular gas. We also compare the morphology and kinematics of the rotational and ro-vibrational lines, finding that the latter can be a good proxy to the global morphology and kinematic properties of the former in strongly turbulent environments. {This demonstrates the potential of using the higher frequency ro-vibrational lines to study turbulent molecular gas. Provided they are bright enough, this enables studies of terbulence in galaxies at intermediate and high redshifts well into the Epoch of Reionisation while most rotational lines are redshifted out of the MIRI bandpass for $z\ge1.5$}.}
    \titlerunning{Warm H$_2$ in 3C~326~N seen with the JWST}
    \authorrunning{Leftley et al.}

   \maketitle
%

\section{Introduction}
\label{sec:introduction}

Active Galactic Nuclei (AGN) residing in the centres of galaxies have been prime suspects of regulating star formation in galaxies for more than two decades. \citet[][]{silk_quasars_1998} were first to recognise that a small fraction of the immense energy output of AGN would suffice to unbind most of the interstellar gas from massive host galaxies -- provided that the AGN energy can efficiently be injected into the gas of the host galaxy over kiloparsec scales. This would resolve several of the long standing open questions of extragalactic astrophysics, including, e.g., the old stellar populations, low gas content, and star formation rates in massive galaxies, the metal enrichment in the intra-cluster gas of massive galaxy clusters, or the tight relationship between the masses of supermassive black holes and, e.g., the masses or central velocity dispersions of the bulges of their host galaxies \citep[e.g.,][]{gebhardt_relationship_2000,ferrarese_fundamental_2000,springel_modelling_2005,croton_many_2006,forman_reflections_2005,kirkpatrick_direct_2009,bharadwaj_extending_2015,schaye_eagle_2015,weinberger_supermassive_2018,werner_hot_2020}. 

Many observational and theoretical studies have since been addressing the question of how the AGN energy is injected into the surrounding gas, and how this affects star formation. Most of these studies have focused on the rapid gas removal through galaxy-wide outflows. However, although many observations have revealed that outflows of molecular, atomic, or ionised gas are common in the host galaxies of powerful AGN and more likely related to the AGN rather than concomitant star formation \citep[e.g.,][]{fiore_agn_2017,veilleux_fast_2013}, uniquely establishing a physical link between the AGN energy injection and the suppression of star formation is still a challenge. For example, many host galaxies of powerful AGN, and even AGN-driven outflows, do not show a concomitant deficit in star formation \citep[][]{stanley_mean_2017,ellison_star_2016}. 

A remarkable exception is a subclass of (mostly) powerful radio galaxies. In about 30\% of nearby ($z<0.2$) galaxies from the 3CRR catalogue \citep{laing_bright_1983} observed with the Infrared Spectrograph (IRS) on-board the Spitzer Space Telescope, \citet{ogle_shocked_2007} and \citet{ogle_jet-powered_2010} identified bright mid-infrared line emission from luminous, warm molecular hydrogen \citep[Molecular Hydrogen Emission Galaxies, "MOHEGS", see also ][]{dicken_spitzer_2012}, while typical AGN and star-formation tracers like forbidden lines, PAHs, or the mid-infrared dust continuum are either weak or undetected. 

Due to its very simple, and highly symmetric structure, the homonuclear H$_2$ molecule does not have a permanent electric dipole moment, and therefore radiates only at high gas temperatures, above of-order 100~K. This makes H$_2$ line emission so elusive in many star forming molecular clouds, which have typical temperatures of-order 10~K. The galaxies found by \citeauthor[][]{ogle_jet-powered_2010} are remarkable, in that their H$_2$ line luminosities can reach up to about~20\% of their total infrared dust luminosity \citep[][]{ogle_shocked_2007}, which requires a continuous gas heating mechanism over long timescales \citep[][]{nesvadba_energetics_2010}. 

 These galaxies show also a marked offset from the Kennicutt-Schmidt relationship between molecular gas mass and star-formation rate surface densities, as seen in multiple star-formation tracers, including PAHs, cold dust observed with Herschel, the UV continuum, or H$\alpha$ \citep[e.g.,][]{nesvadba_energetics_2010, alatalo_suppression_2015, lanz_star_2016, nesvadba_jet-driven_2021,drevet_mulard_star_2023}. In extreme cases, their star-formation rates can be one to two orders of magnitude lower at a given gas-mass surface density than in normal star-forming galaxies, which might be a signature of suppression of star formation through the AGN. 
 
 \citet{ogle_jet-powered_2010} and \citet{nesvadba_energetics_2010} hypothesised that the bright H$_2$ line emission in these galaxies is produced by shocks that are driven into the molecular gas through interactions with the radio jet, as indicated by several line diagnostics comparing the luminosity of warm H$_2$ lines with indicators of AGN and star-formation activity, like PAHs or the mid-infrared continuum \citep[e.g.,][]{nesvadba_dense_2011,guillard_strong_2012}. This is also indicated by enhanced line widths of the warm H$_2$ lines in these galaxies \citep[][]{nesvadba_energetics_2010,guillard_strong_2012}. Comparison with CO observations of cold molecular gas suggest that at least in some of these galaxies, the warm molecular gas component dominates the overall gas budget \citep[][]{nesvadba_energetics_2010,ogle_jet-powered_2010}, with warm molecular gas masses of up to few$\times 10^9$ M$_{\odot}$ \citep[][]{ogle_jet-powered_2010,nesvadba_energetics_2010}. 

 Nearly 15 years after the end of the cold Spitzer mission, the James Webb Space Telescope is now again enabling new mid-infrared observations of AGN, at orders of magnitude higher sensitivity, spectral and spatial resolution, albeit somewhat more restrictive spectral range \citep[e.g.,][]{pereira-santaella_low-power_2022,lai_goals-jwst_2022,armus_goals-jwst_2023,alvarez-marquez_nuclear_2023}. Here we present new observations of one of the best-studied MOHEGs, 3C~326~N, at $z=0.09$, which we obtained through open-time observations with the integral-field units (IFUs) of the JWST/Near-Infrared Spectrograph (NIRSpec) and Mid-Infrared Instrument (MIRI) in Cycle 1. 3C~326~N is a giant FRII radio galaxy with a radio power of $2.5\times 10^{26}$~W~Hz$^{-1}$ at 327~MHz, and two Mpc-sized lobes. \citet{willis_multifrequency_1978} estimated a spectral age of the radio source of 200~Myrs. 3C~326~N has a nearby neighbour, 3C~326~S, making it difficult to identify the source of the radio emission. Both have radio cores. \citet{rawlings_new_1990} argued that 3C~326~N is the better candidate, for its brighter stellar continuum, higher stellar mass of $3\times 10^{11}$ M$_{\odot}$ \citep[][]{nesvadba_energetics_2010}, and for being the only galaxy of the two that has [OIII]5007 line emission consistent with the usual relationship between [OIII]5007 luminosity and radio power. 
 
 3C~326~N was observed with Spitzer/IRS by \citet{ogle_shocked_2007}, who identified unusually bright H$_2$ line emission from the pure-rotational lines between 5 and 30~$\mu$m, whereas many typical AGN lines, and in particular PAHs, were either very faint or unobserved. The galaxy was further studied by \citet{nesvadba_energetics_2010}, who obtained IRAM CO(1--0) observations, and found that the warm gas is most likely heated by shocks, and that the warm molecular gas dominates the overall molecular gas budget, with a warm molecular gas mass of 2$\times 10^9$ M$_{\odot}$. They also found Na~D absorption indicating a (likely jet-driven) outflow of about $30-40$~M$_{\odot}$ yr$^{-1}$ in neutral gas, and a terminal velocity of $-1800$~km s$^{-1}$. 3C~326~N has a very low star-formation rate \citep[SFR$=0.087^{+0.106}_{-0.046}$~M$_{\odot}$~yr$^{-1}$ from Herschel dust photometry;][]{lanz_star_2016}, and a radiatively weak AGN \citep[$L_{X}=10^{40.63}$ erg s$^{-1}$;][]{lanz_star_2016}, so that neither star formation nor AGN are powerful enough to heat the molecular gas \citep[][]{nesvadba_energetics_2010}. Moreover, the Cosmic Ray flux required to heat the observed amount of molecular gas would be high enough to destroy the H$_2$ molecules \citep[][]{nesvadba_energetics_2010}, which leaves mechanical heating through shocks driven by the radio source as the only plausible gas heating mechanism. This was subsequently confirmed through remarkably broad line widths (FWHM=650\,km\,s$^{-1}$) of ro-vibrational emission lines of H$_2$ seen with the VLT/SINFONI near-infrared IFU \citep[][]{nesvadba_dense_2011}, as well as through very high line ratios of H$_2$ to Pa$\alpha$, which are also a good shock indicator \citep[][]{puxley_molecular_1990}. These authors also suggested that the line broadening may be indicative of high gas turbulence, which, if it cascades down to the usual scales of giant molecular clouds ($\sim$100~pc) may suppress the star formation by the molecular gas. By comparing the energy injection that can plausibly be injected through star formation, AGN radiation, or the radio jets, \citet{nesvadba_energetics_2010} concluded that only the radio source is powerful enough to explain the observed gas kinematics. 

{In this work, we aim to provide a more in depth look, through novel modelling methods and improved data quality and wavelength coverage, into the molecular and ionised gas structures of 3C~326~N initially probed in these previous works. We provide a rich set of modelled spectral features which we use to deduce a reliable systemic redshift and investigate the kinematics and heating mechanism of the molecular gas. We also investigate the potential of the ro-vibrational lines, that are produced by hot $>1000$\,K gas which comprises only a small fraction of the total gas mass, as tracers of turbulence in place of pure rotational lines which are produced by the bulk of the molecular gas at cooler $\sim100$\,K temperatures. The rotational lines are redshifted out of the MIRI bandpass at $z\geq1.5$ whereas the ro-vibrational lines are accessible up to the Epoch of Reionisation if they are bright enough.}

 The paper is organised as follows. In Section~\ref{sec:observations} we describe our observations and data reduction for the JWST and ancillary data sets obtained with ALMA and the JVLA. We then explain our methodology and Bayesian fitting routine in Section~\ref{sec:LineFit}, before describing our analysis of the stellar and dust continuum, radio morphology, and kinematics of molecular and warm ionised gas in Section~\ref{sec:analysis}. In Section~\ref{sec:discussion} we discuss our results in the context of multi-phase molecular gas properties, and star formation. We then use this source to investigate what our results imply about observations of distant galaxies in the early Universe, where rotational lines are not observable with MIRI, in Section~\ref{sec:highz}. We summarise our results in Section~\ref{sec:summary}. 

 Throughout the paper we use the flat \citet{planck_collaboration_planck_2016} cosmology, where $H_0=67.8$\,km\,s$^{-1}$ Mpc$^{-1}$ and $\Omega_M=0.308$. In this cosmology, the luminosity distance to 3C~326~N at $z=0.08979$ is $D_L=423.6$~Mpc, and the angular size distance, $D_A=356.8$~Mpc. 1.729~kpc are projected onto 1 arcsec. 

\section{Observations and data reduction}
\label{sec:observations}

\subsection{NIRSpec and MIRI imaging spectroscopy}
\label{ssec:jwstobs}
3C~326~N was observed with the integral-field units (IFUs) of the Near-Infrared Spectrograph (NIRSpec) and the Medium-Resolution Spectrograph (MRS) of the Mid-Infrared Instrument (MIRI) on the James-Webb Space Telescope as part of program GO1-2162 (PI Nesvadba). NIRSpec data were obtained on 4 March 2023 through the filter F170LP and with the grating G235H. They cover a near-continuous spectral range between 1.66~$\mu$m and 3.17~$\mu$m with a gap between 2.40~$\mu$m and 2.56~$\mu$m due to the physical gap between the two NIRSpec detectors in the focal plane of the telescope. The field-of-view of the NIRSpec IFU is 3.0\arcsec$\times$3.0\arcsec, and the dispersion through the G235H grating is $\Delta\lambda=8.7\times10^{-4}$~$\mu$m, which corresponds to a spectral resolution of $85-150$\,km\,s$^{-1}$. 

Data were obtained during one visit with 2976~seconds of on-source observing time at four dither positions, using the dither pattern NRSIRS2RAPID with 50 single-integration groups. We also obtained a background exposure with the same readout pattern, 10 single-integration groups, one dither position, and a total exposure time of 160~seconds.  

MIRI data were obtained on 28 March 2023 with the four channels of the LONG setting. Channels 1 to 4 cover spectral ranges between 6.53~$\mu$m and 7.65~$\mu$m, 10.02~$\mu$m and 11.7~$\mu$m, 15.41~$\mu$m and 17.98~$\mu$m and 24.19~$\mu$m and 27.9~$\mu$m, respectively. The spectral resolving power is R$=$1330-3610, corresponding to a spectral resolution between 83 and 226\,km\,s$^{-1}$. The field-of-view of the MIRI IFU is between 3.2\arcsec$\times$3.7\arcsec\ in Channel~1 and 6.6\arcsec$\times$7.7\arcsec\ in Channel~4.

We used the MRSLONG and MRSSHORT detectors with the LONG(C) grating, FASTR1 readout pattern, 35 single-integration groups, and four dither positions that were optimised for extended sources. The on-source exposure time was 389~seconds per detector. A second visit was used to observe a background  frame with the same setup directly following the on-source observations. 

Both the NIRSpec and MIRI data were reduced with version 1.12.4 of the official pipeline \textsc{python} package: \textsc{jwst}. We used the CRDS context collection \textsc{jwst\_1146} for static calibration file association. Both the datasets required no additional data cleaning beyond what the pipeline performed. In summary, the raw files were processed with Detector1Pipeline to correct for detector level effects. The resulting ramp files were passed to the Spec2Pipeline which performs the WCS correction, flat field correction, background correction, and flux calibration. Furthermore, the pipeline applies an MSA imprint and pathloss correction for the NIRSpec data as well as a fringe correction for the MIRI MRS data. We experimentally applied the fringe correction to the NIRSpec data as well but found no improvement. Finally, the data for NIRSpec and MIRI were resampled and combined with the Spec3Pipeline. We used the drizzle combination method to combine and rotate the NIRSpec data and each MIRI channel separately. This pipeline also performed outlier detection and a master background subtraction for the MIRI MRS data.

\subsection{ALMA interferometry}
\label{ssec:almaobs}
We also obtained archival ALMA data in band~3 from program 2015.1.01120.S, which cover the expected wavelength of $^{12}$CO(1--0) at z=0.0898, 105.8~GHz. Continuum emission from the nucleus is well detected, with a flux density at 115.3~GHz in the rest-frame of $S_{115GHz}=1.6\pm0.01$ mJy bm$^{-1}$. CO(1--0) line emission is not detected. We use this data set to derive an upper limit on the line flux, which has $RMS=0.2$~mJy bm$^{-1}$ in 15.4~MHz wide spectral channels, for a beam size of FWHM=0.46\arcsec$\times$0.26\arcsec\ along the major and minor axis, respectively, and position angle PA=-13.5$^\circ$, measured from North towards East. 

We follow the prescription of \citet{sage_cool_2007} and \citet{young_atlas3d_2011} by setting $\sigma_{line}^2 = \Delta v^2 \sigma^2\ N_l\ (1.+N_l/N_b)$, where $\sigma_{line}$ is the line-integrated standard deviation of the data set, $\Delta\ v$ the line width in\,km\,s$^{-1}$, $N_l$ the sampling of the line, and $N_b=118$ is the number of spectral channels from which $\sigma$ was estimated. Assuming a line width of FWHM=100\,km\,s$^{-1}$, a sampling of 43.26~km~s$^{-1}$ as in our data, and $RMS=0.2$~mJy~bm$^{-1}$, we find $\sigma_{line}=30.7$~mJy\,km\,s$^{-1}$ bm$^{-1}$. Using Equation~3 of \citet{solomon_molecular_1997}, we can translate this flux into a line luminosity, by setting $L^\prime_{CO}=3.25\times 10^7\ S_{CO}\ \Delta V \nu_{obs}^{-2}\ D_L^2 (1+z)^{-3}$, where $S_{CO}\Delta V$ is the integrated line flux in Jy\,km\,s$^{-1}$, $\nu_{obs}$ the observed frequency in GHz, $D_L$ the luminosity distance in Mpc, and $z$ the redshift. To estimate a molecular gas mass from the luminosity, we adopt the standard conversion factor appropriate for, e.g., the Milky Way, $\alpha_{CO} = 4.3 M_{\odot}/$ [K\,km\,s$^{-1}$ pc$^2$] \citep[e.g.,][]{bolatto_co--h2_2013}, and find a $3\sigma$ upper limit of $1.5\times 10^8 M_{\odot}$ bm$^{-1}$. Assuming that CO line emission would extend over the same area also seen with NIRSpec in warm molecular gas, we expect that the total molecular gas mass would be a factor 5.3 larger, i.e., correspond to a total of $M_{H2,CO}= 8\times 10^8$ M$_{\odot}$. This is about a factor~3 less than the $M_{H2}=2\times 10^9 M_{\odot}$ of warm molecular gas previously found with Spitzer \citep[][]{ogle_shocked_2007}, and suggests that the majority of the molecular gas in 3C~326~N is indeed warm. 

\citet{nesvadba_energetics_2010} previously reported weak CO(1--0) line emission detected at $5\sigma$ with the IRAM Plateau de Bure interferometer in the D configuration at 5\arcsec\ beam size. Given the new, deeper observations available now from ALMA with a much larger bandwidth, we consider this previous detection spurious. Although the spectrum was carefully extracted from the data cube, the large beam size and narrow band width made it very challenging to isolate the continuum from the line emission, which can explain the faint putative line detection that was made with the best instrument available at that time. 

\subsection{JVLA interferometry}
\label{ssec:jvalobs}

We also obtained deep radio continuum imaging with the Karl G. Jansky Very Large Array (JVLA) in the A-array in the L, C, and X-bands at 1.52, 7.25, and 9.0~GHz through program 18A$-$285 (PI Nesvadba). We observed the source in four sessions between 2 March 2018 and 9 June 2018 with 24 to 27 antennae and a total allocation of 6.4~hrs. Our main goal was to probe the morphology of the radio core of 3C~326~N, and to search for extended emission on scales comparable to the 3\arcsec\ size of the molecular disk. We used 3C~286 as flux and bandpass calibrator, and J1609+2641 as gain calibrator. In the L-band, we centred the two basebands on 1.264 and 1.776~GHz to obtain 1~GHz of nominal band width. In the C and X-bands, we tuned each baseband to 5.5 and 6.5~GHz, and to 9.5 and 10.5~GHz, respectively, reaching 2~GHz of nominal band width. 

Data were reduced in the standard way using \textsc{casa} Version 5.4.2-5. First, data were flagged, antennae positions corrected, and an initial flux density scaling was derived from our primary calibrator 3C~286 in each band. We then did an initial phase calibration for all bands before solving for antennae-based delays and performing the bandpass calibration in each band, again, using 3C~286. Complex gains were derived from J1609+2641 before we used the flux of our primary calibrator to derive the amplitude gains of J1609+2641 and 3C~326~N in each band. We eventually applied the calibration before reconstructing and cleaning the images, using briggs=0.5. The final RMS obtained in this way is about 5~mJy~bm$^{-1}$ in the C and X-bands, and 90~mJy~bm$^{-1}$ in the L-band, which is strongly affected by RFI. Beam sizes are around 0.2\arcsec-0.3\arcsec\ in the C and X-bands, and 1.1\arcsec\ in L. The central tuning, beam size and position angle, and RMS in each band can be found in Tab.~\ref{tab:radiodata}.

\subsection{Relative alignment}
\label{ssec:alignment}

The JWST/NIRSpec and each channel of the MIRI observations show a small spatial offset with each other and with the ancillary data sets, mainly along right ascension. Relative to the NIRSpec data, MIRI data are offset in RA by -0.7\arcsec, 0.58\arcsec, and 0.32\arcsec in channel~1, channel~2, and channel~3, respectively. The offsets in Dec are subpixel in MIRI and we do not list the offset for channel~4 due to an extremely tentative continuum detection. The positional offsets are significantly larger than the given 0.1\arcsec pointing accuracy of JWST\footnote{\url{https://jwst-docs.stsci.edu/jwst-observatory-characteristics/jwst-pointing-performance}}. We therefore did not simply use the position on the sky as indicated in the file headers, but used the stellar continuum peak in the NIRSpec data cube to align the JWST data relative to the radio data. In aligning the data sets we made the simple astrophysical assumption that the radio core falls on top of the peak in the stellar continuum emission, which is also the location of the highest Pa$\alpha$, [FeII]$\lambda$1.599 an [FeII]$\lambda$1.6440 surface brightness, highlighting the presence of a faint AGN at that position. 

\begin{table*}[]
    \centering
    \begin{tabular}{lcccccc}
\hline
    Band  & central frequency & beam size & PA    & RMS                  & size & flux \\
          & [GHz]             &  [arcsec] & [deg] & [$\mu$Jy bm$^{-1}$]  & [arcsec] & [mJy bm$^{-1}$] \\
\hline
       L  & 1.52              & $1.18\times1.11$ & $-41.8$ & 90     & $1.08\times1.01$ & 2.10$\pm$0.09 \\
       C  & 7.25              & $0.29\times0.26$ & $65.5$  & 4.7    & $0.29\times0.24$ & 1.00$\pm$0.005  \\
       X  & 9.00              & $0.22\times0.20$ & $-32.8$ & 5.3    & $0.21\times0.20$ & 1.15$\pm$0.005 \\
\hline
    \end{tabular}
    \caption{Observational parameters and results of our JVLA observations. Position angle PA is measured from North through East. The most accurate position of the radio core measured in the X-band is RA=15:52:09.112, Dec=20:05:48.2.} 
    \label{tab:radiodata}
\end{table*}

\section{Emission line and stellar continuum fitting}\label{sec:LineFit}

To model the spectral features of the JWST data, we used a Markov Chain Monte Carlo (MCMC) based fitting method. The sampling was performed with the python package \textsc{emcee} \citep{goodman_ensemble_2010} and the data file handling was performed by \textsc{astropy}\citep{astropy_collaboration_astropy:_2013}. 

\subsection{Emission Lines}

Unless otherwise stated, the line modelling was performed on one line at a time, to be able to probe differences in line kinematics or morphology, if present. When modelling a line, we cut the cube in the spectral direction such that the modelled wavelength range is $\lambda_0(1\pm0.01)(1+z_\mathrm{in})$ where $\lambda_0$ is the lines rest frame wavelength and $z_\mathrm{in}$ is our initial redshift guess set to 0.09 based on the measurement of \citet{nesvadba_dense_2011} and visual inspection of the NIRSpec lines in the central 0.5" of the galaxy.

We reproduced the emission line features using a Gaussian line profile coupled with a linear continuum in frequency. We find that the line shape is often complex and requires multiple Gaussians to explain. Therefore, the line profile can be as:
\begin{equation}
    F_\nu(\nu) = m\nu + b + \sum^{N}_{n=1} (1+z_n)A_n\exp{\left(-0.5 \left(\frac{(1+z_n)\nu-\nu_0}{\sigma_n}\right)^2\right)},
\end{equation}
where $A_n$ is the amplitude of Gaussian $n$, $\nu$ is the frequency, $\nu_0$ is the line rest frame frequency, $z_n$ is the redshift of Gaussian $n$, $\sigma_n$ is the standard deviation of line $n$, $m$ is the continuum line gradient, and $b$ is the continuum intercept. Note that $m$ and $b$ are not adjusted for redshift when modelled. We allow up to 2 Gaussians and select the number used by the Bayesian Information Criterion (BIC).

When modelling, we measure the goodness of fit from a parameter set using the maximum likelihood + prior probability distribution. Our maximum likelihood is defined as
\begin{align}
         2\ln{\left(L\right)}&=&-\sum\left[\frac{\left(F_{\nu,ob}\left(\nu\right)-F_\nu\left(\nu,\alpha\right)\right)^2}{\delta}+\ln\left(2\pi\delta\right)\right],\\
         \delta&=&err^2+f^2F_\nu\left(\nu,\alpha\right)^2
\end{align}\label{eq:likelyhood}
where $err$ is the data uncertainty, $f$ is the fraction by which the uncertainty is underestimated, and $F_{\nu,ob}$ is the observed flux. We define the prior probability function for all parameters to be uniform between the boundaries and $-\inf$ otherwise. We select nonrestrictive boundaries of -5--5, -1--1\,Jy, 0--10\,mJy, 0.1--300\,GHz, 0.088--0.0915 for $m$, $b$, $A_n$, $\sigma_n$, and $z_n$, respectively For $z$ we define an additional prior probability $P_z$. $P_z$ is a normal distribution such that
\begin{equation}
    P_z = \prod_{n=1}^N\left(\frac{1}{\sigma_z\sqrt{2\pi}}\exp{\left(-0.5 \left(\frac{z_n-<z>}{\sigma_z}\right)^2\right)}\right),
\end{equation}
where $<z>$ is the mean $z$ for all $N$ Gaussians in the model and the standard deviation of $z$ separation probability $\sigma_z=0.001$. We base this probability distribution on the fact that the separation between Gaussians for one line can differ but greater separations are less likely to be part of the same line instead of two unrelated features. We choose the $\sigma_z$ to be 0.001 to place a 95\% chance that a line emitting source is not separated from the average velocity of emitting structures in one spaxel by more than $\sim$550\,km\,s$^{-1}$ which is similar to the maximum velocity offset between components for \textit{any} single line in 3C~326~N from \citet{nesvadba_dense_2011}. We find this significantly improves multi-line fits without being overly constraining on the velocity separation of coincidental projected structures. We also set the rule that $z_n<z_{n+1}$ to prevent the degenerate solution of components switching parameters. {These additional rules for $z$ have no effect in the one component case.}

\begin{figure*}
    \includegraphics[width=\textwidth]{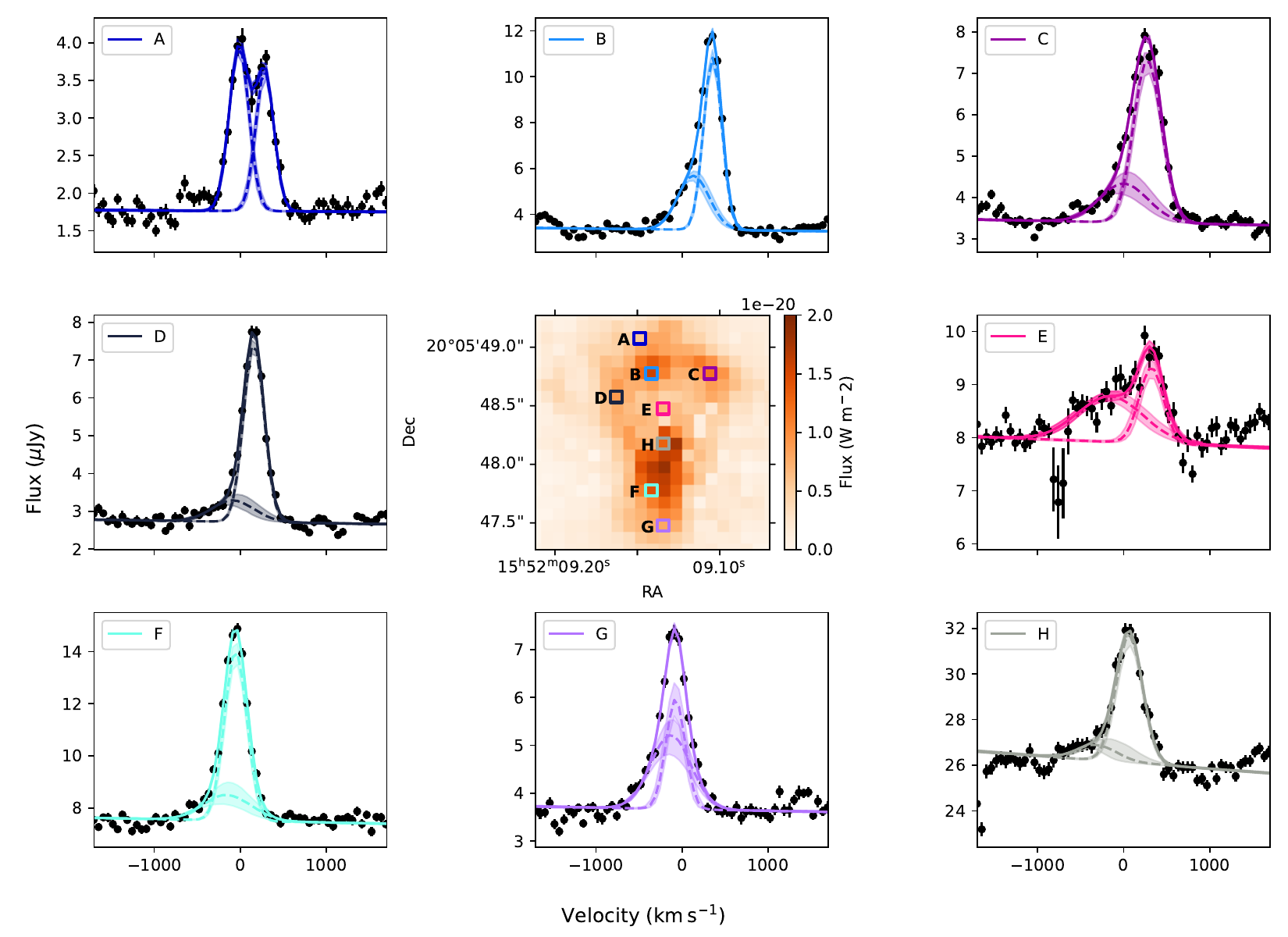}
    \caption{Centre: Integrated model flux for  H$_2$1--0 S(3). Outside: Gaussian model fit to the H$_2$1--0 S(3) line for different spaxels of the central image. The coloured squares in the image represent the position of the plotted spaxel of the same colour. Dashed lines are individual Gaussian components and the solid line is the summed Gaussians. Shaded regions are 1\,$\sigma$ errors. These spectra area also a good illustration of the high quality of the date and the automatic fitting routine.}\label{fig:IFUlineComp}
\end{figure*}

All Gaussian components are then modelled simultaneously for the line and each spaxel is modelled independently. We use a uniform initial distribution between the boundaries of 0--1\,mJy and 1--100\,GHz for the amplitude and standard deviation, respectively, and a normal distribution around $0.09\pm0.001$ for z. We use 5000 steps and 200 walkers. We then perform a second run starting from the best fit parameters by likelihood of the first run with a normal distribution with a 0.1\% standard deviation for the initial positions. The second run also uses 5000 steps and 200 walkers. An example result for a 2 Gaussian fit to the H$_2$1--0 S(3) line for different regions of the galaxy can be seen in Fig.\,\ref{fig:IFUlineComp}.

When reporting Gaussian line component amplitudes and widths, we account for the instrumental line spread function. These vary slightly between NIRSpec and the different MIRI channels. The resolution for NIRSpec IFU and MIRI MRS long channel 1, 2, 3, and 4 is 2700, 3100–-3610, 2860-–3300, 1980–-2790, and 1630–-1330, respectively\footnote{values from \url{https://jwst-docs.stsci.edu/jwst-near-infrared-spectrograph/nirspec-observing-modes/nirspec-ifu-spectroscopy} and \url{https://jwst-docs.stsci.edu/jwst-mid-infrared-instrument/miri-observing-modes/miri-medium-resolution-spectroscopy} for NIRSpec IFU and MIRI MRS, respectively}.

\begin{table*}[]
    \centering
    \begin{tabular*}{\textwidth}{@{\extracolsep{\fill}} c c c c c c c c c c c}\hline
    Label&F$_\mathrm{blue}$&F$_\mathrm{red}$&F$_\mathrm{Total}$&$A_\mathrm{blue}$&$A_\mathrm{red}$&$\sigma_\mathrm{blue}$&$\sigma_\mathrm{blue}$&$Vel_\mathrm{blue}$&$Vel_\mathrm{red}$\\\hline
    &[$10^{-20}\,$W$\,$m$^{-2}$]&[$10^{-20}\,$W$\,$m$^{-2}$]&$[10^{-20}\,$W$\,$m$^{-2}$]&[$\mu$Jy]&$[\mu$Jy]&[km\,s$^{-1}$]&[km\,s$^{-1}$]&[km\,s$^{-1}$]&[km\,s$^{-1}$]\\\hline\hline
A&$0.32^{0.022}_{0.024}$&$0.253^{0.023}_{0.022}$&$0.573^{0.013}_{0.013}$&$2.42^{0.12}_{0.12}$&$2.09^{0.1}_{0.1}$&$112.0^{8.9}_{11.0}$&$103.0^{9.1}_{9.7}$&$-15.6^{8.4}_{8.9}$&$283.0^{9.7}_{10.0}$\\[3pt]
B&$0.553^{0.067}_{0.068}$&$1.01^{0.065}_{0.066}$&$1.56^{0.019}_{0.02}$&$2.45^{0.2}_{0.22}$&$8.39^{0.33}_{0.37}$&$192.0^{14.0}_{12.0}$&$102.0^{3.9}_{3.6}$&$133.0^{24.0}_{23.0}$&$356.0^{3.3}_{3.3}$\\[3pt]
C&$0.343^{0.08}_{0.089}$&$0.813^{0.094}_{0.078}$&$1.15^{0.023}_{0.023}$&$0.948^{0.23}_{0.31}$&$4.11^{0.29}_{0.23}$&$302.0^{32.0}_{34.0}$&$168.0^{10.0}_{8.8}$&$2.71^{74.0}_{58.0}$&$276.0^{8.9}_{10.0}$\\[3pt]
D&$0.182^{0.047}_{0.049}$&$0.713^{0.049}_{0.045}$&$0.894^{0.018}_{0.018}$&$0.586^{0.12}_{0.18}$&$5.15^{0.19}_{0.18}$&$258.0^{36.0}_{32.0}$&$118.0^{5.3}_{4.5}$&$-83.3^{75.0}_{63.0}$&$156.0^{3.6}_{3.4}$\\[3pt]
E&$0.425^{0.062}_{0.068}$&$0.253^{0.045}_{0.055}$&$0.683^{0.04}_{0.041}$&$0.878^{0.094}_{0.091}$&$1.52^{0.18}_{0.17}$&$412.0^{49.0}_{54.0}$&$143.0^{21.0}_{22.0}$&$-150.0^{70.0}_{60.0}$&$320.0^{16.0}_{16.0}$\\[3pt]
F&$0.355^{0.11}_{0.12}$&$1.09^{0.13}_{0.1}$&$1.43^{0.028}_{0.031}$&$1.0^{0.32}_{0.49}$&$7.0^{0.45}_{0.38}$&$286.0^{45.0}_{46.0}$&$132.0^{7.6}_{7.1}$&$-150.0^{88.0}_{39.0}$&$-54.0^{4.0}_{4.3}$\\[3pt]
G&$0.457^{0.09}_{0.077}$&$0.333^{0.075}_{0.092}$&$0.79^{0.017}_{0.018}$&$1.55^{0.39}_{0.35}$&$2.56^{0.36}_{0.32}$&$250.0^{18.0}_{21.0}$&$112.0^{15.0}_{16.0}$&$-131.0^{26.0}_{18.0}$&$-75.6^{6.7}_{7.2}$\\[3pt]
H&$0.205^{0.075}_{0.14}$&$1.1^{0.14}_{0.078}$&$1.31^{0.045}_{0.046}$&$0.782^{0.19}_{0.27}$&$5.86^{0.45}_{0.27}$&$242.0^{80.0}_{68.0}$&$158.0^{10.0}_{8.7}$&$-315.0^{110.0}_{160.0}$&$65.1^{7.7}_{8.4}$\\[3pt]\hline
    \end{tabular*}
    \caption{The parameters of the Gaussians shown in Fig.\,\ref{fig:IFUlineComp} with a reference to the fitted spaxel in the figure.}
    \label{tab:IFUlineComp}
\end{table*}

\subsection{CO band head absorption}

\begin{figure*}
    \centering
    \includegraphics[width=\textwidth]{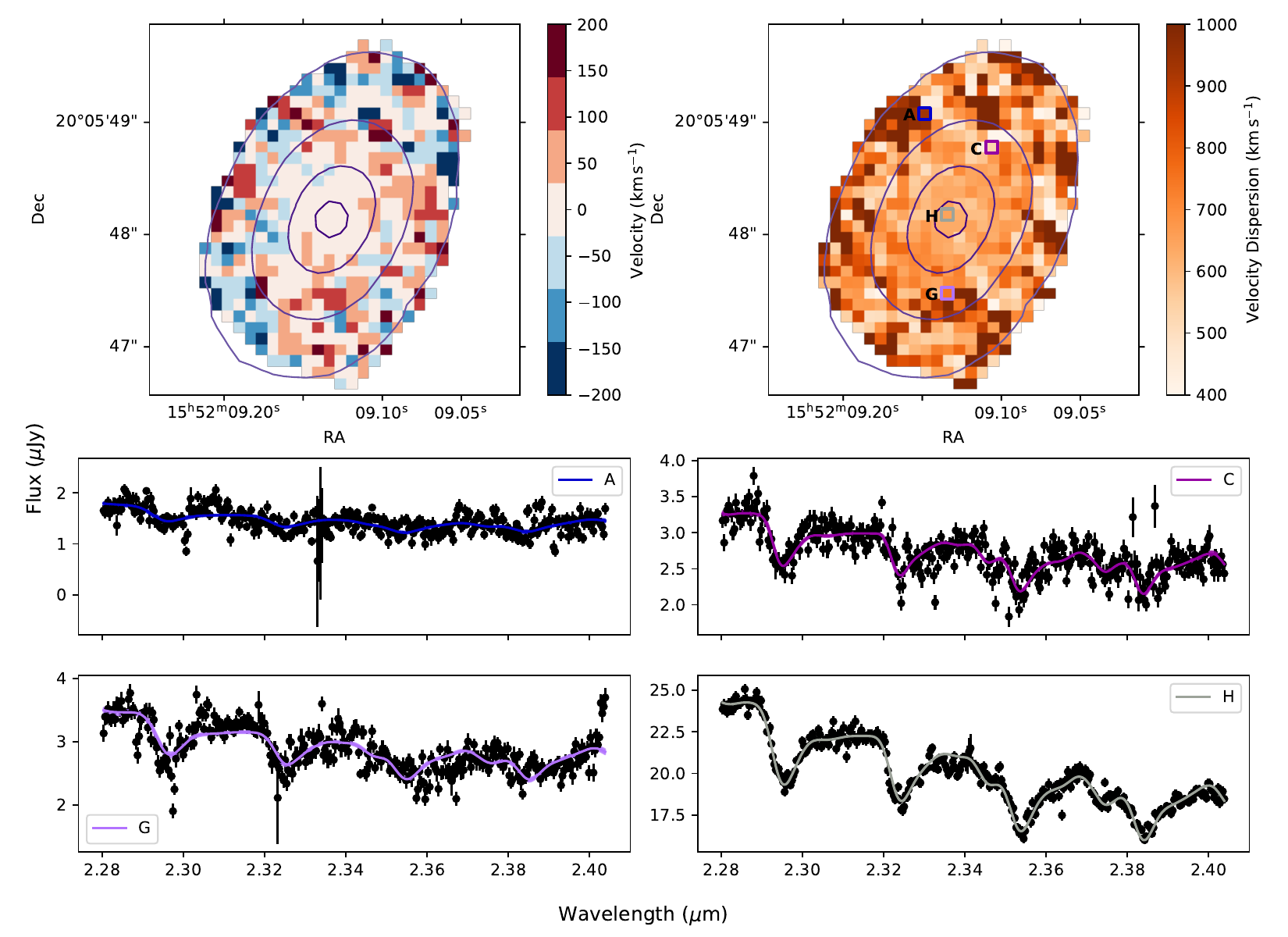}

    \caption{Top left: The CO band head velocity map from our template fit. For clarity, while we model all spaxels, here we only plot the velocities for spaxels with a average SNR $>10$. Top right: The FWHM of the 
    velocity dispersion for the CO features, the spaxels with an average SNR $<10$ 
    are omitted. Overplotted as coloured squares are the positions of the four spaxels displayed beneath. The spaxel positions and labels match Fig.\,\ref{fig:IFUlineComp}. Overplotted on both top plots in purple are contours of the near-IR continuum emission in Fig.\,\ref{fig:IRcontmorphology}. Bottom: The model fit to the CO features in four selected spaxels. The filled regions are the 1$\,\sigma$ uncertainties.}
    \label{fig:coFit}
\end{figure*}

To model the CO band head absorption features, we used a template provided by \citet{winge_gemini_2009} instead of a Gaussian line profile. The templates are of single stars; to adjust for a stellar population we assume the distribution of velocities is normal. We convolve the template spectrum with a Gaussian with a standard deviation of the stellar velocity dispersion, $\sigma_\mathrm{SVD}$, and resample it to the spectral resolution of the NIRSpec spectrum using the python package \textsc{spectres} \citep{carnall_spectres_2017}. We then multiply the spectrum by a 2$^\mathrm{nd}$ order polynomial continuum to correct for the normalisation applied by \citet{winge_gemini_2009}. We compare the adjusted template to the data between the observed wavelengths of 2.485$\,\mu$m and 2.62$\,\mu$m which contains only CO band head features and no emission features (Fig.\,\ref{fig:coFit}). Similarly to the emission lines, the instrumental line spread function (R=2700) is accounted for in the given velocity dispersions.

To find the best template to reproduce our data, we compare every template to the highest signal-to-noise {spaxel} in the NIRSpec dataset. We find that the template star of HD\,9138, a K4III star, is a very good reproduction of our data without additional spectral types. We proceed to fit only the HD\,9138 template to every spaxel. We generate an initial position for the walkers using a normal distribution around the best fit parameters of the highest signal-to-noise spaxel fit with a standard deviation of 1\%. We use 2000 steps and 100 walkers when performing the fitting.

An example of the model fitting can be seen for four selected spaxels in Fig.\,\ref{fig:coFit}. It can be seen that the one template can reproduce our observed data very well. 

\section{Analysis}
\label{sec:analysis}

\subsection{Properties of the radio core}
\label{ssec:radiocore}

\begin{figure*}
    \centering
    \includegraphics[width=0.32\textwidth]{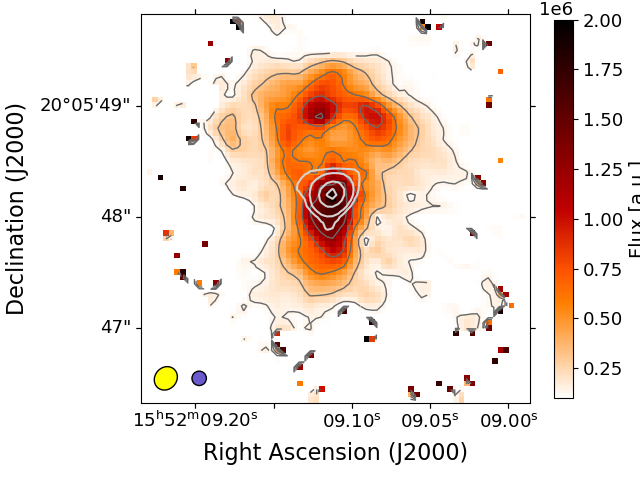}
    \includegraphics[width=0.32\textwidth]{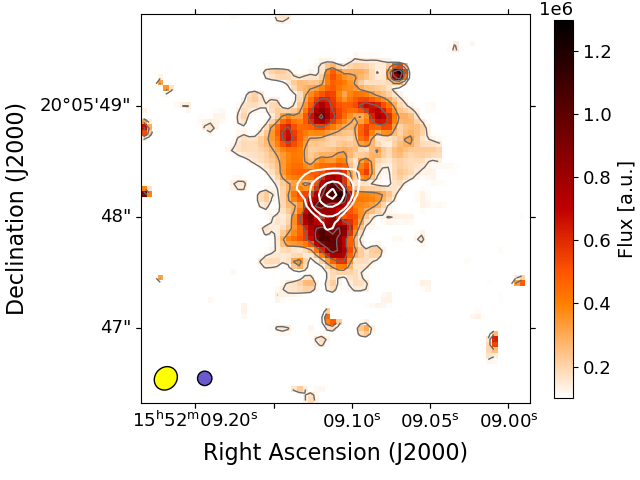}
    \includegraphics[width=0.32\textwidth]{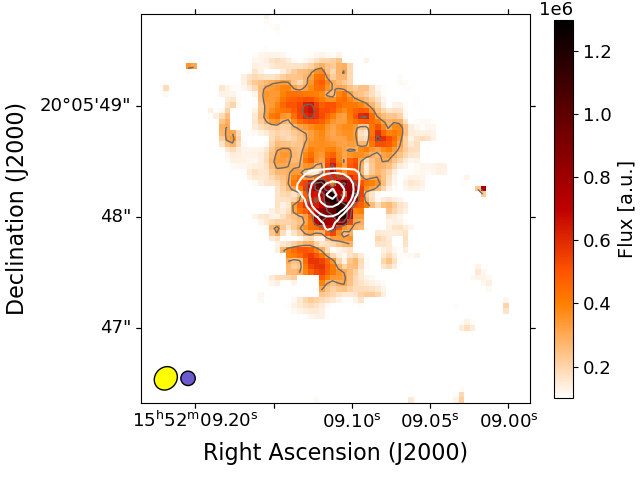}\\
    \includegraphics[width=0.32\textwidth]{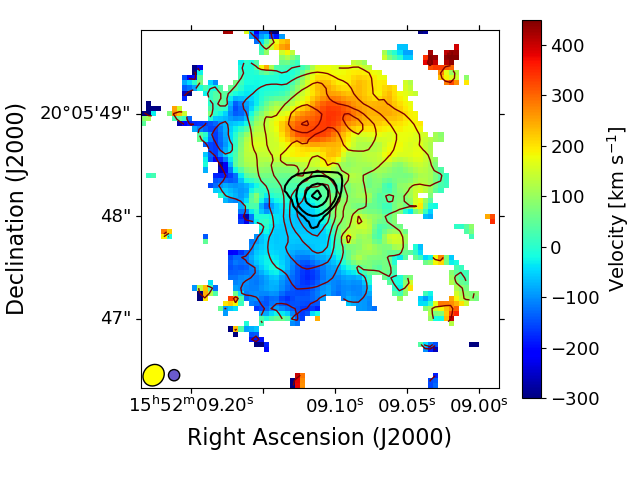}
    \includegraphics[width=0.32\textwidth]{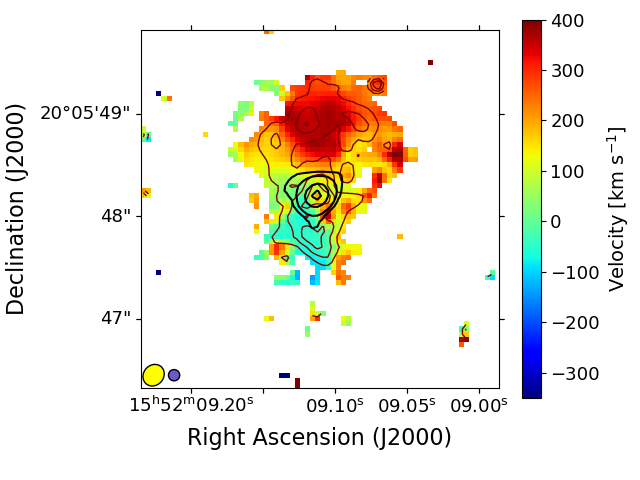}
    \includegraphics[width=0.32\textwidth]{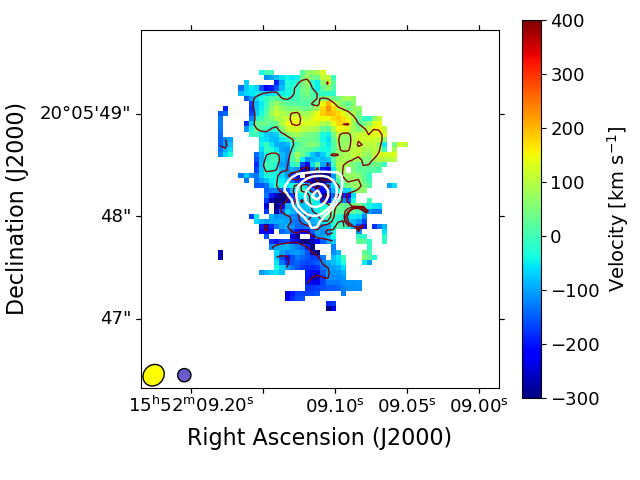}\\
    \includegraphics[width=0.32\textwidth]{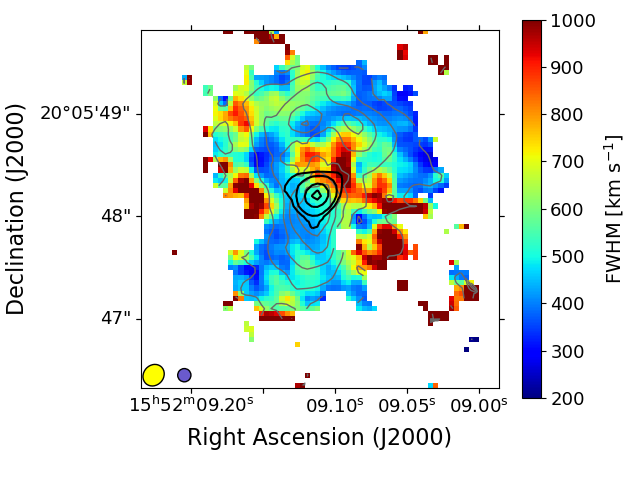}
    \includegraphics[width=0.32\textwidth]{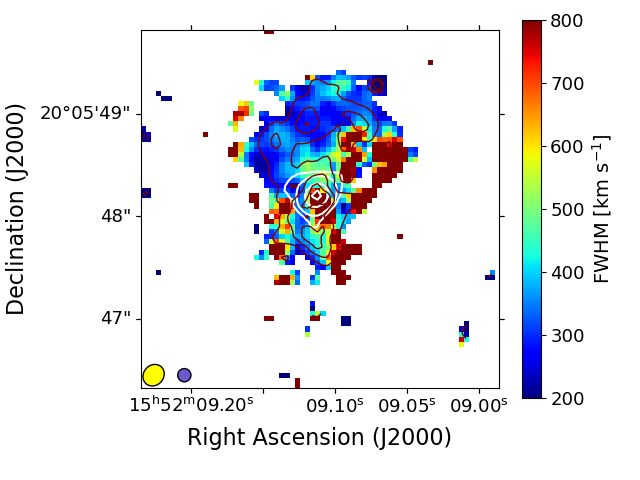}
        \includegraphics[width=0.32\textwidth]{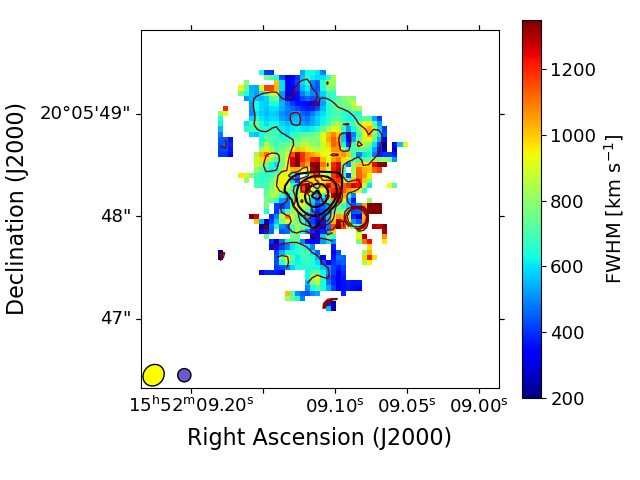}\\
    \caption{H$_2$ 1--0 S(3) emission-line morphology {\it (top)}, velocities {\it (centre)} and FWHM line widths {\it(bottom)} with the JVLA 8.4~GHz radio continuum shown as thick contours near the centre. Thin contours show the emission-line morphology in the top left panel to ease orientation. {\it (left to right):} Single-component fit, systemic line component and blueshifted line component. The emission-line morphology and single-component and systemic velocity maps show an hourglass gas morphology with a strongly blue and redshifted bipolar velocity field, consistent with a pair of bubbles that are being inflated by the radio source at the centre. The yellow ellipse and blue circle in this and subsequent figures indicate the beam size of the JVLA and the PSF of the JWST data set.} 
    \label{fig:emlinemorphology10}
\end{figure*}

The GHz radio morphology of 3C~326~N is compact on scales of few arcsec in all three bands that we observed. As an example, the 10~GHz radio morphology is shown as circumnuclear contours in Fig.~\ref{fig:emlinemorphology10}. Radio synchrotron emission follows a simple power law, where the flux, $f$, is set by the frequency, $\nu$, and a spectral index, $\alpha$,
as $f \propto \nu^{-\alpha}$. With the values for the radio flux and central frequencies given in Tab.~\ref{ssec:jvalobs}, we find a $\alpha=0.21$ between 1.52~GHz and 10.0~GHz, a typical value for the flat-spectrum radio core of a radio AGN. Although we obtained fairly deep data, with RMS=90~$\mu$Jy at 1.52~GHz, and RMS$\sim$5$\mu$Jy at 7.5 and 10~GHz, we did not detect diffuse emission around the nucleus in any band. 

To investigate whether we may have missed a diffuse component, we compare the flux measured at 1.52~GHz, $S_{1.52}=2.10\pm0.09$~mJy, with that of FIRST \citep[][]{becker_first_1995,white_catalog_1997}, $S_{1.4}=3.94\pm0.14$~mJy, finding that 47\% of the flux measured with FIRST at 5.4\arcsec\ beam size are missing in our data obtained with a 1.2\arcsec\ beam. Assuming that such emission would be of uniform surface brightness and beam-filling in the FIRST data, we find that we would need to probe down to a  surface brightness of 95~$\mu$Jy in the L-band, which corresponds to our RMS, so that it is well possible that faint, diffuse emission is present but not detected. In the higher frequency bands, due to the much smaller beam size, and likely steep spectral index, such an extended component would also have been missed. The 5.4\arcsec\ beam of FIRST is slightly larger than the molecular disk in 3C~326~N. 

A similar situation was encountered by \citet{zovaro_jets_2019}, who identified ro-vibrational line emission from molecular Hydrogen in the z=0.06 radio galaxy 4C31.04 using GEMINI/NIFS, which extended over much larger radii than the radio source. Their detailed comparison with the relativistic hydrodynamic simulations of \citet{mukherjee_relativistic_2018} showed a significant component of diffuse radio emission which extends well beyond the brightest regions, but would require a dynamic range much greater than the factor 100 or so possible with current radio facilities to be probed. The same arguments apply to our present case.

\begin{figure*}
    \centering
    \includegraphics[width=\textwidth]{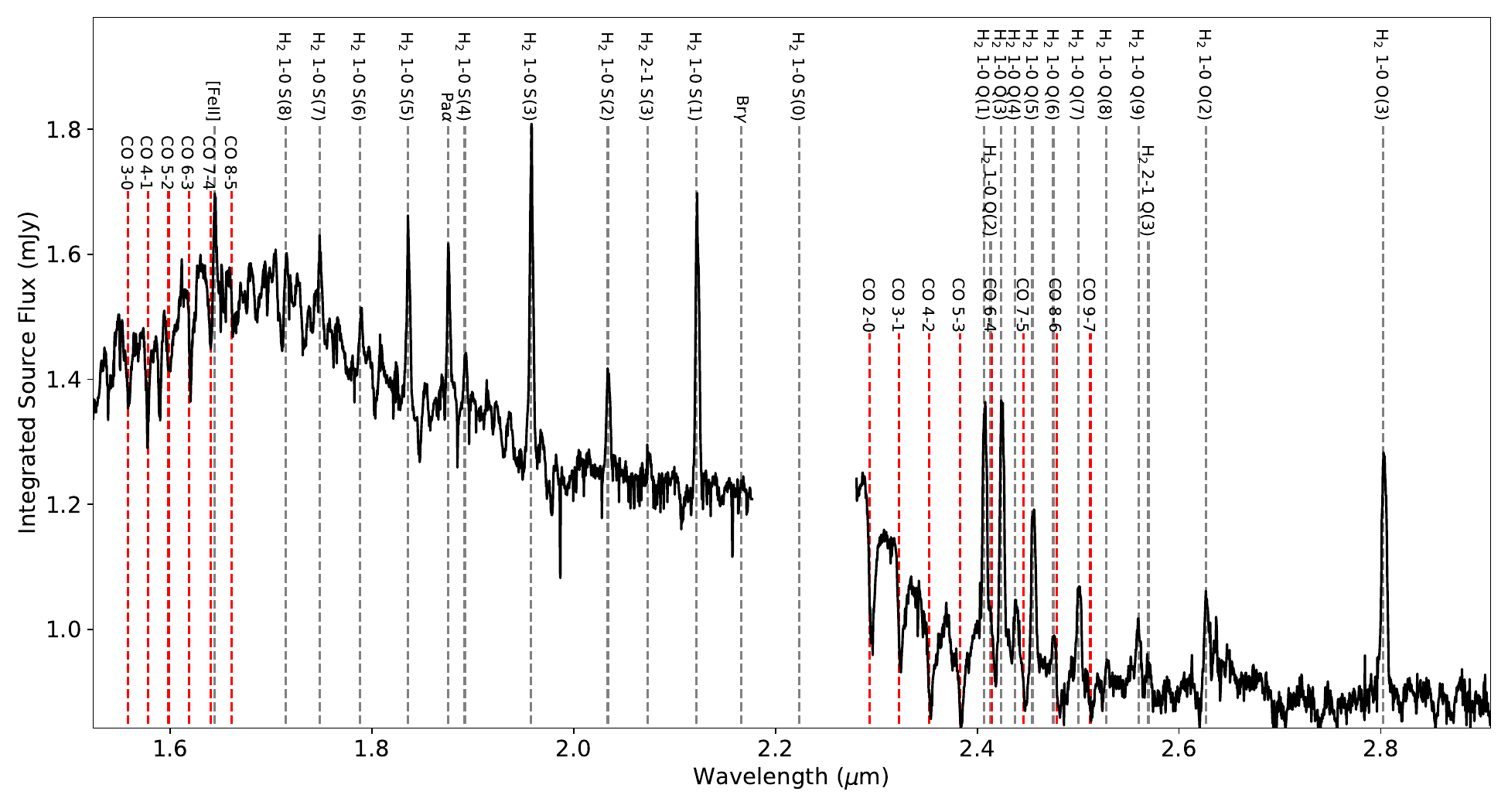}
    \caption{The source-integrated spectrum of 3C~326~N from the NIRSpec observation from 1.5 to 2.9$\mu$m showing most of the lines. The flux was extracted with an elliptical aperture centred on the peak continuum flux with a major axis along Dec of 1.1" and a minor axis of 0.9". Empty pixels were not included and the pixels within the aperture were sigma clipped with a tolerance of 5\,$\sigma$. The spectral gap at 2.2\,$\mu$m is caused by the NIRSpec IFU detector gap.}
    \label{fig:nirspec_intspec}
\end{figure*}

\begin{figure*}
    \centering
    \includegraphics[width=\textwidth]{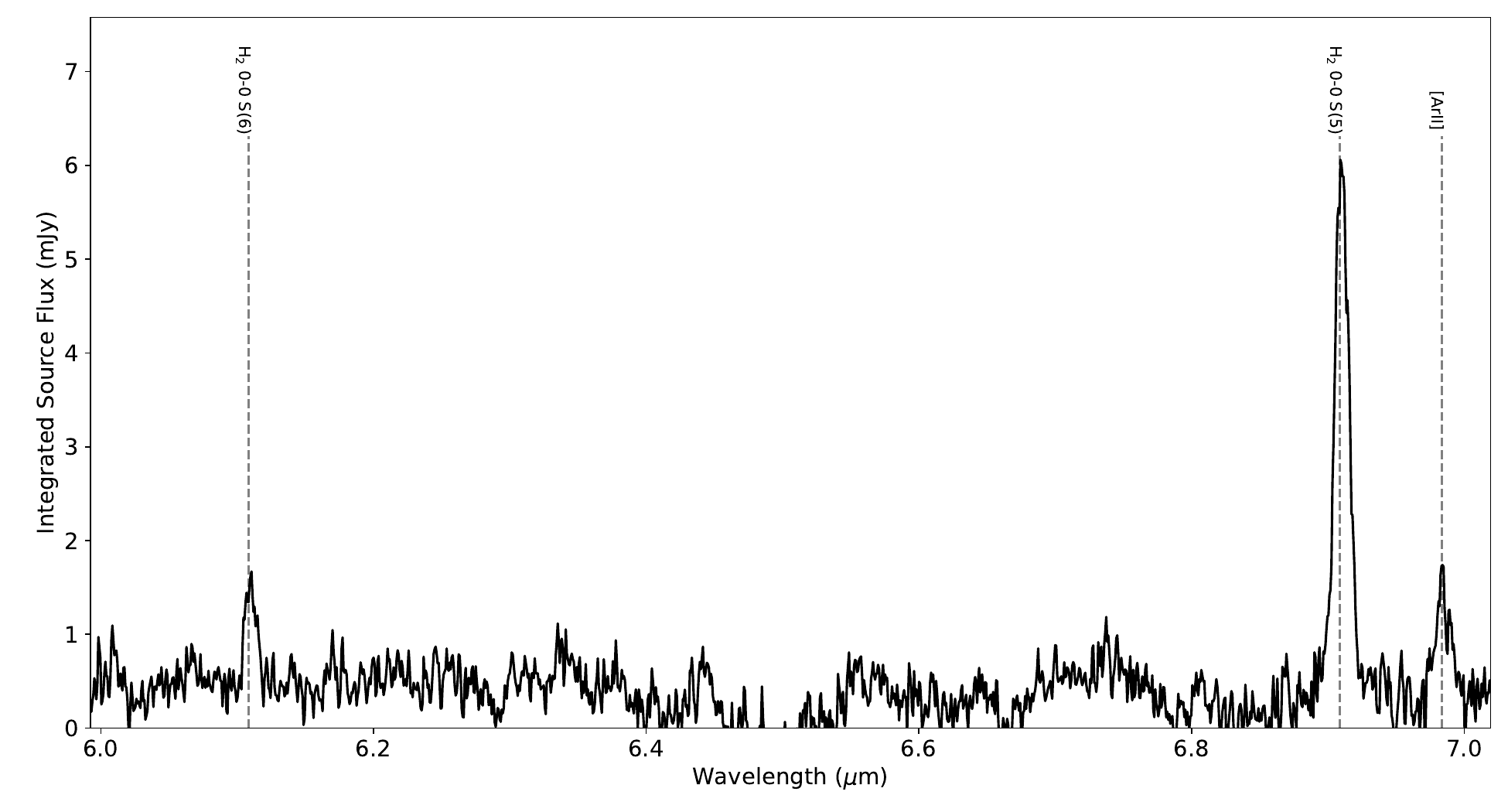}
    \caption{The integrated source spectra from the MIRI MRS Channel 1 observation. The flux was extracted with an elliptical aperture centred on the peak continuum flux with a major axis along Dec of 1.1" and a minor axis of 0.9". Empty pixels were not included and the pixels within the aperture were sigma clipped with a tolerance of 5\,$\sigma$.}
    \label{fig:Ch1_intspec}
\end{figure*}

\begin{figure}
    \centering
    \includegraphics[width=0.48\textwidth]{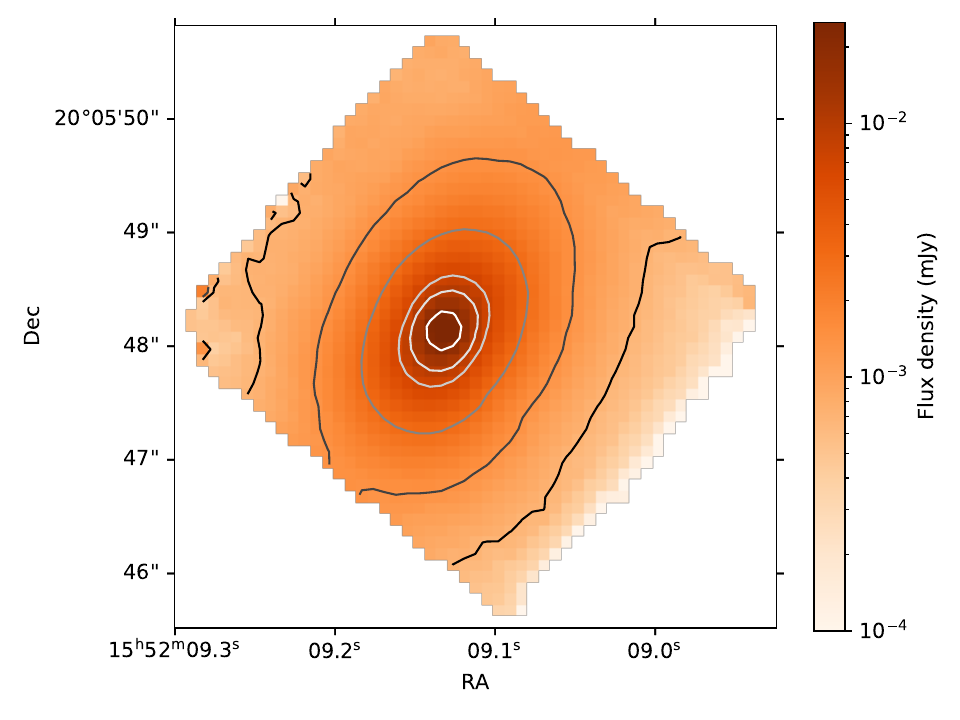}
    \caption{The 1.8$\,\mu$m continuum with contours overplotted.}
    \label{fig:IRcontmorphology}
\end{figure}

\begin{figure*}
    \centering
    \includegraphics[width=0.32\textwidth]{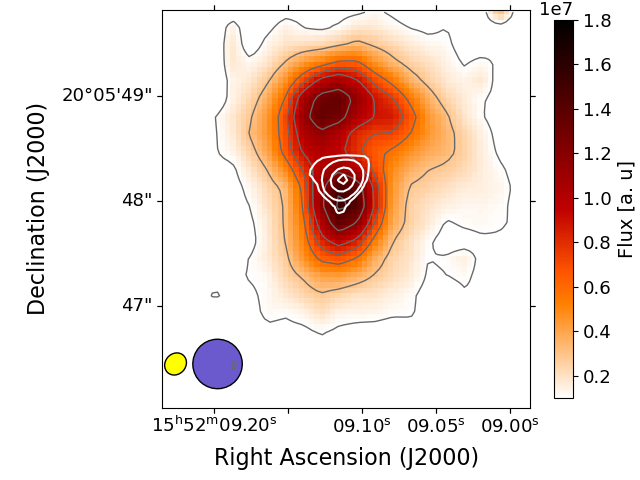}
    \includegraphics[width=0.32\textwidth]{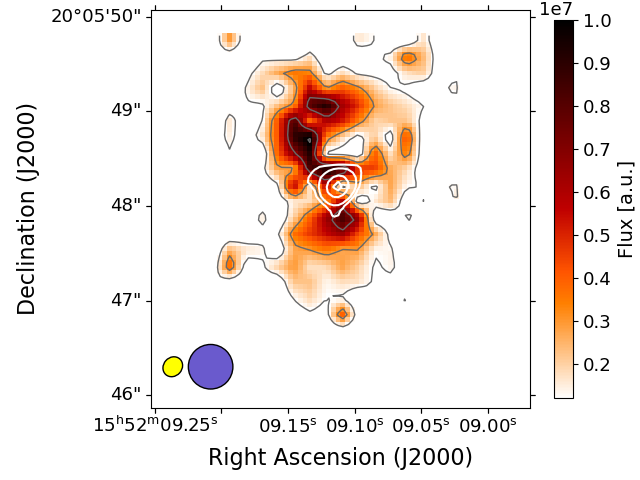}
    \includegraphics[width=0.32\textwidth]{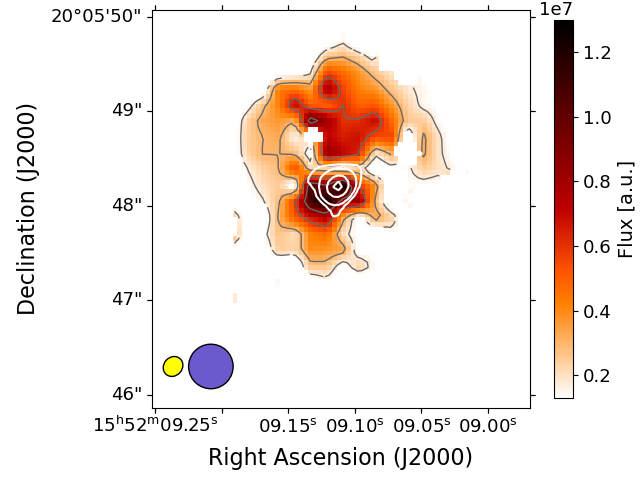}\\
    \includegraphics[width=0.32\textwidth]{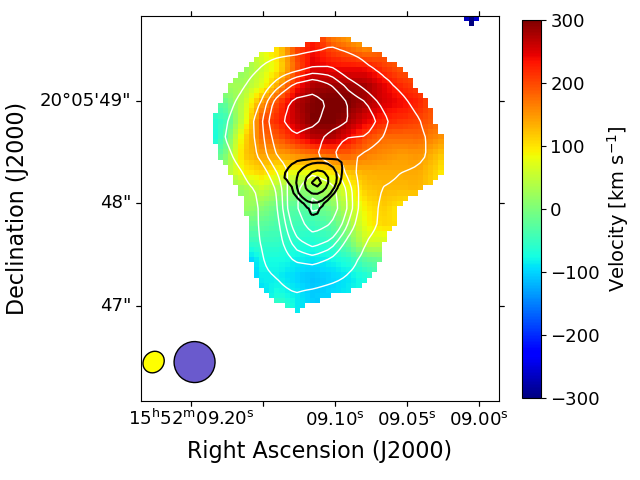}
    \includegraphics[width=0.32\textwidth]{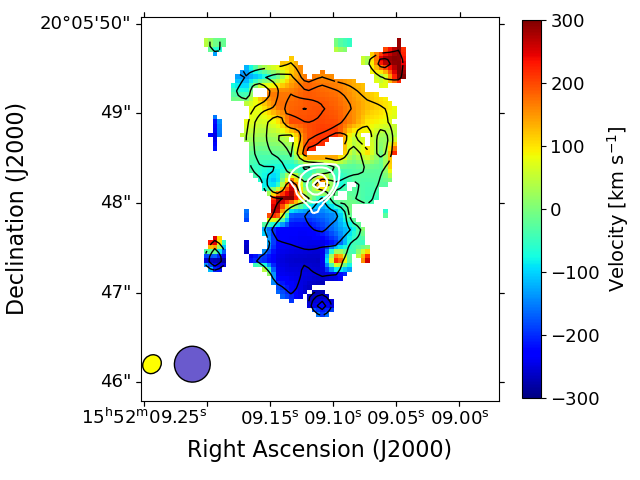}
    \includegraphics[width=0.32\textwidth]{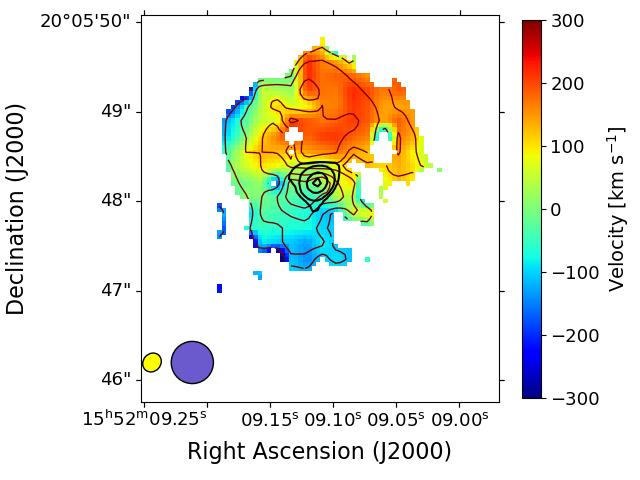}\\
    \includegraphics[width=0.32\textwidth]{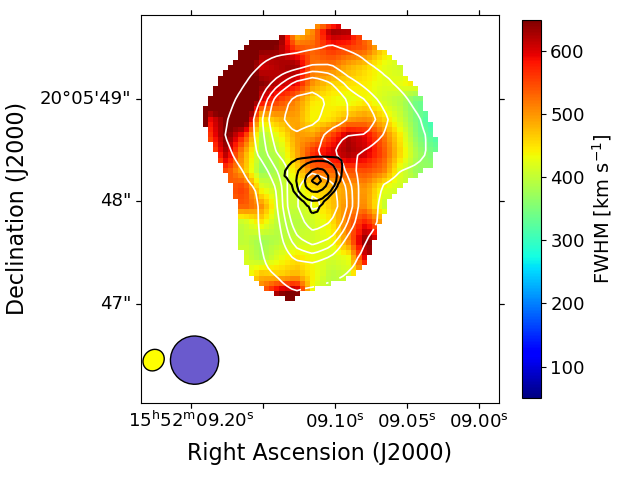}
    \includegraphics[width=0.32\textwidth]{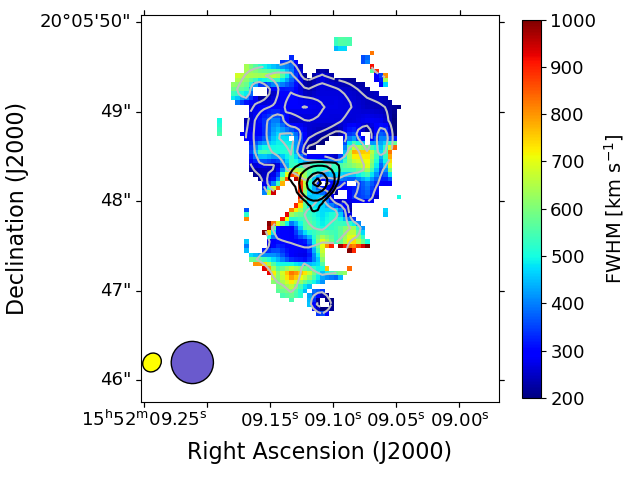}
    \includegraphics[width=0.32\textwidth]{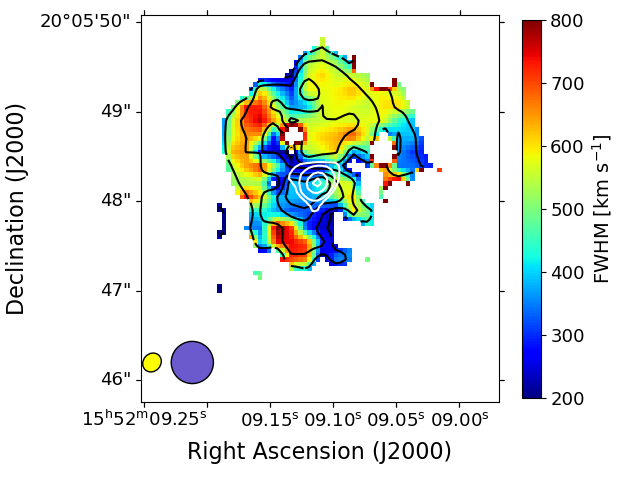}
    \caption{H$_2$ 0--0 S(3) emission-line morphology {\it (top)}, velocities {\it (centre)} and FWHM line widths {\it(bottom)} with the JVLA 8.4~GHz radio continuum shown as thick contours near the centre. Thin contours show the emission-line morphology in the top left panel to ease orientation. {\it (left to right):} Single-component fit, systemic line component and blueshifted line component. The emission-line morphology and single-component and systemic velocity maps show an hour-glass gas morphology with a strongly blue and redshifted bipolar velocity field, consistent with a pair of bubbles that are being inflated by the radio source at the centre. } 
    \label{fig:emlinemorphology00}
\end{figure*}

\begin{figure*}
    \centering
    \includegraphics[width=0.32\textwidth]{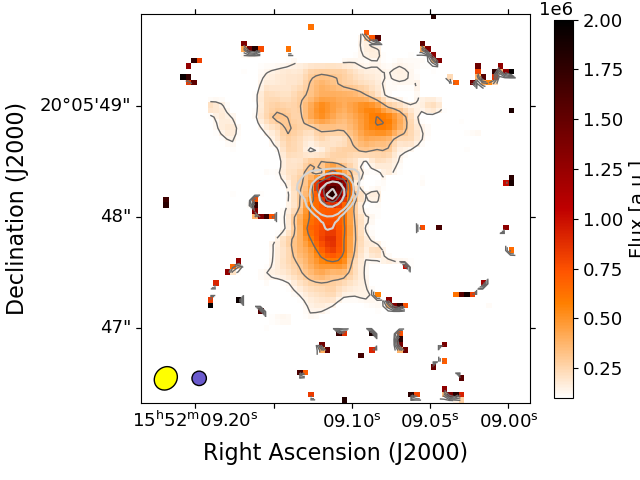}
    \includegraphics[width=0.32\textwidth]{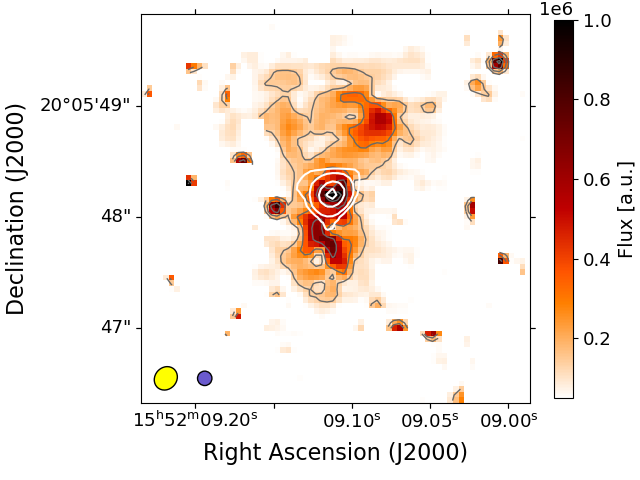}
    \includegraphics[width=0.32\textwidth]{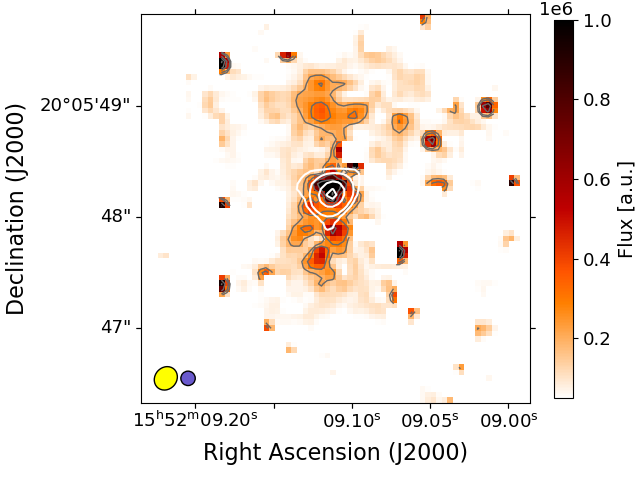}\\
    \includegraphics[width=0.32\textwidth]{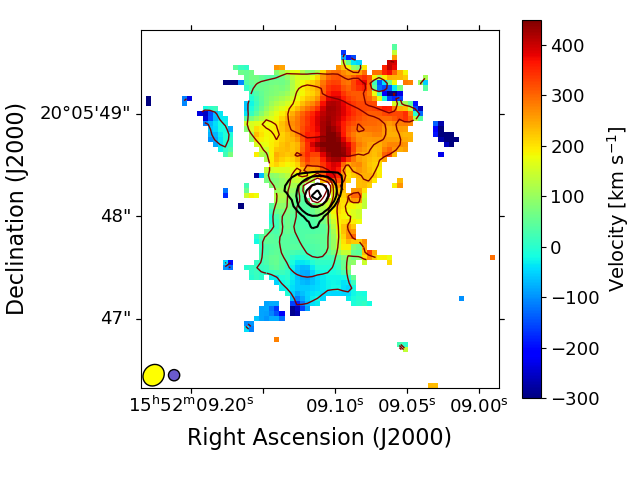}
    \includegraphics[width=0.32\textwidth]{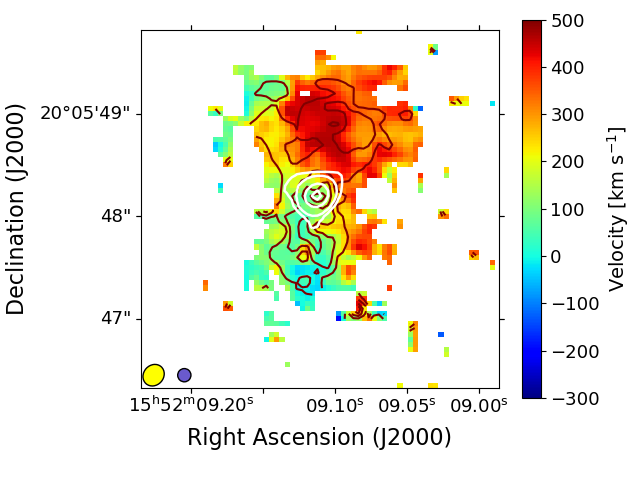}
    \includegraphics[width=0.32\textwidth]{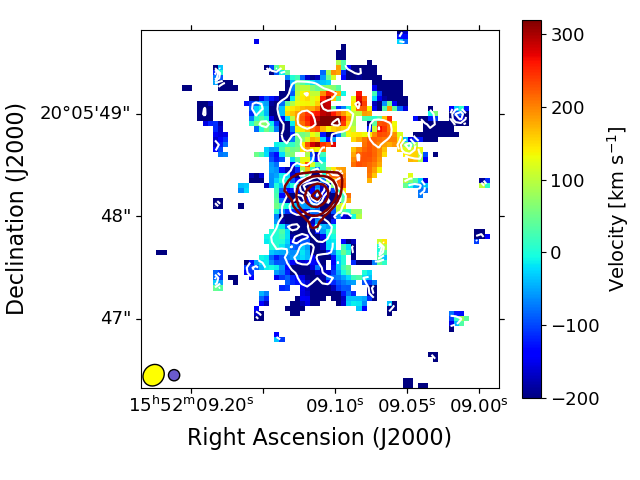}\\
    \includegraphics[width=0.32\textwidth]{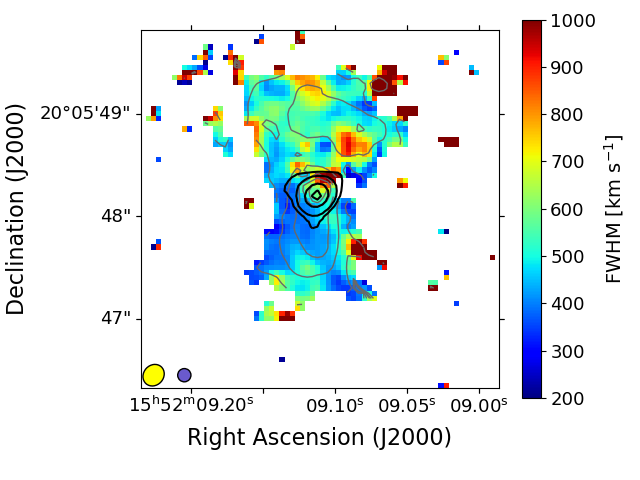}
    \includegraphics[width=0.32\textwidth]{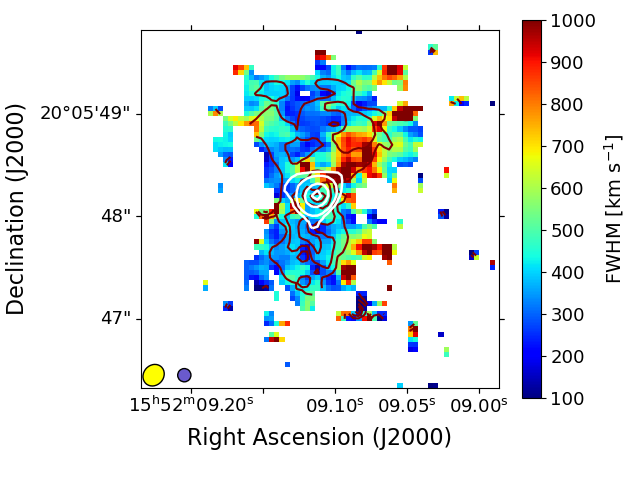}
    \includegraphics[width=0.32\textwidth]{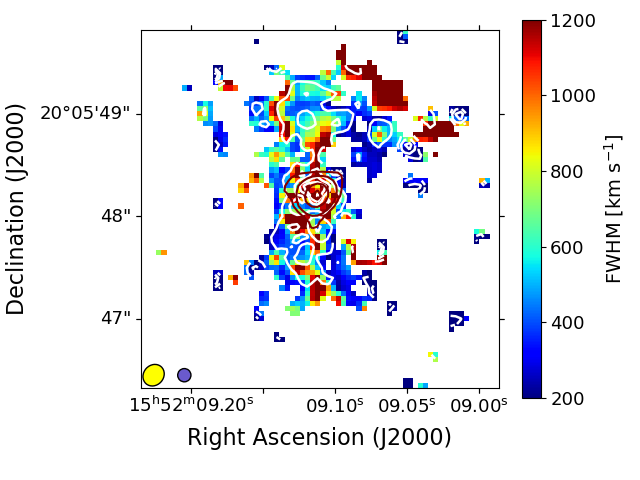}
    \caption{Pa$\alpha$ emission-line morphology {\it (top)}, velocities {\it (centre)} and FWHM line widths {\it(bottom)} with the JVLA 8.4~GHz radio continuum shown as thick contours near the centre. Thin contours show the emission-line morphology in the top left panel to ease orientation. {\it (left to right):} Single-component fit, systemic line component and blueshifted line component. The emission-line morphology and single-component and systemic velocity maps show an hour-glass gas morphology with a strongly blue and redshifted bipolar velocity field, consistent with a pair of bubbles that are being inflated by the radio source at the centre.} 
    \label{fig:emlinemorphologyPa}
\end{figure*}

\begin{figure*}
    \centering
    \includegraphics[width=0.32\textwidth]{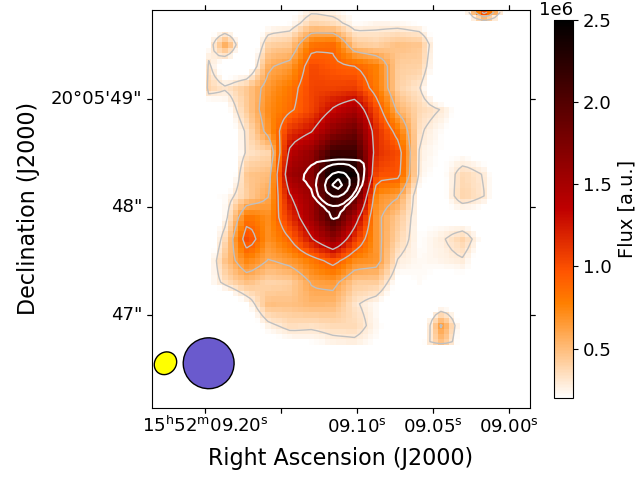}
    \includegraphics[width=0.32\textwidth]{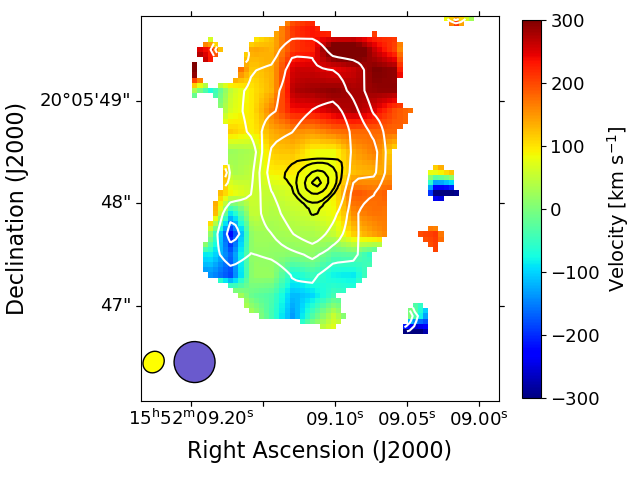}
    \includegraphics[width=0.32\textwidth]{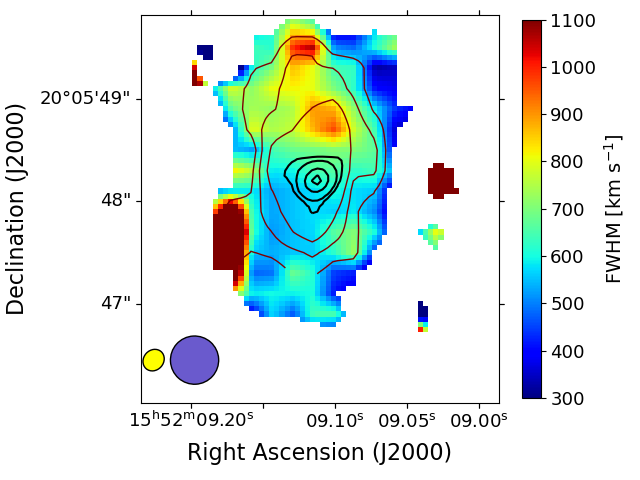}\\
    \caption{{One Component }[NeIII]$\lambda$15.56 emission-line morphology {\it (left)}, velocities {\it (centre)} and FWHM line widths {\it(bottom)} with the JVLA 8.4~GHz radio continuum shown as thick contours. {Due to strong fringing in MIRI channel 3, the blueshifted line, while present, is noisy and difficult show clearly. Therefore, we do not plot the two component fit.}}
    \label{fig:linemapsniii}
\end{figure*}

\begin{figure*}
    \centering
    \includegraphics[width=0.32\textwidth]{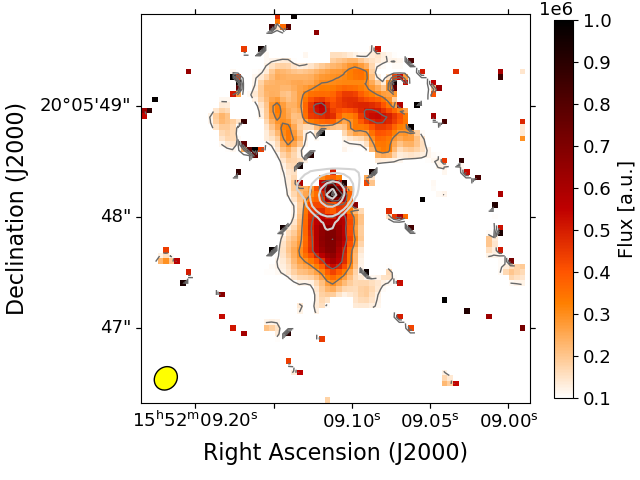}
    \includegraphics[width=0.32\textwidth]{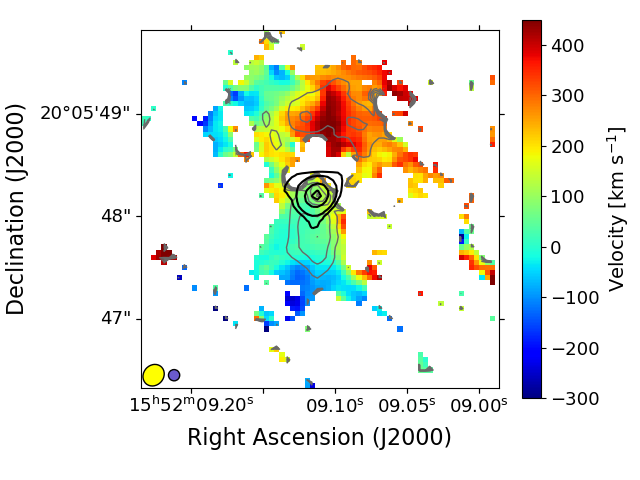}
    \includegraphics[width=0.32\textwidth]{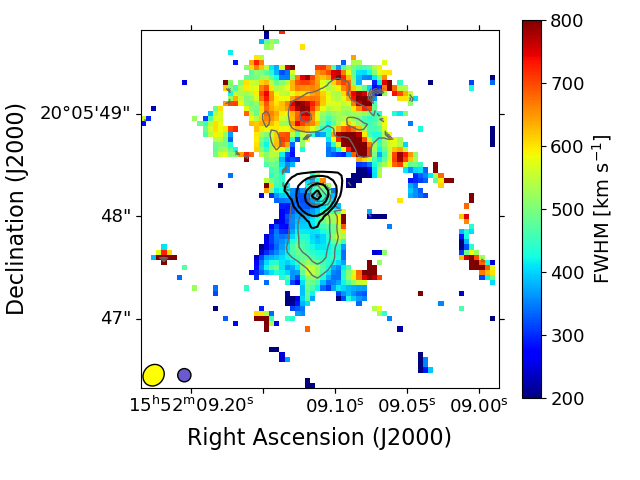}
    \caption{[FeII]$\lambda$1.644 emission-line morphology {\it (top)}, velocities {\it (centre)} and FWHM line widths {\it(bottom)} with the JVLA 8.4~GHz radio continuum shown as thick contours near the centre. Thin contours show the emission-line morphology in the top left panel to ease orientation. {\it (left to right):} Single-component fit, systemic line component and blueshifted line component. The emission-line morphology and single-component and systemic velocity maps show an hour-glass gas morphology with a strongly blue and redshifted bipolar velocity field, consistent with a pair of bubbles that are being inflated by the radio source at the centre.} 
    \label{fig:emlinemorphologyFe}
\end{figure*}

\subsection{Emission line and CO band head detections}
\label{ssec:emlinedetections}

We show the source-integrated spectrum obtained with NIRSpec in Fig.~\ref{fig:nirspec_intspec} and channel 1 of MIRI MRS in Fig.\,\ref{fig:Ch1_intspec}. As has already been pointed out by \citet{nesvadba_dense_2011}, the spectrum is strongly dominated by the ro-vibrational emission lines of molecular Hydrogen. In NIRSpec we cover the 1--0 S and the 1--0 Q series from S(0) to S(7) and Q(1) to Q(9). In NIRSpec, we also identify the 1--0 O(2) and O(3) lines, and potentially the 2--1 S(3) and Q(3) lines. With MIRI we detect the rotational lines, H$_2$ 0--0 S(3), S(5), and S(6), previously observed by \citet{ogle_shocked_2007}, as well as [ArII]$\lambda$6.985 and [NeIII]$\lambda$15.56. \citet{ogle_shocked_2007} detected additional forbidden lines of [FeII]$\lambda5.34$, [SIV]$\lambda$10.51, [NeII]$\lambda$12.81, [SIII]$\lambda$18.71, and [OIV]$\lambda$25.89, that are not covered by our data, but to our knowledge, these are the first detections of [ArII]$\lambda$6.985 and [NeIII]$\lambda$15.56. Other lines of the 0--0 S series are not covered by our setup. 

Moreover, with NIRSpec we identify the warm ionised gas line Pa$\alpha$ as well as the iron line [FeII]$\lambda$1.644. [FeII]$\lambda$1.599 is tentatively detected but falls within the strong CO 5--2 band head absorption feature. Br$\gamma$ is covered, but not detected, probably because the interstellar line emission is filling in the stellar Br$\gamma$ absorption line. At R$\sim$2700, the spectral resolving power of NIRSpec does not allow us to robustly isolate these two components.

In addition to these emission lines originating from interstellar gas, we also detect the $\Delta v = 2$ and $\Delta v =3$ stellar absorption-line series of $^{12}$C$^{16}$O, i.e., eight lines between $v=2-0$ and $v=9-7$ at 2.29 to 2.51~$\mu$m, and six absorption lines between $v=3-0$ to $v=8-5$ at 1.56 to 1.66~$\mu$m. We subsequently use the absorption lines between 2.3$\,\mu$m and 2.4$\,\mu$m to determine the systemic redshift and search for evidence of rotation of the stellar component. We do not use the redward lines due to contamination from the H$_2$ 1--0 Q series of emission lines and our templates do not cover the bluer absorption lines.

While a thorough analysis of all detected emission lines is beyond the scope of this paper, we provide the source integrated fluxes from our spatially resolved 1--2 Gaussian component model fitting described Sect.\,\ref{sec:LineFit} in Tab.\,\ref{tab:emissionlinesflux}. {In Tab.\,\ref{tab:upperlims}, we state 5\,$\sigma$ upper limits to Br$\gamma$ and the 6.2$\,\mu$m PAH feature which were covered by our data but not detected. As a courtesy to the reader we also perform single Gaussian fits to the source integrated spectrum and provide the source-integrated redshift and FWHM line widths of all emission lines except H$_2$ 1--0 Q(2) and H$_2$ 1--0 Q(8) in Tab.~\ref{tab:IntPar}. The lines are more complex than a single Gaussian and, therefore, the given values are only to give the reader an estimate of the integrated line properties. For fluxes, we recommend only those from Tab.\,\ref{tab:emissionlinesflux} be used. The 2 omitted Q series lines become blurred with other spectral features in the source integrated spectrum. However, they are distinct in the spatially resolved spectrum. }

\subsection{Continuum morphology and stellar absorption bands}
\label{ssec:continuum}

3C~326~N has a well extended continuum morphology clearly observable in the near-IR. We plot the continuum morphology at 1.8$\,\mu$m in Fig.\,\ref{fig:IRcontmorphology}. The extended emission is elliptical with a major axis position angle of $\sim155^\circ$. 

Since NIRSpec covers the $\Delta v = 2$ and $\Delta v =3$ series of CO band heads, we can use these absorption features to derive the systemic redshift and stellar velocity dispersion. Stellar absorption lines are the most robust tracers of the systemic redshift, because the stellar component dominates the mass budget within the central few kpc of early-type galaxies \citep[$3\times 10^{11}$~M$_{\odot}$ in 3C~326~N compared to $2\times 10^9$~M$_{\odot}$ in molecular gas][]{nesvadba_energetics_2010}, and because it is not affected by hydrodynamic processes like AGN or starburst-driven winds, which may significantly impact the gas kinematics. This is a particular worry in case of radio galaxies like 3C~326~N. 

When we model the CO band heads, we find no evidence of a velocity gradient across the galaxy (see Fig.\,\ref{fig:coFit}). This suggests that the galaxy stellar population is not rotationally dominated. Thus, 3C~326~N resembles a classical "slow rotator" as is typical for a massive galaxy of few $10^{11} M_{\odot}$ of stellar mass \citep[e.g.][]{emsellem_atlas3d_2011}. We also find no significant change in the FWHM of the model stellar velocity distribution (SVD) with radius. Therefore, we can find the systemic redshift by merging all posterior probability distributions of each spaxel fit. To reduce uncertainty, we merge the parameter probability distributions for all spaxels with an average wavelength bin SNR $>10$ in the modelled data to find the systemic redshift and its uncertainty. 

We find a systemic redshift of $z_{syst}=0.08979\pm0.0003$, consistent with the value of $z=0.09$ from \citet{guillard_exceptional_2015} measured from [CII]. \citet{ogle_shocked_2007} previously measured $z=0.089\pm0.001$ with Spitzer from the pure-rotational lines of warm H$_2$. \citet{nesvadba_dense_2011} found $z=0.0903\pm0.0002$ from ro-vibrational lines and Pa$\alpha$ measured with VLT/SINFONI. We use our value as the systemic redshift for all other velocity plots. From our CO modelling, we also find that the FWHM of the CO band head velocity dispersion is $5.1 \pm0.8$~nm, equivalent to a velocity of $660\pm 100$~km s$^{-1}$.

\subsection{Emission-line morphology and kinematics}
\label{ssec:emlinemorphology}
\subsubsection{Ro-vibrational line emission of H$_2$}

We will start our discussion of the emission-line morphology and kinematics with the properties of the ro-vibrational H$_2$ 1--0 lines observed with NIRSpec, using the S(3) line as reference for this set of lines. S(3) is amongst the brightest lines we detected, not {noticeably} contaminated by other nearby lines, and in the blue part of the spectral range covered by our data, thus maximising spatial resolution. {H$_2$ 1--0 S(3) can be contaminated by [SiVI]$\lambda$1.962; however, we note that we find similar modelling results in all high bright H$_2$ lines, including in MIRI, and that we find no obvious contamination in this line. Fig.\,\ref{fig:0010ratio} shows comparisons between different H$_2$ lines and shows no significant variations in line profile for H$_2$ 1--0 S(3).} Other H$_2$ 1--0 lines {also} show the same spatial distribution and kinematics within the measurement uncertainties. 

We find that the molecular hydrogen lines show a clear North-South morphology shown in Fig.\,\ref{fig:emlinemorphology10} roughly centred on the galaxy nucleus as indicated by the peak in surface brightness of the stellar continuum emission. Bright line emission is probed over 2.0\arcsec\ in North-South direction, and over 0.5\arcsec\ and 1.2\arcsec, in East-West direction in the Southern and Northern part of the source, respectively. North of the nucleus we find a cavity-like structure delineated by several bright, unresolved knots of line emission at the far (Northern) end. This cavity and the structure of the components themselves are not mirrored in the continuum supporting the idea that it is not caused by stellar emission. 
South of the nucleus, the emission-line morphology is more regular and featureless, and extends to about 0.8\arcsec\ South from the nucleus.

We fit Gaussian line profiles to the line emission in all spaxels as detailed in Sect.\,\ref{sec:LineFit}. Examining the line profiles (e.g.,~Fig.~\ref{fig:IFUlineComp}), we find that fits with two Gaussians are preferred in almost all regions of the galaxy over single-Gaussian fits. {The exception to this is} the lowest surface-brightness regions where the signal-to-noise is not sufficient to fit two lines components. Some areas close to the nucleus can also be explained by a single component where the systemic redder component becomes dominant. For simplicity, we focus most of our discussion on the two-component modelling. 
{In this discussion, we} distinguish the two components as a redder (higher z) systemic (or "primary") component and a bluer (lower z, or "secondary"){, often broader,} component. The line profiles, quality of the spectra and Gaussian fits can be seen in Fig.~\ref{fig:IFUlineComp}.

Generally speaking, the gas velocities in both Gaussian components show a monotonous gradient from blueshifted in the South towards more redshifted in the North (Fig.~\ref{fig:emlinemorphology10}). The {systemic} Southern component shows redshifted velocities along its Western edge. As can be seen from the left panel of Fig.~\ref{fig:emlinemorphology10}, the same trend can be seen in the single-component fit, although the gradient is smaller and less pronounced, because of the lower quality of the fit stemming from fitting a double-peaked profile with a single Gaussian. 

The secondary component, which is always blueshifted relative to the first, shows the same trend between blueshifted gas South, and redshifted gas North, but is blueshifted by a value of {{100-200\,km\,s$^{-1}$} relative to the first component. Typical line FWHM are $\sim400-1000$\,km\,s$^{-1}$ for the bluer component and $\sim200-400$\,km\,s$^{-1}$ for the redder. Other H$_2$ 1--0 lines have similar morphologies and kinematics within the measurement uncertainties. 

\subsubsection{Rotational line emission of H$_2$}
The pure rotational lines of H$_2$ were observed with MIRI at about $3-4\times$ less spatial resolution than the ro-vibrational lines with NIRSpec, i.e., with a PSF of 0.3-0.42\arcsec. We will also discuss these lines on the example of the H$_2$ $0-0$ S(3) line, which is at a rest-frame wavelength of $\lambda=9.66\ \mu$m. These lines are interesting to study in parallel to the H$_2$
1--0 lines, because they probe the dominant molecular gas component in this galaxy (Sect.~\ref{ssec:gasmass}).

The integrated emission-line morphology is shown in Fig.~\ref{fig:emlinemorphology00}. H$_2$ $0-0$ S(3) line emission is also clearly elongated along North-South direction with a projected size of 2.0\arcsec. In the Northern and Southern part of the source, the East-West extent is 1.0\arcsec, and 0.5\arcsec, respectively. Most of the line emission is originating from the blob South of the nucleus and a bright emission-line region North of it. There is a region of much fainter line emission between the nucleus and the Northern emission-line region, which corresponds to the cavity already seen in H$_2$ 1--0 S(3). Overall, the morphology  and gas kinematics follow closely those already seen in the H$_2$ 1--0 S(3) line, although the blending due to the lower spatial resolution of the data tend to lower the velocity offsets, enhance the line widths, and generally, smooth out small-scale features. We will do a more quantitative comparison between the two lines, including the line kinematics, in Sect.~\ref{sec:highz}, that takes into account the effect of the different beam sizes. We do not observe significant differences between the morphology of the H$_2$ 0--0 S(3) line described here, and the 0--0 S(5) and S(6) lines, which we also covered. 

\subsubsection{Warm ionised gas lines}

NIRSpec and MIRI have also observed several warm ionised gas lines. We discuss here the lines of Pa$\alpha$, [FeII]$\lambda$1.644, and [NeIII]$\lambda$15.56. While it has already been shown conclusively by \citet{ogle_shocked_2007}, \citet{nesvadba_energetics_2010}, and \citet{nesvadba_dense_2011} that the emission lines of warm H$_2$ are mainly heated by shocks driven by the radio source into the gas, the ionisation mechanism of the other lines is not clear. Pa$\alpha$ can be either heated by shocks, AGN photoionisation or UV photons from young stellar populations \citep[for a review, see, e.g.][]{kewley_understanding_2019}, [FeII]$\lambda$1.644 can be either heated by shocks or AGN photoionisation \citep[e.g.][]{rodriguez-ardila_molecular_2004}], and [NeIII]$\lambda$15.56, which has a high ionisation potential of $E_{ion}=40.96$~eV \citep[][]{levesque_ne_2014}, either by AGN photoionisation, or, in extreme cases, star formation \citep[][]{ho_mid-infrared_2007}.  
\citet{spoon_discovery_2009} and \citet{spoon_high-velocity_2009} found that [NeIII]$\lambda$15.56 line emission likely traces circumnuclear gas in nearby ULIRGs that is strongly perturbed through interactions with the AGN.

We can use the Pa$\alpha$ and [FeII]$\lambda$1.644 morphology to conclude that both lines are most likely heated by shocks. Their morphology follows very closely that of the H$_2$ 1--0 and H$_2$ 0--0 lines. A contribution from AGN photoionisation near the nucleus, where Pa$\alpha$ and [FeII]$\lambda$1.644 are brighter than the H$_2$ 1--0 S(3) line, is possible. They follow the North and South components seen in H$_2$ 1--0 S(3), but the North component is comparatively fainter relative to the South and the North component is brighter on the western side. This can be seen in the single component fits or the systemic component of the 2 Gaussian fit in Figs.\,\ref{fig:emlinemorphology10},\ref{fig:emlinemorphology00}, \ref{fig:emlinemorphologyPa}, and \ref{fig:emlinemorphologyFe}; we will discuss this further in Sect.\,\ref{sec:discussion}. 

The [NeIII]$\lambda$15.56 emission, as shown in Fig.\,\ref{fig:linemapsniii}, does extend North-South but it does not show the same Northern bright spot, and its morphology is more symmetric around the nucleus than, e.g., H$_2$ 0--0 S(3), which shows a depression between the nucleus and the Northern clumps co-spatial with the cavity already identified in the H$_2$ 1--0 S(3) line. 

\subsection{Mass budgets}
\label{ssec:gasmass}
3C~326~N has already several detailed mass estimates of different phases of warm and cold gas. \citet{ogle_shocked_2007} and \citet{nesvadba_energetics_2010} estimated a mass of warm H$_2$ from the pure rotational line emission finding M$_{H2,rot}=1.1\times 10^9 M_{\odot}$ and M$_{H2,rot}=1.3-2.7 \times 10^9 M_{\odot}$, respectively. We will in the following use the latter value, which is appropriate for gas at temperatures $T\ge 150$~K, and has taken explicitly into account that the gas is heated by shocks. \citet{nesvadba_energetics_2010} also found M$_{ion}=2\times 10^7$~M$_{\odot}$ for warm ionised gas probed by H$\alpha$. \citet{guillard_exceptional_2015} estimated M$_{CNM}=9.5\times 10^8$~M$_{\odot}$, for gas with $T=100$~K and $n_H=1000$~cm$^{-3}$. \citet{lanz_star_2016} estimated a dust mass of $M_{dust}=6.05^{+0.376}_{-0.218}\times 10^7$ M$_{\odot}$. To complete these different mass estimates, we now also provide an estimate of the warmest H$_2$ observed through the ro-vibrational lines. 

We use the approach of \citet{scoville_velocity_1982} and \citet{mazzalay_molecular_2013} to estimate the mass of very warm H$_2$ from the H$_2$1--0 S(1) line. They set $M_{H2,ro-vib}=5.0875\times 10^{13}\ D^2\ F_{1-0 S(1)}\ 10^{0.4 A_{2.2}}$, where D is the luminosity distance in Mpc, $F_{1-0 S(1)}$ the flux of the H$_2$ 1--0 S(1) line in erg s$^{-1}$ cm s$^{-2}$, and $A_{2.2}$ the extinction at 2.2~$\mu$m. Their estimate is appropriate for gas with T=2000~K, a transition probability of $A_{S(1)} = 3.47 \times 10^{-7}$ s$^{-1}$, and a population fraction in the upper level of 1.22\%. 

We have no estimate of A$_{2.2}$, as Pa$\alpha$ is the only line of warm ionised gas that is detected in our data sets, however, \citet{nesvadba_energetics_2010} obtained a global estimate of $A_V=0.17$~mag from stellar population modelling in the optical, assuming a \citet{calzetti_dust_2000} extinction law. This suggests that extinction in the near- to mid-infrared can be safely ignored, at least for source-integrated estimates like the one we are doing here. We will therefore adopt A$_{2.2}$ = 0~mag.

With these assumptions, and for an integrated H$_2$ 1--0 S(1) line flux of 1.5$\times 10^{-18}$\,W\,m$^{-2}$, we find $M_{H2,ro-vib}=1.4\times 10^{4}$~M$_{\odot}$. This is thus a very small mass component compared to the other estimates, in spite of the bright line luminosity. {The hot gas is expected to only be a small fraction of the total gas mass because it needs to be heated to relatively high temperatures, for example by an AGN or by shocks, and the gas cools rapidly.}

It is also interesting to compare the warm and cold molecular gas mass in 3C~326~N with that 
of other galaxies, to highlight the unusual environment that this galaxy provides. \citet{mazzalay_molecular_2013} {additionally} propose an empirical calibration of warm to cold gas probed by CO(1--0), which was derived empirically by comparing both masses in about 50 low-redshift galaxies spanning a wide range of types. Based on their relationship, we would expect to find a cold molecular gas mass of $9.87\times10^{9}$\,M$_\odot$. In Sect.~\ref{sec:observations} we have estimated a 3$\sigma$ upper limit on the cold molecular gas mass in 3C~326~N of $M_{H2,CO}=8\times 10^8 M_{\odot}$, assuming that cold molecular gas extends over the same area as the warm molecular gas. This is more than an order of magnitude less that what would be expected in normally star-forming galaxies. 

Likewise, from the dust mass estimated by \citet{lanz_star_2016} from Herschel SPIRE observations, $M_{dust}=6.05^{+0.376}_{-0.218}\times 10^7$ M$_{\odot}$, we would expect a cold molecular gas mass of about $6\times 10^9 M_{\odot}$, assuming a typical gas-to-dust mass ratio of $G/D=100$, as would be typical for massive early-type galaxies with solar or super-solar metallicity \citep[][]{remy-ruyer_gas--dust_2014}. This is again about a factor 10 greater than what is observed from CO line emission for cold molecular gas. Both estimates fall however within factors of a few from the warm H$_2$ mass observed in 3C~326~N.  

\begin{table}
\centering
\begin{tabular}{lcc}
\hline
Line            & $\lambda_{rest}$ & 
Flux Upper Limit\\
                & [$\mu$m]         
                & [W m$^{-2}$]\\
\hline
Br$\gamma$      & 2.1660 & 2$\times10^{-19}$  \\
PAH & 6.2 &1.6$\times10^{-18}$\\
\hline
\end{tabular}
\caption{5$\sigma$ upper limits to non-detected emission lines. Please note that Br$\gamma$ is in a particularly noisy part of the spectrum. \label{tab:upperlims}}
\end{table}

\begin{table*}
\centering
\begin{tabular}{lcccccccc}
\hline
Line&$\lambda$&F$_\mathrm{T}$&F$_\mathrm{S}$&F$_\mathrm{C}$&F$_\mathrm{N}$&F$_\mathrm{NE}$&F$_\mathrm{NW}$&F$_\mathrm{Bub}$\\\hline
&[$\mu$m]&[$10^{-20}\,$W$\,$m$^2$]&[$10^{-20}\,$W$\,$m$^2$]&[$10^{-20}\,$W$\,$m$^2$]&[$10^{-20}\,$W$\,$m$^2$]&[$10^{-20}\,$W$\,$m$^2$]&[$10^{-20}\,$W$\,$m$^2$]&[$10^{-20}\,$W$\,$m$^2$]\\\hline\hline
[Fe2]&1.64355&$88.0^{0.91}_{1.0}$&$9.6^{0.23}_{0.22}$&$6.2^{0.45}_{0.46}$&$7.6^{0.15}_{0.13}$&$3.7^{0.15}_{0.15}$&$7.2^{0.11}_{0.11}$&$6.7^{0.36}_{0.37}$\\[3pt]
1-  0 S(  8)&1.7146&$39.0^{0.79}_{0.71}$&$3.4^{0.21}_{0.18}$&$4.6^{0.24}_{0.29}$&$2.2^{0.12}_{0.12}$&$1.3^{0.095}_{0.095}$&$1.6^{0.1}_{0.11}$&$6.0^{0.29}_{0.26}$\\[3pt]
1-  0 S(  7)&1.748&$85.0^{1.8}_{1.8}$&$11.0^{0.54}_{0.81}$&$10.0^{0.94}_{1.2}$&$6.6^{0.28}_{0.22}$&$2.7^{0.12}_{0.099}$&$4.5^{0.14}_{0.17}$&$12.0^{0.72}_{0.58}$\\[3pt]
1-  0 S(  6)&1.7879&$54.0^{1.4}_{1.5}$&$5.0^{0.54}_{0.42}$&$3.8^{0.75}_{0.61}$&$3.7^{0.09}_{0.095}$&$2.1^{0.16}_{0.19}$&$2.9^{0.13}_{0.13}$&$6.6^{0.54}_{0.48}$\\[3pt]
1-  0 S(  5)&1.8357&$120.0^{0.87}_{0.88}$&$16.0^{0.33}_{0.3}$&$12.0^{0.49}_{0.39}$&$12.0^{0.14}_{0.28}$&$5.4^{0.087}_{0.09}$&$8.1^{0.14}_{0.16}$&$17.0^{0.22}_{0.27}$\\[3pt]
Pa$\alpha$&1.875&$96.0^{0.87}_{0.86}$&$13.0^{0.22}_{0.26}$&$9.5^{0.31}_{0.27}$&$8.1^{0.12}_{0.14}$&$3.5^{0.084}_{0.082}$&$7.8^{0.095}_{0.12}$&$11.0^{0.24}_{0.28}$\\[3pt]
1-  0 S(  4)&1.8919&$38.0^{0.55}_{0.57}$&$4.2^{0.094}_{0.12}$&$3.4^{0.32}_{0.3}$&$3.4^{0.076}_{0.072}$&$1.6^{0.086}_{0.081}$&$1.9^{0.066}_{0.076}$&$3.7^{0.16}_{0.19}$\\[3pt]
1-  0 S(  3)&1.9575&$180.0^{1.1}_{1.2}$&$26.0^{0.18}_{0.23}$&$19.0^{0.43}_{0.41}$&$18.0^{0.084}_{0.09}$&$8.4^{0.062}_{0.069}$&$13.0^{0.14}_{0.12}$&$24.0^{0.71}_{0.6}$\\[3pt]
1-  0 S(  2)&2.0337&$65.0^{0.8}_{0.74}$&$8.2^{0.17}_{0.21}$&$6.4^{0.28}_{0.26}$&$6.0^{0.091}_{0.12}$&$2.9^{0.1}_{0.13}$&$4.2^{0.075}_{0.077}$&$7.8^{0.26}_{0.24}$\\[3pt]
1-  0 S(  1)&2.1217&$150.0^{1.0}_{1.1}$&$21.0^{0.36}_{0.86}$&$14.0^{0.49}_{0.89}$&$15.0^{0.17}_{0.24}$&$7.2^{0.078}_{0.066}$&$11.0^{0.079}_{0.074}$&$18.0^{0.2}_{0.2}$\\[3pt]
1-  0 Q(  1)&2.4065&$95.0^{0.76}_{0.81}$&$13.0^{0.096}_{0.11}$&$8.3^{0.15}_{0.12}$&$9.8^{0.073}_{0.43}$&$4.7^{0.079}_{0.073}$&$6.8^{0.065}_{0.064}$&$9.7^{0.23}_{0.19}$\\[3pt]
1-  0 Q(  2)&2.4133&$89.0^{2.2}_{2.3}$&$6.8^{0.89}_{1.4}$&$6.8^{0.98}_{1.1}$&$9.5^{0.22}_{0.33}$&$5.1^{0.21}_{0.23}$&$6.8^{0.37}_{0.44}$&$9.3^{0.4}_{0.4}$\\[3pt]
1-  0 Q(  3)&2.4236&$91.0^{0.64}_{0.79}$&$13.0^{0.091}_{0.13}$&$9.3^{0.12}_{0.14}$&$9.0^{0.25}_{0.2}$&$3.8^{0.072}_{0.051}$&$6.4^{0.06}_{0.065}$&$10.0^{0.11}_{0.13}$\\[3pt]
1-  0 Q(  4)&2.4374&$21.0^{0.47}_{0.4}$&$2.2^{0.064}_{0.07}$&$1.5^{0.092}_{0.083}$&$1.3^{0.053}_{0.061}$&$0.78^{0.055}_{0.061}$&$0.92^{0.073}_{0.076}$&$1.9^{0.11}_{0.098}$\\[3pt]
1-  0 Q(  5)&2.4547&$69.0^{0.51}_{0.55}$&$9.3^{0.084}_{0.078}$&$5.9^{0.088}_{0.11}$&$7.6^{0.059}_{0.058}$&$2.9^{0.1}_{0.096}$&$5.1^{0.058}_{0.056}$&$7.9^{0.11}_{0.1}$\\[3pt]
1-  0 Q(  6)&2.4754&$23.0^{0.55}_{0.57}$&$2.1^{0.061}_{0.059}$&$1.8^{0.086}_{0.088}$&$1.3^{0.053}_{0.055}$&$0.61^{0.064}_{0.061}$&$1.1^{0.074}_{0.061}$&$1.4^{0.14}_{0.13}$\\[3pt]
1-  0 Q(  7)&2.5&$59.0^{1.3}_{1.0}$&$6.9^{0.55}_{0.39}$&$5.3^{0.41}_{0.38}$&$5.2^{0.085}_{0.14}$&$2.3^{0.093}_{0.084}$&$3.8^{0.14}_{0.15}$&$6.3^{0.23}_{0.25}$\\[3pt]
1-  0 Q(  8)&2.5277&$23.0^{0.74}_{0.59}$&$1.1^{0.072}_{0.075}$&$1.1^{0.19}_{0.14}$&$1.9^{0.096}_{0.089}$&$1.0^{0.096}_{0.086}$&$0.87^{0.086}_{0.079}$&$2.5^{0.16}_{0.18}$\\[3pt]
1-  0 Q(  9)&2.5599&$40.0^{0.78}_{0.9}$&$3.6^{0.077}_{0.093}$&$3.2^{0.17}_{0.16}$&$2.8^{0.17}_{0.17}$&$1.5^{0.12}_{0.12}$&$2.5^{0.1}_{0.12}$&$3.4^{0.23}_{0.24}$\\[3pt]
1-  0 O(  2)&2.6268&$51.0^{1.1}_{0.87}$&$5.8^{0.13}_{0.13}$&$5.1^{0.14}_{0.14}$&$3.8^{0.23}_{0.19}$&$1.8^{0.19}_{0.11}$&$3.2^{0.21}_{0.11}$&$5.2^{0.21}_{0.19}$\\[3pt]
1-  0 O(  3)&2.8024&$100.0^{0.85}_{1.1}$&$14.0^{0.18}_{0.5}$&$9.4^{0.29}_{0.6}$&$11.0^{0.12}_{0.13}$&$5.1^{0.15}_{0.18}$&$6.2^{0.081}_{0.08}$&$11.0^{0.39}_{0.33}$\\[3pt]
0-  0 S(  6)&6.1086&$160.0^{2.1}_{2.3}$&$13.0^{0.21}_{0.21}$&$13.0^{0.35}_{0.41}$&$15.0^{0.2}_{0.24}$&$5.4^{0.4}_{0.42}$&$10.0^{0.3}_{0.33}$&$18.0^{0.37}_{0.38}$\\[3pt]
0-  0 S(  5)&6.9089&$590.0^{3.2}_{3.3}$&$71.0^{0.32}_{0.42}$&$67.0^{0.52}_{1.7}$&$62.0^{0.38}_{0.33}$&$33.0^{0.39}_{0.36}$&$35.0^{0.55}_{0.42}$&$83.0^{0.53}_{0.88}$\\[3pt]
Ar2&6.98337&$140.0^{4.4}_{3.7}$&$14.0^{0.3}_{0.23}$&$15.0^{0.44}_{0.36}$&$7.3^{0.28}_{0.28}$&$3.6^{0.54}_{0.6}$&$9.9^{0.57}_{0.61}$&$15.0^{1.3}_{1.3}$\\[3pt]
0-  0 S(  3)&9.6645&$910.0^{2.2}_{10.0}$&$96.0^{0.24}_{11.0}$&$86.0^{0.24}_{4.8}$&$83.0^{0.14}_{0.17}$&$51.0^{0.14}_{0.16}$&$52.0^{0.12}_{0.14}$&$120.0^{0.21}_{0.19}$\\[3pt]

[NeIII]&15.5509&$93.0^{0.82}_{0.89}$&$9.1^{0.11}_{0.17}$&$11.0^{0.14}_{0.19}$&$6.1^{0.44}_{0.39}$&$3.3^{0.081}_{0.08}$&$3.8^{0.071}_{0.073}$&$15.0^{0.16}_{0.16}$\\[3pt]
\hline
\end{tabular}
\caption{Integrated fluxes in the apertures of Fig.\,\ref{fig:0010ratio} for modelled emission lines. Errors are derived from modelling uncertainties. \label{tab:emissionlinesflux}}
\end{table*}


\begin{table*}[]
    \centering
    \caption{Properties of a one Gaussian fit to the integrated spectrum for all detected emission lines.}
    \label{tab:IntPar}
    \begin{tabular*}{\textwidth}{@{\extracolsep{\fill}}lccccc}\hline
Line&$\lambda$&F$_\mathrm{Int}$&FWHM&$z$&V$_\mathrm{off}$\\\hline&[$\mu$m]&[$10^{-20}\,$W$\,$m$^2$]&[km\,s$^{-1}$]&&[km\,s$^{-1}$]\\\hline\hline$[$FeII$]$&1.644&$55.8^{0.6}_{0.62}$&$573.0^{5.8}_{5.5}$&$0.09006^{9.9e-06}_{9.8e-06}$&$75.1^{2.7}_{2.7}$\\[3pt]
H$_2$ 1--0 S(8)&1.7146&$19.7^{0.48}_{0.51}$&$523.0^{14.0}_{13.0}$&$0.09022^{1.9e-05}_{1.7e-05}$&$119.0^{5.1}_{4.7}$\\[3pt]
H$_2$ 1--0 S(7)&1.748&$46.2^{0.7}_{0.67}$&$822.0^{17.0}_{15.0}$&$0.08994^{1.7e-05}_{1.7e-05}$&$41.6^{4.7}_{4.8}$\\[3pt]
H$_2$ 1--0 S(6)&1.7879&$26.6^{0.51}_{0.47}$&$630.0^{11.0}_{11.0}$&$0.09054^{1.6e-05}_{1.5e-05}$&$206.0^{4.5}_{4.0}$\\[3pt]
H$_2$ 1--0 S(5)&1.8357&$95.2^{0.51}_{0.48}$&$713.0^{4.1}_{3.7}$&$0.08998^{4.8e-06}_{5.1e-06}$&$52.4^{1.3}_{1.4}$\\[3pt]
Pa$\alpha$&1.8751&$65.5^{0.47}_{0.47}$&$571.0^{4.7}_{4.7}$&$0.09026^{5.5e-06}_{5.4e-06}$&$130.0^{1.5}_{1.5}$\\[3pt]
H$_2$ 1--0 S(4)&1.8919&$22.3^{0.33}_{0.34}$&$516.0^{8.3}_{7.8}$&$0.09022^{1.2e-05}_{1.2e-05}$&$118.0^{3.2}_{3.2}$\\[3pt]
H$_2$ 1--0 S(3)&1.9575&$141.0^{0.39}_{0.35}$&$625.0^{1.9}_{1.8}$&$0.09005^{2.4e-06}_{2.3e-06}$&$70.3^{0.65}_{0.64}$\\[3pt]
H$_2$ 1--0 S(2)&2.0337&$48.8^{0.42}_{0.42}$&$636.0^{6.0}_{6.5}$&$0.09009^{7.1e-06}_{7.5e-06}$&$82.7^{2.0}_{2.1}$\\[3pt]
H$_2$ 2-1 S(3)&2.0734&$14.4^{0.38}_{0.39}$&$669.0^{19.0}_{17.0}$&$0.09048^{2.5e-05}_{2.6e-05}$&$188.0^{7.0}_{7.0}$\\[3pt]
H$_2$ 1--0 S(1)&2.1217&$118.0^{0.35}_{0.36}$&$623.0^{1.9}_{2.0}$&$0.0902^{2.6e-06}_{2.6e-06}$&$112.0^{0.71}_{0.7}$\\[3pt]
H$_2$ 1--0 Q(1)&2.4065&$65.6^{0.58}_{0.58}$&$507.0^{4.6}_{4.2}$&$0.09013^{6.6e-06}_{6.7e-06}$&$92.4^{1.8}_{1.9}$\\[3pt]
H$_2$ 1--0 Q(3)&2.4235&$101.0^{0.69}_{0.64}$&$623.0^{3.0}_{2.8}$&$0.09022^{3.2e-06}_{3.1e-06}$&$118.0^{0.88}_{0.85}$\\[3pt]
H$_2$ 1--0 Q(4)&2.4374&$27.4^{0.45}_{0.45}$&$729.0^{9.7}_{11.0}$&$0.09067^{1.6e-05}_{1.6e-05}$&$241.0^{4.4}_{4.4}$\\[3pt]
H$_2$ 1--0 Q(5)&2.4547&$65.4^{0.37}_{0.34}$&$620.0^{3.4}_{3.1}$&$0.09025^{4.5e-06}_{4.5e-06}$&$128.0^{1.2}_{1.3}$\\[3pt]
H$_2$ 1--0 Q(6)&2.4754&$14.8^{0.27}_{0.27}$&$487.0^{8.6}_{8.8}$&$0.08996^{1.5e-05}_{1.4e-05}$&$46.9^{4.1}_{3.9}$\\[3pt]
H$_2$ 1--0 Q(7)&2.5&$46.2^{0.59}_{0.59}$&$758.0^{12.0}_{11.0}$&$0.09018^{1.1e-05}_{1.1e-05}$&$107.0^{3.1}_{3.1}$\\[3pt]
H$_2$ 1--0 O(2)&2.6268&$25.8^{0.34}_{0.35}$&$544.0^{8.1}_{7.6}$&$0.09001^{1e-05}_{1.1e-05}$&$60.9^{2.9}_{3.0}$\\[3pt]
H$_2$ 1--0 O(3)&2.8024&$85.8^{0.37}_{0.4}$&$655.0^{3.1}_{3.1}$&$0.0902^{4e-06}_{3.8e-06}$&$113.0^{1.1}_{1.1}$\\[3pt]
H$_2$ 0--0 S(6)&6.1086&$79.3^{1.2}_{1.0}$&$484.0^{7.2}_{6.7}$&$0.09003^{9.8e-06}_{1.1e-05}$&$65.7^{2.7}_{3.0}$\\[3pt]
H$_2$ 0--0 S(5)&6.9089&$416.0^{1.2}_{1.3}$&$560.0^{2.0}_{1.8}$&$0.09^{2.6e-06}_{2.4e-06}$&$58.4^{0.7}_{0.65}$\\[3pt]
$[$ArII$]$ &6.98337&$89.4^{1.4}_{1.3}$&$616.0^{9.4}_{10.0}$&$0.08995^{1.2e-05}_{1.3e-05}$&$44.2^{3.3}_{3.5}$\\[3pt]
H$_2$ 0--0 S(3)&9.6645&$643.0^{0.62}_{0.6}$&$684.0^{0.67}_{0.65}$&$0.09022^{9.5e-07}_{1e-06}$&$118.0^{0.26}_{0.28}$\\[3pt]
$[$NeIII$]$&15.5509&$74.0^{0.43}_{0.43}$&$608.0^{3.7}_{3.7}$&$0.09012^{4.9e-06}_{5.3e-06}$&$92.1^{1.3}_{1.5}$\\\hline
    \end{tabular*}
    
\end{table*}

\section{Discussion}
\label{sec:discussion}

\subsection{The nature of the emission-line regions in 3C~326~N}
\label{ssec:natureemlineregions}
In the previous section we have described the properties of the molecular disk in 3C~326~N, which is roughly centred on the nucleus and extends to a radius of about 1.5~kpc. This disk is strongly kinematically disturbed with dual-component line profiles throughout the disk, and large line widths, typically 200-400\,km\,s$^{-1}$ in the systemic, and up to 1000\,km\,s$^{-1}$ in the secondary, bluer component. Both components show a velocity gradient that increases from South to North, and which resembles disk rotation when seen at low spatial resolution \citep[][]{nesvadba_dense_2011}. The two components are typically offset from each other by about $80-100$\,km\,s$^{-1}$. 

The excellent spatial resolution of NIRSpec reveals a large cavity in the Northern part of the disk, that is 0.6\arcsec$\times0.8$\arcsec\ across along its minor and major axis, respectively, corresponding to 1.0~kpc$\times$1.4~kpc at the redshift of 3C~326~N (Fig.~\ref{fig:emlinemorphology10}). This cavity is delineated by three bright clumps in the far North. Inside the cavity the gas is much fainter (but still detected) and reaches the highest velocities relative to the nucleus ($<380$ km s$^{-1}$), which are best seen in the central panel of Fig.~\ref{fig:emlinemorphology10}. 

Such a cavity may either be produced by an expanding bubble driven by AGN or star-formation feedback \citep[e.g.,][]{capetti_origin_1999, mayya_stellar_2023}, or indicate the presence of a massive dust lane across the galaxy. Star-formation can be ruled out given the low star-formation rate estimated with Herschel, $SFR=0.087^{+0.106}_{-0.046}$~M$_{\odot}$ \citep[][]{lanz_star_2016}. We do not identify any bright, warm dust or PAH emission at this location, and in the optical, the global extinction is very low, $A_V=0.17$~mag \citep[][]{nesvadba_energetics_2010}, so that the presence of a massive dust lane able to obscure near and mid-infrared emission in the H$_2$ 1--0 S(3) and 0--0 S(3) lines, seems unlikely. We thus conclude that we are seeing an inflating, AGN-driven bubble expanding through the molecular gas, reaching an expansion velocity of up to 380~km s$^{-1}$. 

The single-line fits plotted in the left panel of Fig.~\ref{fig:emlinemorphology10} do not show the true projected gas velocities, because the complex line profiles are not well represented with a single Gaussian, but they probe deeper than the dual-Gaussian fit. They show a clear hour-glass shaped velocity distribution, which in the North is associated with the cavity, and surrounded by gas at more moderate velocities. The same feature is identified in the systemic (redder) component in the dual-Gaussian fits. 

In the South, we do not identify an analogous blueshifted bubble, but the global gas morphology and velocity field of the systemic component are clearly hourglass shaped, implying a symmetric underlying structure. The onset of this structure can be seen in the right panel of Fig.~\ref{fig:emlinemorphology10}, which shows the blueshifted gas component of H$_2$ 1--0 S(3), with velocities that are offset by up to about $-300$~km s$^{-1}$ from the systemic redshift. The structure is better seen in H$_2$ 0--0 S(3) (Fig.~\ref{fig:emlinemorphology00}), and also in warm ionised gas, in particular Pa$\alpha$ (Fig.~\ref{fig:emlinemorphologyPa}).

{The gas velocities are} more than previously estimated from the seeing-limited VLT/SINFONI data \citep[][]{nesvadba_dense_2011} at 1\arcsec\ spatial resolution, and also significantly more than the rotation velocity expected for a massive elliptical galaxy. Turnover velocities of rotating gas disks in elliptical galaxies with $3\times 10^{11}$ M$_{\odot}$ in stellar mass, similar to 3C~326~N, are $\sim 150$ km s$^{-1}$, with turnover radii around 1.5~kpc, comparable to the size of the molecular disk in 3C~326~N \citep[][]{yoon_rotation_2021}. 

The most likely explanation of these features is that we are seeing a bubble expanding within a massive, rotating, molecular disk, where the observed velocity gradient is most likely a combination of disk rotation and outflowing motion. The excellent  0.1\arcsec\ resolution provided by NIRSpec is critical to identify the complex kinematic features of such a configuration, and to measure gas velocities robustly. Nonetheless, the overall complexity of the gas kinematics and morphology make it difficult to clearly separate a kinematically perturbed from the quiescently rotating part of the disk, and to derive, e.g., a robust estimate of the gas energetics in a simple, unique, and analytic way. In a companion paper, we will therefore use hydrodynamic simulations to characterise the interactions between AGN and gas disk in 3C326~N in a more robust way (Shende et al., 2024, in prep.)

Given that star formation rates and the bolometric luminosity of the AGN are very low, and are not sufficient to either power the observed emission-line fluxes or drive outflows or major gas motions \citep[][]{nesvadba_energetics_2010, nesvadba_dense_2011}, the only possible culprit that could inflate this bubble is the radio jet.

The radio core is compact in all frequency bands in which we observed 3C~326~N with the VLA (Sect.~\ref{ssec:jvalobs}), corresponding to beam sizes between 0.2\arcsec\ and 1.2\arcsec, respectively, at 9~GHz and 1.4~GHz, respectively.
However, we also observe 47\% of missing flux at 1.4~GHz compared to the flux measured at 5.4\arcsec\ beam size with FIRST \citep[][Sect.~\ref{ssec:radiocore}]{becker_first_1995}. This suggests that significant amounts of radio emission must be present at scales between about 1 and 4.6~kpc around the nucleus. This is likely the radio plasma that is permeating the gas disk in 3C~326~N, stirring up the gas, creating turbulence, and inflating the cavity seen in the Northern hemisphere. A similar situation has previously been found by \citet{zovaro_jets_2019} in the radio galaxy 4C~31.04.

Qualitatively similar, bipolar gas motions have already been predicted by \citet{meenakshi_modelling_2022} from hydrodynamic simulations of the passage of radio jets through a surrounding gas disk (their Fig.~5). The jet energy of their simulation, $P_j=10^{44}$ erg s$^{-1}$, is somewhat less than that estimated by \citet{nesvadba_energetics_2010} for 3C~326~N, $P_{jet,326}=2-13\times 10^{44}$~erg s$^{-1}$, obtained with different methods ranging from X-ray cavity inflation \citep[][]{birzan_radiative_2008} to the 5~GHz core luminosity \citep[][]{merloni_measuring_2007}. Nonetheless, the simulated bubbles of \citet{meenakshi_modelling_2022} reach gas velocities of up to $\sim 300$\,km\,s$^{-1}$, slightly lower than the 380~km s$^{-1}$ that we observe (albeit not correcting for inclination), illustrating that the jet in 3C~326~N may well cause the outstanding gas properties that we observe.

We also note that warm ionised gas in the hourglass component has a general redshift of about $100-200$~km s$^{-1}$ relative to H$_2$. This effect is only seen in the systemic gas component, which shows this hourglass structure, not in the blueshifted component. We are not aware of any systematic effect that could cause this behaviour in only one line component, and which is also seen in other warm ionised gas lines (Fig.~\ref{fig:LineProfiles}). We might be seeing an effect of stratification, where the warm ionised gas is accelerated to somewhat different velocities, or where different parts of the same structure are particularly bright in the warm ionised and molecular gas {(i.e. different components of the same strcture.)}. 
 
\subsection{An extended, turbulent, multi-phase interstellar medium throughout the galaxy}
\label{ssec:comparisongasphases}

The spatially resolved line ratios in 3C~326~N provide interesting information about the gas conditions and heating mechanism of the gas in this galaxy. While a detailed analysis of the physical conditions of the warm molecular gas is beyond the scope of this paper, we can make several interesting conclusions about these mechanisms from comparing the line ratios in the different components of the galaxy. 

\citet{nesvadba_energetics_2010} and \citet{nesvadba_dense_2011} already showed that the source-integrated rotational and ro-vibrational line emission of this galaxy is heated by shocks; however, their data did not have the spatial resolution necessary to resolve the different components of this galaxy, and to search for pockets of star formation that could be embedded in a globally shock heated interstellar medium. 

As first argued by \citet{puxley_molecular_1990}, ro-vibrational lines in the NIR are an excellent shock tracer, when comparing their line fluxes with those of warm ionised gas lines like Br$\gamma$ or Pa$\alpha$. \citet{puxley_molecular_1990} used 44 HII regions observed in 30 galaxies to infer that line ratios between H$_2$ S(1) and Br$\gamma$ are typically between 0.1 and 1.5 in star-forming regions, with an average of $0.5\pm0.3$.  We note that this empirical ratio does not take into account potential effects from shocks present in these star-forming regions. Outliers in their sample with much brighter H$_2$ emission include in particular NGC~6240 and NGC 3079, known to have large amounts of shocked gas \citep[e.g.,][]{keel_shock_1990,veilleux_nuclear_1994,meijerink_evidence_2013,medling_tracing_2021}. 

Instead of Br$\gamma$ and S(1), we will in the following use Pa$\alpha$ and S(3) for our comparison. Pa$\alpha$ is significantly brighter than Br$\gamma$, but falls in between the atmospheric windows for nearby galaxies observed from the ground. For a galaxy at $z\sim0.1$ like 3C~326~N and observations from space like with JWST, this is however not a concern. Moreover, Pa$\alpha$ is less affected by underlying absorption lines from stellar photospheres, which can artificially lower the measured Br$\gamma$ flux in medium-resolution spectra like those provided by the NIRSpec IFU. 

The decrement between Br$\gamma$ and Pa$\alpha$ for gas with an electron temperature, T$_e=10^4$~K, is F$_{Pa\alpha}$/F$_{Br\gamma}=11.2$ \citep[e.g.,][]{dopita_internal_2003}, and the line ratio between H$_2$~1--0 S(3) and S(1) in 3C~326~N is on average F$_{S(3)}/$F$_{(S1)}=1.2$ (Tab.~\ref{tab:emissionlinesflux}). This implies that the \citet{puxley_molecular_1990} criterion of F$_{Br\gamma}/$F$_{H_2 1-0 S(3)}\sim 1$ for line emission from shocked gas translates into a criterion of F$_{H_2 1-0 S(3)} >0.1$~F$_{Pa\alpha}$. 

Line emission throughout 3C~326~N is well above this threshold. 
When we take the ratio of total source Pa$\alpha$ to H$_2$ 1--0 S(3) we find the same low value 0.5 as in \citet{nesvadba_dense_2011}. If we take the ratios within the apertures defined in Fig.\,\ref{fig:0010ratio}, we find a ratio of 0.5 for aperture S,
of 0.6 in aperture NW, 0.45 in N, and 0.4 in NE. 

This implies that shocks dominate the gas heating of molecular gas throughout the galaxy not only in a global sense, but also for individual features seen in the line maps of Fig.~\ref{fig:emlinemorphology10}. This includes in particular the compact clumps in the North of the galaxy, which from their morphology are reminiscent of young star-forming clusters, as are commonly found in other galaxies. The line ratios however, suggest that they are not heated by UV photons from young stars, but from shocks like the ambient gas, otherwise Pa$\alpha$ should be an order of magnitude brighter compared to the S(3) line than what is observed. They are thus more likely gas that has fragmented along the rim of the jet-inflated bubble. 

Shocks are also the dominant heating mechanism of the pure-rotational line emission of H$_2$ in 3C~326~N, as already shown by \citet{nesvadba_energetics_2010}. They constructed a diagnostic diagram based on PAH, H$_2$ 0--0 S(1), CO(1--0), and 24~$\mu$m dust emission, and compared with expectations from theoretical models of star-forming, photon-dominated regions (PDRs), to find that the source-integrated S(1) flux of 3C~326~N exceeded the gas heating possible with UV photons from young stars by more than an order of magnitude. They also disfavoured cosmic ray heating, arguing that the required cosmic ray flux densities would be so high that the H$_2$ molecule would be destroyed, leaving shocks as only plausible gas heating source. 

\citet{nesvadba_energetics_2010} also compared the source-integrated flux measurements of H$_2$ 0--0 S(0) to S(7), and fitted the shock models of \citet{flower_excitation_2010}, finding that at least three different shock velocities between $v_s=4$ and 40\,km\,s$^{-1}$ were needed for C-type shocks to explain the excitation diagram, for two assumptions on gas density, respectively ($n_{H2} = 10^3$ and $10^4$ cm$^{-3}$, respectively). From their Fig.~3 we can read off that the H$_2$ 0--0 S(3) line is most sensitive to gas heated by intermediate-velocity shocks, whereas the S(5) line is dominated by the highest shock velocities, and representative of the least important mass component. This is one of the reasons why we focus the discussion of the warm molecular gas in 3C~326~N onto the S(3) component. 

\begin{figure*}
    \centering
    \includegraphics[width=\linewidth]{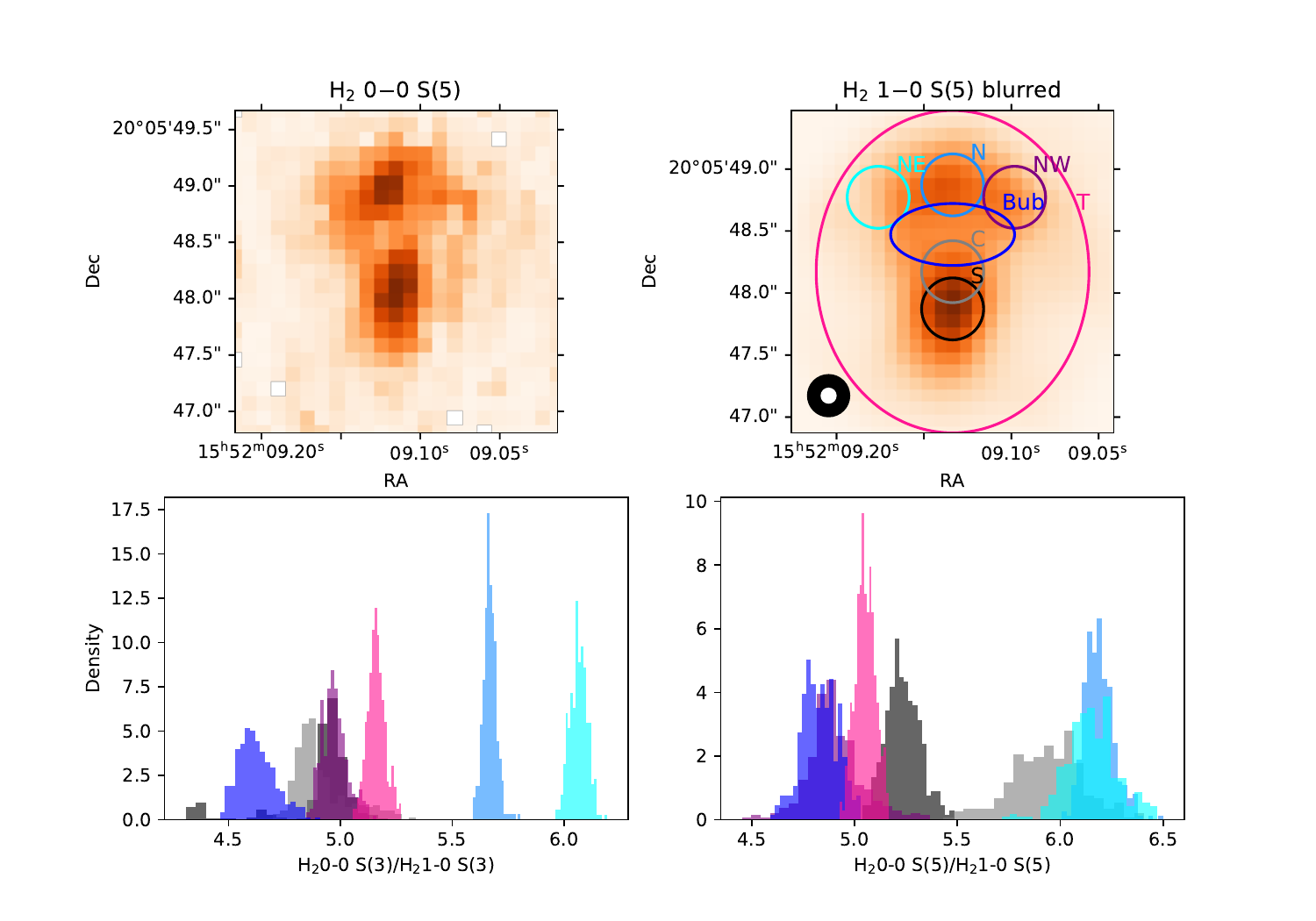}
    \caption{Top: Integrated flux for the H$_2$0--0 S(5) line and the H$_2$1--0 S(5) line as titled. The H$_2$1--0 S(5) flux image has been convolved to the PSF at the H$_2$0--0 S(5) line. The original PSF for H$_2$1--0 S(5) and the post-convolution PSF are the white and black filled circles in the H$_2$1--0 S(5) plot, respectively. Bottom: Ratio of 1--0 to 0--0 line fluxes for the apertures shown in the top image.}
    \label{fig:0010ratio}
\end{figure*}

\begin{figure*}
    \includegraphics[width=\textwidth]{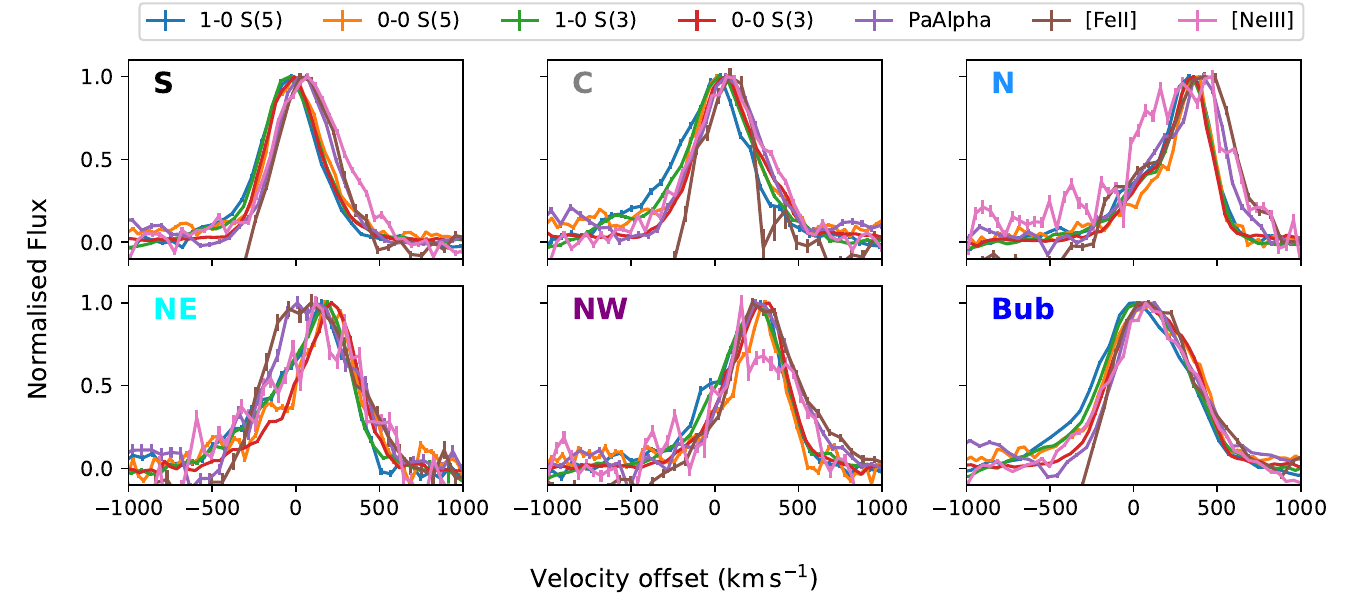}
    \caption{The extracted line profiles for a selection of observed lines. The zero velocity is set by the CO band head absorption lines. The lines are extracted using the apertures in Fig.\,\ref{fig:0010ratio} as indicated by the coloured marker in the corner of each subplot.}\label{fig:LineProfiles}
\end{figure*}

We also note that, on top of the similar gas morphologies in different lines of molecular and warm ionised gas highlighted in Sect.~\ref{sec:analysis}, the line profiles and velocities of the H$_2$ 1--0 S(3), H$_2$ 1--0 S(5), H$_2$ 0--0 S(3), and H$_2$ 0--0 S(5) all agree very well. [FeII]$\lambda$1.644 and Pa$\alpha$ also agree with each other and relatively well with the molecular line profiles. This can be seen by comparing the line maps shown in Figs.\ref{fig:emlinemorphology10}, \ref{fig:emlinemorphology00}, \ref{fig:emlinemorphologyPa}, and \ref{fig:emlinemorphologyFe}, and also by comparing the line profiles extracted from several apertures within 3C~326~N shown in Fig.~\ref{fig:LineProfiles}. 

Fig.~\ref{fig:LineProfiles} shows again the offset already discussed in Sect.~\ref{ssec:natureemlineregions} between the velocities of the ionised and molecular emission. This offset is most apparent in the systemic line component which is redshifted with respect to the molecular lines by 100-200\,km\,s$^{-1}$ in the southern component, the galaxy centre, and the centre of the northern component (aperture N in Fig.\,\ref{fig:0010ratio}). The systemic component is then in agreement in velocity with the molecular lines in the Western wing of the northern component. However, in the Eastern wing of the northern component (aperture NW) both ionised emission lines are double peaked with both peaks being comparable in amplitude. The second peak is blueshifted with respect to the molecular lines by 200\,km\,s$^{-1}$.

The one line with a strongly different morphology is [NeIII]$\lambda$15.56, which does not follow the same morphology seen in the other lines, but is fairly symmetric about the nucleus. [NeIII]$\lambda$15.56 can be photoionised by AGN, star formation in extreme cases, or shocks. \citet{nesvadba_energetics_2010} estimated a line ratio of [NeII]$\lambda$12.81/[NeIII]$\lambda$15.56$\ \sim 2.2$ for shock conditions favoured by the molecular gas lines. We can use the [NeII] flux measured by \citet{ogle_shocked_2007} of F([NeII]$\lambda$12.82)$ = 4\times 10^{-19}$ W m$^{-2}$, and the source-integrated [NeIII]$\lambda$15.56 flux in Table~\ref{tab:emissionlinesflux}, to measure a source-integrated line flux of [NeII]$\lambda$12.81/[NeIII]$\lambda$15.56$=4.3$, which is about twice as large as the value expected for shocked gas. This would be consistent with having a second line component present, presumably heated by AGN photoionisation.

In conclusion, the ionised and molecular lines follow globally the same approximate bulk velocity, although local differences do exist. 
Similar gas kinematics in different gas phases are expected in multi-phase media, where the different gas phases do not probe different components of a galaxy, but are co-spatial and undergoing a constant mass and energy exchange \citep[][]{guillard_h2_2009}. This is characteristic for gas in non-equilibrium, rapidly changing conditions, e.g., in rapidly cooling post-shock gas following the passage of a shock front, e.g., from an expanding radio source \citep[][]{mukherjee_relativistic_2018}. 

\subsection{3C~326~N in the context of turbulence-driven star formation}
\label{ssec:turbdrivensf}

\subsubsection{Kennicutt-Schmidt diagram}
One of the outstanding properties of radio galaxies with large amounts of warm molecular Hydrogen are their low star-formation rates compared to their molecular gas content. \citet{schmidt_space_1968} and \citet{kennicutt_star_1989} showed that gas-mass surface densities are a major determinant of the star-formation rate density in galaxies. A subset of MOHEGs show a significant offset from the typical relationship between gas mass and star-formation rate surface densities in normal star-forming galaxies by factors 10-100. This offset is seen even in spiral galaxies with normal stellar mass surface densities and moderately bright radio jets \citep[][]{nesvadba_jet-driven_2021}. 

The offset of radio galaxies in the Kennicutt-Schmidt diagram was first noticed by \citet{nesvadba_energetics_2010}, who compared the molecular gas mass surface densities of these galaxies with star formation traced by PAH emission. PAHs are an imperfect tracer of star formation, as PAHs are relatively easily destroyed by shocks \citep[][]{micelotta_polycyclic_2010,egorov_phangs-jwst_2023}. Subsequently, this observation was confirmed with Herschel/SPIRE photometry \citep[][]{lanz_star_2016}. The star-formation rate of 3C~326~N that these measurements found is SFR$=0.087^{+0.106}_{-0.046}$~M$_{\odot}$~yr$^{-1}$. 

We have now presented new ALMA observations (Sect.~\ref{ssec:almaobs}), which have revealed that the previous CO(1--0) detection of \citet{nesvadba_energetics_2010}, based on shallower data observed with the IRAM Plateau de Bure Interferometer in 2009 could not be confirmed. These data had much smaller bandwidth, and a large, 5\arcsec\ beam, making it impossible to separate the central continuum emission from putative more extended line emission. The new ALMA data place an upper limit of $8\times 10^8$~M$_{\odot}$ on the cold molecular gas mass derived from the CO(1--0) line flux for a Milky-Way like H$_2$-to-CO conversion factor of $\alpha_{CO}=4.3 M_{\odot}$/ (K\,km\,s$^{-1}$ pc$^2$).

In Fig.~\ref{fig:ks}, we show the Kennicutt-Schmidt diagram with the updated position of 3C~326~N, taking these new measurements into account. We estimated a gas mass surface density of cold molecular gas by using the upper limit on the molecular gas derived in Sect.~\ref{ssec:almaobs}, and assumed that the cold molecular gas morphology follows that of H$_2$ 0--0 S(3). We also assume that star formation is distributed over the same area. This is a conservative estimate, as we do not observe any signatures of star formation in our data. With these estimates and assumptions, 3C~326~N falls a factor 23 below the usual Kennicutt-Schmidt relationship for normally star-forming galaxies. {\citet{villar_martin_limited_2023} show that cold gas morphology can be smaller than the H$_2$; however, a smaller morphology would only increase the discrepancy.}

For comparison, we also show where the galaxy falls when using the observed H$_2$ 0--0 S(3) mass surface densities. To estimate these, we adopted the total mass estimate of warm molecular hydrogen of $2\times 10^9 M_{\odot}$ \citep[][]{nesvadba_energetics_2010} and assumed that changes in gas excitation are not large enough to strongly alter the overall mass-to-light ratio of the gas. This allowed us to estimate a range of gas-mass surface densities, $\Sigma=200-680$ M$_{\odot}$ pc$^{-2}$, from the surface-brightness distribution of the line shown in Fig.~\ref{fig:emlinemorphology00}. Showing the position of 3C~326~N for warm molecular Hydrogen in this diagram bears of course a large risk of systematic errors, but allows us also to estimate where this galaxy would approximately fall if the gas heating that is required to keep this gas warm, was absent. It can be seen that the galaxy falls about 2 orders of magnitude below normal star-forming galaxies in this case. 

\begin{figure*}
\centering
    \includegraphics[width=0.7\textwidth]{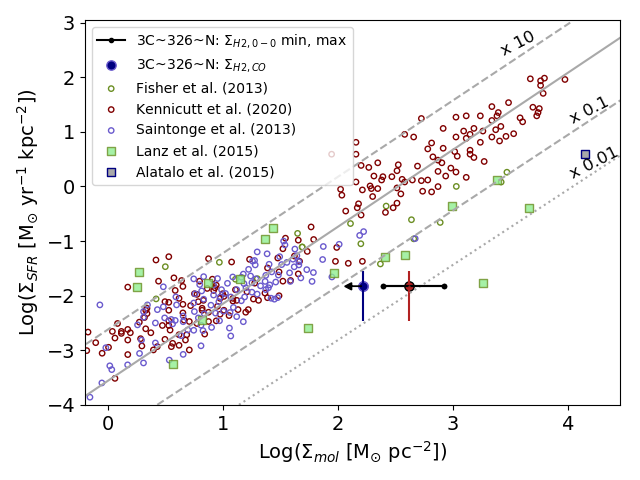}
    \caption{Kennicutt-Schmidt law between star-formation rate and molecular gas mass surface density. The large blue and red dot show the position of 3C~326~N when using the mass surface density of cold molecular gas probed by CO(1--0) and H$_2$ 0--0 lines, respectively. For the latter estimate, the black line shows the range of surface-brightness observed with MIRI. Small empty circles show several sets of normally star-forming galaxies in the nearby Universe taken from \citet{fischer_determining_2013}, \citet{kennicutt_star_2020}, and \citet{saintonge_validation_2013}. Green and grey squares show the MOHEGs of \citet{ogle_jet-powered_2010} and \citet{alatalo_suppression_2015}.} \label{fig:ks}
\end{figure*}

\subsubsection{Turbulent energy in the warm molecular gas}

The large line widths observed with NIRSpec and MIRI MRS show clearly that the gas in 3C~326~N is strongly stirred up and likely made turbulent through the interactions with the radio jet \citep[][]{nesvadba_energetics_2010, nesvadba_dense_2011}. The 0.11\arcsec\ size of the PSF corresponds to a projected size scale of 190~pc, which demonstrates that the large line widths that have been previously observed are not due to velocity offsets between different gas components on kpc scales, but are present even at size scales typical of giant molecular clouds.

This is an important observation, because interstellar turbulence is now recognised as a major regulation mechanism of star formation \citep[][]{krumholz_general_2005, federrath_comparing_2010,padoan_star_2011,hennebelle_analytical_2011,hennebelle_turbulent_2012}. In 3C~326~N, the only energy source that is powerful enough to power velocity offsets and line widths as observed, is the radio jet. We know also from hydrodynamic simulations that radio jets can make surrounding gas turbulent \citep[e.g.,][]{mukherjee_relativistic_2018}, and impact star formation, by locally compressing, but globally dispersing the gas \citep[][Mandal et al. 2024 In prep.]{mandal_impact_2021}. 

The virial parameter, $\alpha_{vir}$ allows to quantify the amount of turbulent energy compared to gravitational binding energy in molecular clouds \citep[][]{bertoldi_pressure-confined_1992,sun_cloud-scale_2018}. We set $\alpha_{vir}=5 \sigma^2_{cl} (\pi\ G\ R_{cl}\ \Sigma_{gas})^{-1}$, following \citet{bertoldi_pressure-confined_1992}. $\sigma_{cl}$ is the line width dominated by turbulent motion, $\Sigma_{gas}$ the gas-mass surface density, $R_{cl}$ the size of the cloud, and $G$ the gravitational constant. Giant molecular clouds in the Milky Way, as well as nearby spiral galaxies are found to be star-forming for $\alpha_{vir}\sim 1$. In gas with much higher turbulent velocity, star formation is suppressed. \citet{nesvadba_dense_2011} showed with low-resolution SINFONI data, that the high interstellar turbulence in 3C~326~N caused by the radio jet can suppress star formation, provided that high turbulent velocities prevail down to the typical scales of giant molecular clouds, and cloud complexes. 

\begin{figure}
    \centering
    \includegraphics[width=0.48\textwidth]{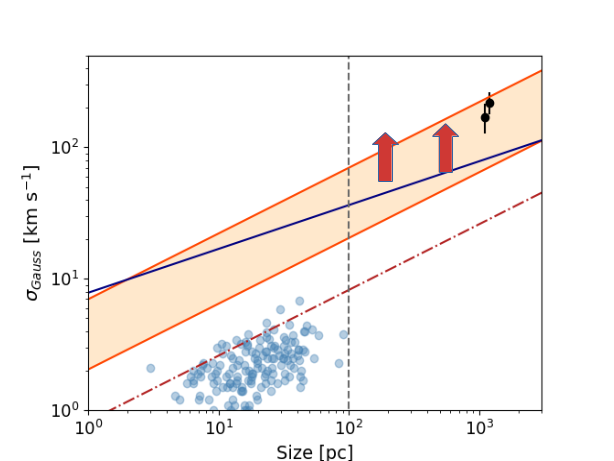}
    \caption{Gaussian line width, $\sigma$, of warm molecular gas in 3C~326~N as a function of radius. The minimal line widths observed in H$_2$ 1--0 S(3) and H$_2$ 0--0 S(3) at spatial resolutions corresponding to 190~pc and 530~pc, respectively, are shown as large, upward-pointing arrows. The broadest line widths observed in both lines are $>1000$\,km\,s$^{-1}$, and fall well outside the plot. Black dots show previous SINFONI observations of H$_2$ 1--0 S(3) at kpc-scale resolution. The red dotted line shows the position of gas clouds with $\alpha_{vir}=1$, for the gas-mass surface densities observed in 3C~326~N. Light blue circles show molecular clouds in the Milky Way. The yellow band indicates the range where the data should fall if the \citet{larson_turbulence_1981} scaling relations of turbulent clouds apply. The blue line indicates where the lines should fall for typical mass-to-light ratios of gas cooling through turbulent dissipation. See text for details.}
    \label{fig:alphavir}
\end{figure}

With our NIRSpec and MRS data, we are now able to test this prediction, by measuring the line broadening at the spatial resolution limit of our data, i.e., 0.11\arcsec\ for H$_2$ 1--0 S(3), corresponding to 190~pc, and 0.3\arcsec\ for H$_2$ 0--0 S(3), corresponding to 530~pc. In particular, we identified the narrowest widths of both lines, FWHM=170\,km\,s$^{-1}$ for H$_2$ 1--0 S(3) and FWHM=147\,km\,s$^{-1}$ for H$_2$ 0--0 S(3). 

Both line widths are shown in Fig.~\ref{fig:alphavir} as large, upward-pointing arrows to indicate that these values are the lowest of a set of measurements, where the highest values are $>1000$\,km\,s$^{-1}$. The position of these measurements is compared with several expected distributions. The red dotted line indicates where clouds fall with $\alpha_{vir}=1$, for a gas mass surface density of $\Sigma=250$ M$_{\odot}$ pc$^{-2}$ appropriate for 3C~326~N. Molecular clouds in the Milky Way fall slightly below that line because the Galactic value is somewhat lower \citep{heyer_re-examining_2009}. The yellow band indicates the range implied by Larson's scaling relationship between radius and turbulent velocity for turbulent molecular clouds \citep[][]{larson_turbulence_1981}, extrapolating from the measured values on 1~kpc scales downward to the scales probed by our new data, and beyond. We allow for a range of about a factor~8 in variation. The blue line supposes that the H$_2$ line emission observed in 3C~326~N is powered by the dissipation of turbulent energy on sub-parsec scales, which implies that $L_{H2}/M_{H2}=3/2 f_{H2} \sigma^3_{cl}/R_{cl}$ \citep[][]{mckee_theory_2007}. $L_{H2}$ and $M_{H2}$ are the luminosity and mass of warm $H_2$, respectively, with $L_{H2}/M_{H2}=0.06\ L_{\odot}/M_{\odot}$ \citep[][]{nesvadba_energetics_2010, nesvadba_dense_2011}. We allow for a fiducial fraction of about 50\% to the total radiative gas cooling from other gas phases, parameterised by the correction factor $f_{H2}=0.5$.

We find that the new observations fall well within the range predicted by the previous VLT/SINFONI observations obtained on kpc scales, which reinforces the previous statement of \citet{nesvadba_dense_2011} that the absence of star formation in 3C~326~N is well explained by the high turbulent gas velocities. One could argue that the high temperature rather than the mechanical energy could be the cause, however, since mechanical interactions of the radio jet with the gas are the primary heating mechanism in this galaxy, the gas heating is a concomitant effect of the same process. 

The good agreement of our new and old data with the Larson scaling relationship and radiative efficiency of H$_2$ line emission produced through the dissipation of turbulent energy adds further credence that the interactions between radio jet and gas are indeed raising the turbulent energy in the gas of 3C~326~N. Thanks to the high spatial resolution of the new data, we can also rule out that small pockets of less turbulent gas exist in some regions of 3C~326~N. Very broad line widths are found to be ubiquitous in the molecular gas throughout the galaxy, at least down to scales of 200~pc. 

\section{On using the H$_2$ 1--0 lines as proxy to the H$_2$ 0--0 lines in distant galaxies} 
\label{sec:highz}
The rotational H$_2$ lines probe the by far dominant mass component of warm, shocked Hydrogen, and are thus very interesting tracers of interstellar turbulence. This makes them important lines to study in particular in distant galaxies at intermediate to high redshifts, where massive galaxies were forming stars at high rates from interstellar gas that had significantly higher mass surface densities and turbulent velocities than today \citep[e.g.,][]{guillard_exceptional_2015}. Unfortunately, the rotational lines are redshifted out of the MIRI bandpass at $z=1.89$ for H$_2$ 0--0 S(3), and at z=3.04 for H$_2$ 0--0 S(5) for the nominal spectral coverage of the MRS. The redshift limit becomes even more constraining when taking into account the loss in count rate in Channel~4 observed during the first year of JWST operations. For example, without including Channel~4C, the limiting redshifts are $z=0.83$ for H$_2$ 0--0 S(3), and $z=1.56$ for H$_2$ 0--0 S(5). 

The situation is much better for the ro-vibrational lines which fall into the MRS bandpass out to, e.g., redshift $z=12.1$ and $z=7.3$, respectively, for observations of 1--0 H$_2$ S(3) with and without Channel~4C. However, it is not clear to what degree observations of the ro-vibrational lines can be used as a proxy of the pure-rotational lines. The temperature range they probe is much higher than for the rotational lines (T$\sim 1000$ K instead to T$\sim 100$~K), and in some environments, where the mid-infrared spectrum is strongly dominated by rotational line emission from warm shocked molecular Hydrogen, like Stephan's Quintet \citep[][]{appleton_powerful_2006}, the ro-vibrational lines are very faint.  

Our MIRI and NIRSpec observations provide us with an excellent opportunity to study in a very clean environment not contaminated by star formation or strong AGN emission, to what degree information from ro-vibrational lines heated by shocks can be trusted to infer the basic properties of the gas probed by pure-rotational lines. For different regions of the galaxy we compare the 1--0 and 0--0 S(3) and S(5) lines to demonstrate whether the 1--0 lines are good proxies for the 0--0 lines in the case of galaxies with strong shocks. To this end, we convolved the data cubes of H$_2$ 1--0 S(3) and S(5) with a Gaussian PSF to produce data cubes that have the same spatial resolution as the data cubes of the mid-infrared lines. 

We blur the 1--0 lines with a Gaussian using a FWHM based on the quadratic difference between the 1--0 and 0--0 PSFs. Both 1--0 lines are covered by NIRSpec and therefore can be explained by a PSF of $0.115$" in our observing mode \citep{deugenio_fast-rotator_2023}. The PSF for the 0--0 lines can be derived by setting $FWHM_{PSF}= 0.033\lambda(\mu\mathrm{m}) +0.106$ from \citet{law_3d_2023}. When extracting line fluxes, we used the apertures in Fig.\,\ref{fig:0010ratio} discarding pixels that have a flux of more than 5\,$\sigma$ away from the average of the 4$\times$4 pixel median filtered image where $\sigma$ is the standard deviation of the difference between the model flux image and the median filtered image. The spectral resolving power is similar with both instruments, so that no further correction is required in spectral direction. This allows us to compare directly whether both sets of lines probe similar morphologies and gas kinematics. 

This can be seen in the images of Fig.\,\ref{fig:0010ratio}.  For example, there is a bright blob in the Northern part of the galaxy that is less pronounced in the 1--0 line flux than that of the 0--0 line. 
Overall, the ratio for both the 0--0 and 1--0 S(3) and S(5) lines differ slightly between the Northern and Southern components but remain consistent within 15\%. The average 0--0/1--0 ratio for all apertures in Fig.\,\ref{fig:0010ratio}, not including the cavity or total source (blue and green apertures), is $5.5\pm0.6$. The Southern component has a lower line ratio than the Northern component for both S(3) and S(5) lines.

These variations indicate that the shock conditions are not exactly the same everywhere in the galaxy, but that the uncertainties that this implies, e.g., for determining global emission-line morphologies and to isolate individual structures like shocks and clumps within the galaxy, are not larger than other systematic uncertainties in this kind of work, e.g., due to the stellar initial mass function or the CO-to-H$_2$ conversion factor. Of course, for a more detailed analysis of the gas conditions, measurements of the 0--0 lines are still required. 

The gas kinematics probed by the rotational and ro-vibrational lines are also very similar throughout the galaxy. Fig.~\ref{fig:LineProfiles} shows a comparison of the line profiles of H$_2$ 1--0 S(3) and S(5), H$_2$ 0--0 S(3) and S(5), of Pa$\alpha$, [FeII]$\lambda$1.644, and [NeIII]$\lambda$15.56. This comparison is made for the six apertures shown in Fig.~\ref{fig:0010ratio}. 

The systemic components of the 1--0 and 0--0 lines follow each other very well in velocity and line widths for both S(3) and (5). The blueshifted component is more enhanced in the 1--0 than in the 0--0 lines in some apertures, however, the differences are small. This suggests that turbulent line broadening and global outflow properties can be estimated from the 1--0 lines when the 0--0 lines are not available. We conclude that the shocked molecular Hydrogen in galaxies dominated by turbulent gas motion shows globally similar kinematics and morphologies in the 1--0 and 0--0 lines. Differences present in blue wings and line ratios however suggest that dedicated observations of the 0--0 lines are required to derive robust estimates, e.g., of outflow masses or gas excitation conditions.  

\section{Summary}
\label{sec:summary}
We have obtained new Cycle-1 open time observations with the JWST NIRSpec and MIRI integral-field units, to study the interactions of the radio jet and molecular gas in the nearby (z=0.09) radio galaxy 3C~326~N. This galaxy is one of the best-studied examples of a subset of radio galaxies characterised by bright line emission from warm molecular Hydrogen previously seen with the Spitzer space telescope \citep[][]{ogle_shocked_2007, ogle_jet-powered_2010}, while typical star-formation tracers like PAHs or a dust continuum are either weak or undetected. Subsequently, \citet{nesvadba_energetics_2010} showed that some of these galaxies fall well below the usual star-forming sequence in the Kennicutt-Schmidt diagram, suggesting that star formation may proceed at lower rates than in galaxies without radio jets. The main results of our new study are as follows: 
\begin{itemize}
\item 
We find bright line emission of 19 ro-vibrational lines of the S, Q, and O series of warm molecular Hydrogen, and three lines of pure-rotational line emission, H$_2$ 0--0 S(3), S(5), and S(6). We also obtain a robust systemic redshift estimate of $z=0.08979$ from the CO band heads observed with NIRSpec, using a K4III giant as well matching template. 
\item 
The gas probed by these emission lines is in a 3~kpc sized disk, already identified by \citet{nesvadba_dense_2011} with seeing-limited data. At 10$\times$ higher spatial resolution with NIRSpec, we identify a kpc-sized cavity in the Northern hemisphere of the disk, delineated by three bright knots of line emission at the far end. We interpret this feature as the signature of an expanding, jet-inflated bubble within a rotating gas disk. Similar morphologies are seen in ro-vibrational and rotational lines, as well as the warm ionised gas lines [FeII]$\lambda$1.644, Pa$\alpha$ and [ArII]$\lambda$6.98. Only [NeIII]$\lambda$15.56 has a distinct morphology more centred on the nucleus. 
\item 
Line profiles are complex, with at least two line components throughout the disk seen in the ro-vibrational and rotational lines of warm H$_2$ as well as the warm ionised gas lines Pa$\alpha$ and [FeII]. We developed a new Bayesian line fitting technique based on a Markov Chain Monte Carlo approach, which allows us to robustly fit both line components throughout the data set. The cavity is associated with the redshifted, Northern side of a clear hourglass-shaped velocity distribution in the Southern and Northern part of the disk, with maximal velocities of 380~km s$^{-1}$. We interpret this velocity pattern as the signature of two expanding bubbles, given the large velocity gradient and line widths, more moderate velocities in fainter surrounding gas, and the clear association with the cavity in the North. 
\item 
We also present new JVLA radio observations at 1.5~GHz, 7.3~GHz, and 9~GHz, identifying a compact radio core. Comparison with FIRST shows 47\% of missing flux at 1.5~GHz on scales between 1 and 4.6~kpc from the nucleus. This is likely emission from the radio plasma that is inflating the cavity, stirring up the gas, and making it turbulent. 
\item
Line emission of H$_2$ and most warm ionised lines is heated by shocks. This has already been shown previously from source-integrated data \citep[][]{nesvadba_energetics_2010}, however, we can now also rule out the presence of smaller areas dominated by heating from young stars, down to scales of 190~pc. 
\item 
From archival ALMA CO(1--0) observations we found that a previous detection of cold molecular gas done with the IRAM Plateau de Bure Interferometer in 2009 was spurious, and we now place an upper limit of $8\times 10^8$ M$_{\odot}$, about $3\times$ less than the cold molecular gas mass of \citet{nesvadba_energetics_2010}. This highlights the outstanding nature of this galaxy. 3C~326~N falls still below the Kennicutt-Schmidt law if the new upper limit is not very far from the actual gas mass. When taking into account that maintaining the observed high gas mass requires near-constant gas heating through the radio jet \citep[][]{nesvadba_energetics_2010}, we also argue that, in absence of that heating, 3C~326~N would fall about a factor 100 below the Kennicutt-Schmidt relation of star forming galaxies, an argument derived from using the warm instead of the cold molecular gas mass for this comparison.
\item
The broad line widths and bright line emission can be taken as indicative of enhanced turbulent gas motion and rapid turbulent energy dissipation. We compare observed line widths and size scales with the characteristics of molecular clouds and the line emission produced by turbulent dissipation on sub-pc scales, finding that our data agree with this scenario from kpc scales and down to scales of 190~pc. Star-formation theory postulates that enhanced turbulence can suppress star formation for line widths as observed in 3C~326~N. This reinforces previous arguments that the turbulence injected by radio jets into the interstellar gas of their host galaxies may indeed inhibit star formation, as postulated by cosmological simulations of galaxy evolution. 
\end{itemize}
3C~326~N is a very clear environment where shocks dominate the properties of the interstellar gas. We compare the morphology and line widths of the ro-vibrational and rotational lines of warm H$_2$ to infer whether the former can be used as a proxy to derive the properties of the latter at redshifts where rotational lines have moved out of the MIRI bandpass, e.g., at Cosmic Noon. Rotational lines probe by far the dominant mass component of warm molecular gas. Galaxies in the early Universe often show significant line broadening attributed to enhanced levels of turbulence, and often contain high column densities of molecular gas, which makes observations of warm H$_2$ particular interesting. Our results suggest that, while local differences do exist due to a variation in excitation conditions, the ratio of the S(3) to S(5) line fluxes in both phases agree to within about 60\%, and the line widths within the measurement uncertainties. This is comparable to other systematic uncertainties when studying distant galaxies, e.g., due to the stellar initial mass function or the CO-to-H$_2$ conversion factor, and suggests that ro-vibrational line emission may be an interesting tracer to study the turbulent interstellar medium of distant galaxies, e.g., at Cosmic Noon, and up to the Epoch of Reionization for particularly bright galaxies and gravitational lenses.

\bibliographystyle{aa}
\bibliography{references} 

\begin{thebibliography}{93}
\expandafter\ifx\csname natexlab\endcsname\relax\def\natexlab#1{#1}\fi

\bibitem[{Alatalo {et~al.}(2015)Alatalo, Lacy, Lanz, Bitsakis, Appleton, Nyland, Cales, Chang, Davis, de~Zeeuw, Lonsdale, Martín, Meier, \& Ogle}]{alatalo_suppression_2015}
Alatalo, K., Lacy, M., Lanz, L., {et~al.} 2015, The Astrophysical Journal, 798, 31, aDS Bibcode: 2015ApJ...798...31A

\bibitem[{Appleton {et~al.}(2006)Appleton, Xu, Reach, Dopita, Gao, Lu, Popescu, Sulentic, Tuffs, \& Yun}]{appleton_powerful_2006}
Appleton, P.~N., Xu, K.~C., Reach, W., {et~al.} 2006, The Astrophysical Journal, 639, L51, aDS Bibcode: 2006ApJ...639L..51A

\bibitem[{Armus {et~al.}(2023)Armus, Lai, U, Larson, Diaz-Santos, Evans, Malkan, Rich, Medling, Law, Inami, Muller-Sanchez, Charmandaris, van~der Werf, Stierwalt, Linden, Privon, Barcos-Muñoz, Hayward, Song, Appleton, Aalto, Bohn, Böker, Brown, Finnerty, Howell, Iwasawa, Kemper, Marshall, Mazzarella, McKinney, Murphy, Sanders, \& Surace}]{armus_goals-jwst_2023}
Armus, L., Lai, T., U, V., {et~al.} 2023, The Astrophysical Journal, 942, L37, aDS Bibcode: 2023ApJ...942L..37A

\bibitem[{{Astropy Collaboration} {et~al.}(2013){Astropy Collaboration}, Robitaille, Tollerud, Greenfield, Droettboom, Bray, Aldcroft, Davis, Ginsburg, Price-Whelan, Kerzendorf, Conley, Crighton, Barbary, Muna, Ferguson, Grollier, Parikh, Nair, Unther, Deil, Woillez, Conseil, Kramer, Turner, Singer, Fox, Weaver, Zabalza, Edwards, Azalee~Bostroem, Burke, Casey, Crawford, Dencheva, Ely, Jenness, Labrie, Lim, Pierfederici, Pontzen, Ptak, Refsdal, Servillat, \& Streicher}]{astropy_collaboration_astropy:_2013}
{Astropy Collaboration}, Robitaille, T.~P., Tollerud, E.~J., {et~al.} 2013, Astronomy and Astrophysics, 558, A33

\bibitem[{Becker {et~al.}(1995)Becker, White, \& Helfand}]{becker_first_1995}
Becker, R.~H., White, R.~L., \& Helfand, D.~J. 1995, The Astrophysical Journal, 450, 559, aDS Bibcode: 1995ApJ...450..559B

\bibitem[{Bertoldi \& McKee(1992)}]{bertoldi_pressure-confined_1992}
Bertoldi, F. \& McKee, C.~F. 1992, The Astrophysical Journal, 395, 140, aDS Bibcode: 1992ApJ...395..140B

\bibitem[{Bharadwaj {et~al.}(2015)Bharadwaj, Reiprich, Lovisari, \& Eckmiller}]{bharadwaj_extending_2015}
Bharadwaj, V., Reiprich, T.~H., Lovisari, L., \& Eckmiller, H.~J. 2015, Astronomy and Astrophysics, 573, A75, aDS Bibcode: 2015A\&A...573A..75B

\bibitem[{Bolatto {et~al.}(2013)Bolatto, Wolfire, \& Leroy}]{bolatto_co--h2_2013}
Bolatto, A.~D., Wolfire, M., \& Leroy, A.~K. 2013, Annual Review of Astronomy and Astrophysics, 51, 207, aDS Bibcode: 2013ARA\&A..51..207B

\bibitem[{Bîrzan {et~al.}(2008)Bîrzan, McNamara, Nulsen, Carilli, \& Wise}]{birzan_radiative_2008}
Bîrzan, L., McNamara, B.~R., Nulsen, P. E.~J., Carilli, C.~L., \& Wise, M.~W. 2008, The Astrophysical Journal, 686, 859, aDS Bibcode: 2008ApJ...686..859B

\bibitem[{Calzetti {et~al.}(2000)Calzetti, Armus, Bohlin, Kinney, Koornneef, \& Storchi-Bergmann}]{calzetti_dust_2000}
Calzetti, D., Armus, L., Bohlin, R.~C., {et~al.} 2000, The Astrophysical Journal, 533, 682, aDS Bibcode: 2000ApJ...533..682C

\bibitem[{Capetti {et~al.}(1999)Capetti, Axon, Macchetto, Marconi, \& Winge}]{capetti_origin_1999}
Capetti, A., Axon, D.~J., Macchetto, F.~D., Marconi, A., \& Winge, C. 1999, The Astrophysical Journal, 516, 187, aDS Bibcode: 1999ApJ...516..187C

\bibitem[{Carnall(2017)}]{carnall_spectres_2017}
Carnall, A.~C. 2017, {SpectRes}: {A} {Fast} {Spectral} {Resampling} {Tool} in {Python}, arXiv:1705.05165 [astro-ph]

\bibitem[{Croton {et~al.}(2006)Croton, Springel, White, De~Lucia, Frenk, Gao, Jenkins, Kauffmann, Navarro, \& Yoshida}]{croton_many_2006}
Croton, D.~J., Springel, V., White, S. D.~M., {et~al.} 2006, Monthly Notices of the Royal Astronomical Society, 365, 11, aDS Bibcode: 2006MNRAS.365...11C

\bibitem[{D'Eugenio {et~al.}(2023)D'Eugenio, Perez-Gonzalez, Maiolino, Scholtz, Perna, Circosta, Uebler, Arribas, Boeker, Bunker, Carniani, Charlot, Chevallard, Cresci, Curtis-Lake, Jones, Kumari, Lamperti, Looser, Parlanti, Rix, Robertson, Rodriguez Del~Pino, Tacchella, Venturi, \& Willott}]{deugenio_fast-rotator_2023}
D'Eugenio, F., Perez-Gonzalez, P., Maiolino, R., {et~al.} 2023, A fast-rotator post-starburst galaxy quenched by supermassive black-hole feedback at z=3, publication Title: arXiv e-prints ADS Bibcode: 2023arXiv230806317D

\bibitem[{Dicken {et~al.}(2012)Dicken, Tadhunter, Axon, Morganti, Robinson, Kouwenhoven, Spoon, Kharb, Inskip, Holt, Ramos~Almeida, \& Nesvadba}]{dicken_spitzer_2012}
Dicken, D., Tadhunter, C., Axon, D., {et~al.} 2012, The Astrophysical Journal, 745, 172, aDS Bibcode: 2012ApJ...745..172D

\bibitem[{Dopita {et~al.}(2003)Dopita, Groves, Sutherland, \& Kewley}]{dopita_internal_2003}
Dopita, M.~A., Groves, B.~A., Sutherland, R.~S., \& Kewley, L.~J. 2003, The Astrophysical Journal, 583, 727, aDS Bibcode: 2003ApJ...583..727D

\bibitem[{Drevet~Mulard {et~al.}(2023)Drevet~Mulard, Nesvadba, Meenakshi, Mukherjee, Wagner, Bicknell, Neumayer, Combes, Zovaro, Janssen, Bagchi, Dabhade, \& Prunet}]{drevet_mulard_star_2023}
Drevet~Mulard, M., Nesvadba, N. P.~H., Meenakshi, M., {et~al.} 2023, Astronomy and Astrophysics, 676, A35, aDS Bibcode: 2023A\&A...676A..35D

\bibitem[{Egorov {et~al.}(2023)Egorov, Kreckel, Sandstrom, Leroy, Glover, Groves, Kruijssen, Barnes, Belfiore, Bigiel, Blanc, Boquien, Cao, Chastenet, Chevance, Congiu, Dale, Emsellem, Grasha, Klessen, Larson, Liu, Murphy, Pan, Pessa, Pety, Rosolowsky, Scheuermann, Schinnerer, Sutter, Thilker, Watkins, \& Williams}]{egorov_phangs-jwst_2023}
Egorov, O.~V., Kreckel, K., Sandstrom, K.~M., {et~al.} 2023, The Astrophysical Journal, 944, L16, aDS Bibcode: 2023ApJ...944L..16E

\bibitem[{Ellison {et~al.}(2016)Ellison, Teimoorinia, Rosario, \& Mendel}]{ellison_star_2016}
Ellison, S.~L., Teimoorinia, H., Rosario, D.~J., \& Mendel, J.~T. 2016, Monthly Notices of the Royal Astronomical Society, 458, L34, aDS Bibcode: 2016MNRAS.458L..34E

\bibitem[{Emsellem {et~al.}(2011)Emsellem, Cappellari, Krajnović, Alatalo, Blitz, Bois, Bournaud, Bureau, Davies, Davis, de~Zeeuw, Khochfar, Kuntschner, Lablanche, McDermid, Morganti, Naab, Oosterloo, Sarzi, Scott, Serra, van~de Ven, Weijmans, \& Young}]{emsellem_atlas3d_2011}
Emsellem, E., Cappellari, M., Krajnović, D., {et~al.} 2011, Monthly Notices of the Royal Astronomical Society, 414, 888, aDS Bibcode: 2011MNRAS.414..888E

\bibitem[{Federrath {et~al.}(2010)Federrath, Roman-Duval, Klessen, Schmidt, \& Mac~Low}]{federrath_comparing_2010}
Federrath, C., Roman-Duval, J., Klessen, R.~S., Schmidt, W., \& Mac~Low, M.~M. 2010, Astronomy and Astrophysics, 512, A81, aDS Bibcode: 2010A\&A...512A..81F

\bibitem[{Ferrarese \& Merritt(2000)}]{ferrarese_fundamental_2000}
Ferrarese, L. \& Merritt, D. 2000, The Astrophysical Journal, 539, L9, aDS Bibcode: 2000ApJ...539L...9F

\bibitem[{Fiore {et~al.}(2017)Fiore, Feruglio, Shankar, Bischetti, Bongiorno, Brusa, Carniani, Cicone, Duras, Lamastra, Mainieri, Marconi, Menci, Maiolino, Piconcelli, Vietri, \& Zappacosta}]{fiore_agn_2017}
Fiore, F., Feruglio, C., Shankar, F., {et~al.} 2017, Astronomy and Astrophysics, 601, A143, aDS Bibcode: 2017A\&A...601A.143F

\bibitem[{Fischer {et~al.}(2013)Fischer, Crenshaw, Kraemer, \& Schmitt}]{fischer_determining_2013}
Fischer, T.~C., Crenshaw, D.~M., Kraemer, S.~B., \& Schmitt, H.~R. 2013, The Astrophysical Journal Supplement Series, 209, 1

\bibitem[{Flower \& Pineau Des~Forêts(2010)}]{flower_excitation_2010}
Flower, D.~R. \& Pineau Des~Forêts, G. 2010, Monthly Notices of the Royal Astronomical Society, 406, 1745, aDS Bibcode: 2010MNRAS.406.1745F

\bibitem[{Forman {et~al.}(2005)Forman, Nulsen, Heinz, Owen, Eilek, Vikhlinin, Markevitch, Kraft, Churazov, \& Jones}]{forman_reflections_2005}
Forman, W., Nulsen, P., Heinz, S., {et~al.} 2005, The Astrophysical Journal, 635, 894, aDS Bibcode: 2005ApJ...635..894F

\bibitem[{Gebhardt {et~al.}(2000)Gebhardt, Bender, Bower, Dressler, Faber, Filippenko, Green, Grillmair, Ho, Kormendy, Lauer, Magorrian, Pinkney, Richstone, \& Tremaine}]{gebhardt_relationship_2000}
Gebhardt, K., Bender, R., Bower, G., {et~al.} 2000, The Astrophysical Journal, 539, L13, aDS Bibcode: 2000ApJ...539L..13G

\bibitem[{Goodman \& Weare(2010)}]{goodman_ensemble_2010}
Goodman, J. \& Weare, J. 2010, Communications in Applied Mathematics and Computational Science, 5, 65

\bibitem[{Guillard {et~al.}(2015)Guillard, Boulanger, Lehnert, Forêts, Combes, Falgarone, \& Bernard-Salas}]{guillard_exceptional_2015}
Guillard, P., Boulanger, F., Lehnert, M.~D., {et~al.} 2015, Astronomy \& Astrophysics, 574, A32, publisher: EDP Sciences

\bibitem[{Guillard {et~al.}(2009)Guillard, Boulanger, Pineau Des~Forêts, \& Appleton}]{guillard_h2_2009}
Guillard, P., Boulanger, F., Pineau Des~Forêts, G., \& Appleton, P.~N. 2009, Astronomy and Astrophysics, 502, 515, aDS Bibcode: 2009A\&A...502..515G

\bibitem[{Guillard {et~al.}(2012)Guillard, Ogle, Emonts, Appleton, Morganti, Tadhunter, Oosterloo, Evans, \& Evans}]{guillard_strong_2012}
Guillard, P., Ogle, P.~M., Emonts, B. H.~C., {et~al.} 2012, The Astrophysical Journal, 747, 95, aDS Bibcode: 2012ApJ...747...95G

\bibitem[{Hennebelle \& Chabrier(2011)}]{hennebelle_analytical_2011}
Hennebelle, P. \& Chabrier, G. 2011, The Astrophysical Journal, 743, L29, aDS Bibcode: 2011ApJ...743L..29H

\bibitem[{Hennebelle \& Falgarone(2012)}]{hennebelle_turbulent_2012}
Hennebelle, P. \& Falgarone, E. 2012, Astronomy and Astrophysics Review, 20, 55, aDS Bibcode: 2012A\&ARv..20...55H

\bibitem[{Heyer {et~al.}(2009)Heyer, Krawczyk, Duval, \& Jackson}]{heyer_re-examining_2009}
Heyer, M., Krawczyk, C., Duval, J., \& Jackson, J.~M. 2009, The Astrophysical Journal, 699, 1092, aDS Bibcode: 2009ApJ...699.1092H

\bibitem[{Ho \& Keto(2007)}]{ho_mid-infrared_2007}
Ho, L.~C. \& Keto, E. 2007, The Astrophysical Journal, 658, 314, aDS Bibcode: 2007ApJ...658..314H

\bibitem[{Keel(1990)}]{keel_shock_1990}
Keel, W.~C. 1990, The Astronomical Journal, 100, 356, aDS Bibcode: 1990AJ....100..356K

\bibitem[{Kennicutt(2020)}]{kennicutt_star_2020}
Kennicutt, R. 2020, 235, 362.01, conference Name: American Astronomical Society Meeting Abstracts \#235 ADS Bibcode: 2020AAS...23536201K

\bibitem[{Kennicutt(1989)}]{kennicutt_star_1989}
Kennicutt, Jr., R.~C. 1989, The Astrophysical Journal, 344, 685, aDS Bibcode: 1989ApJ...344..685K

\bibitem[{Kewley {et~al.}(2019)Kewley, Nicholls, \& Sutherland}]{kewley_understanding_2019}
Kewley, L.~J., Nicholls, D.~C., \& Sutherland, R.~S. 2019, Annual Review of Astronomy and Astrophysics, 57, 511, aDS Bibcode: 2019ARA\&A..57..511K

\bibitem[{Kirkpatrick {et~al.}(2009)Kirkpatrick, Gitti, Cavagnolo, McNamara, David, Nulsen, \& Wise}]{kirkpatrick_direct_2009}
Kirkpatrick, C.~C., Gitti, M., Cavagnolo, K.~W., {et~al.} 2009, The Astrophysical Journal, 707, L69, aDS Bibcode: 2009ApJ...707L..69K

\bibitem[{Krumholz \& McKee(2005)}]{krumholz_general_2005}
Krumholz, M.~R. \& McKee, C.~F. 2005, The Astrophysical Journal, 630, 250, aDS Bibcode: 2005ApJ...630..250K

\bibitem[{Lai {et~al.}(2022)Lai, Armus, U, Díaz-Santos, Larson, Evans, Malkan, Appleton, Rich, Müller-Sánchez, Inami, Bohn, McKinney, Finnerty, Law, Linden, Medling, Privon, Song, Stierwalt, Werf, Barcos-Muñoz, Smith, Togi, Aalto, Böker, Charmandaris, Howell, Iwasawa, Kemper, Mazzarella, Murphy, Brown, Hayward, Marshall, Sanders, \& Surace}]{lai_goals-jwst_2022}
Lai, T. S.-Y., Armus, L., U, V., {et~al.} 2022, The Astrophysical Journal Letters, 941, L36, publisher: The American Astronomical Society

\bibitem[{Laing {et~al.}(1983)Laing, Riley, \& Longair}]{laing_bright_1983}
Laing, R.~A., Riley, J.~M., \& Longair, M.~S. 1983, Monthly Notices of the Royal Astronomical Society, 204, 151, aDS Bibcode: 1983MNRAS.204..151L

\bibitem[{Lanz {et~al.}(2016)Lanz, Ogle, Alatalo, \& Appleton}]{lanz_star_2016}
Lanz, L., Ogle, P.~M., Alatalo, K., \& Appleton, P.~N. 2016, The Astrophysical Journal, 826, 29, aDS Bibcode: 2016ApJ...826...29L

\bibitem[{Larson(1981)}]{larson_turbulence_1981}
Larson, R.~B. 1981, Monthly Notices of the Royal Astronomical Society, 194, 809, aDS Bibcode: 1981MNRAS.194..809L

\bibitem[{Law {et~al.}(2023)Law, E.~Morrison, Argyriou, Patapis, Álvarez Márquez, Labiano, \& Vandenbussche}]{law_3d_2023}
Law, D.~R., E.~Morrison, J., Argyriou, I., {et~al.} 2023, The Astronomical Journal, 166, 45, aDS Bibcode: 2023AJ....166...45L

\bibitem[{Levesque \& Richardson(2014)}]{levesque_ne_2014}
Levesque, E.~M. \& Richardson, M. L.~A. 2014, The Astrophysical Journal, 780, 100, aDS Bibcode: 2014ApJ...780..100L

\bibitem[{Mandal {et~al.}(2021)Mandal, Mukherjee, Federrath, Nesvadba, Bicknell, Wagner, \& Meenakshi}]{mandal_impact_2021}
Mandal, A., Mukherjee, D., Federrath, C., {et~al.} 2021, Monthly Notices of the Royal Astronomical Society, 508, 4738, aDS Bibcode: 2021MNRAS.508.4738M

\bibitem[{Mayya {et~al.}(2023)Mayya, Alzate, Lomelí-Núez, Zaragoza-Cardiel, Gómez-González, Silich, Fernández-Arenas, Vega, Ovando, Rodríguez, Rosa-González, Luna, Zamora-Avilés, \& Rosales-Ortega}]{mayya_stellar_2023}
Mayya, Y.~D., Alzate, J.~A., Lomelí-Núez, L., {et~al.} 2023, Monthly Notices of the Royal Astronomical Society, 521, 5492, aDS Bibcode: 2023MNRAS.521.5492M

\bibitem[{Mazzalay {et~al.}(2013)Mazzalay, Saglia, Erwin, Fabricius, Rusli, Thomas, Bender, Opitsch, Nowak, \& Williams}]{mazzalay_molecular_2013}
Mazzalay, X., Saglia, R.~P., Erwin, P., {et~al.} 2013, Monthly Notices of the Royal Astronomical Society, 428, 2389, aDS Bibcode: 2013MNRAS.428.2389M

\bibitem[{McKee \& Ostriker(2007)}]{mckee_theory_2007}
McKee, C.~F. \& Ostriker, E.~C. 2007, Annual Review of Astronomy and Astrophysics, 45, 565, aDS Bibcode: 2007ARA\&A..45..565M

\bibitem[{Medling {et~al.}(2021)Medling, Kewley, Calzetti, Privon, Larson, Rich, Armus, Allen, Bicknell, Díaz-Santos, Heckman, Leitherer, Max, Rupke, Treister, Messias, \& Wagner}]{medling_tracing_2021}
Medling, A.~M., Kewley, L.~J., Calzetti, D., {et~al.} 2021, The Astrophysical Journal, 923, 160, aDS Bibcode: 2021ApJ...923..160M

\bibitem[{Meenakshi {et~al.}(2022)Meenakshi, Mukherjee, Wagner, Nesvadba, Bicknell, Morganti, Janssen, Sutherland, \& Mandal}]{meenakshi_modelling_2022}
Meenakshi, M., Mukherjee, D., Wagner, A.~Y., {et~al.} 2022, Monthly Notices of the Royal Astronomical Society, 516, 766, aDS Bibcode: 2022MNRAS.516..766M

\bibitem[{Meijerink {et~al.}(2013)Meijerink, Kristensen, Weiß, van~der Werf, Walter, Spaans, Loenen, Fischer, Israel, Isaak, Papadopoulos, Aalto, Armus, Charmandaris, Dasyra, Diaz-Santos, Evans, Gao, González-Alfonso, Güsten, Henkel, Kramer, Lord, Martín-Pintado, Naylor, Sanders, Smith, Spinoglio, Stacey, Veilleux, \& Wiedner}]{meijerink_evidence_2013}
Meijerink, R., Kristensen, L.~E., Weiß, A., {et~al.} 2013, The Astrophysical Journal, 762, L16, aDS Bibcode: 2013ApJ...762L..16M

\bibitem[{Merloni \& Heinz(2007)}]{merloni_measuring_2007}
Merloni, A. \& Heinz, S. 2007, Monthly Notices of the Royal Astronomical Society, 381, 589, aDS Bibcode: 2007MNRAS.381..589M

\bibitem[{Micelotta {et~al.}(2010)Micelotta, Jones, \& Tielens}]{micelotta_polycyclic_2010}
Micelotta, E.~R., Jones, A.~P., \& Tielens, A. G. G.~M. 2010, Astronomy and Astrophysics, 510, A36, aDS Bibcode: 2010A\&A...510A..36M

\bibitem[{Mukherjee {et~al.}(2018)Mukherjee, Bicknell, Wagner, Sutherland, \& Silk}]{mukherjee_relativistic_2018}
Mukherjee, D., Bicknell, G.~V., Wagner, A.~Y., Sutherland, R.~S., \& Silk, J. 2018, Monthly Notices of the Royal Astronomical Society, 479, 5544, aDS Bibcode: 2018MNRAS.479.5544M

\bibitem[{Nesvadba {et~al.}(2011)Nesvadba, Boulanger, Lehnert, Guillard, \& Salome}]{nesvadba_dense_2011}
Nesvadba, N. P.~H., Boulanger, F., Lehnert, M.~D., Guillard, P., \& Salome, P. 2011, Astronomy and Astrophysics, 536, L5, aDS Bibcode: 2011A\&A...536L...5N

\bibitem[{Nesvadba {et~al.}(2010)Nesvadba, Boulanger, Salomé, Guillard, Lehnert, Ogle, Appleton, Falgarone, \& Pineau Des~Forets}]{nesvadba_energetics_2010}
Nesvadba, N. P.~H., Boulanger, F., Salomé, P., {et~al.} 2010, Astronomy and Astrophysics, 521, A65, aDS Bibcode: 2010A\&A...521A..65N

\bibitem[{Nesvadba {et~al.}(2021)Nesvadba, Wagner, Mukherjee, Mandal, Janssen, Zovaro, Neumayer, Bagchi, \& Bicknell}]{nesvadba_jet-driven_2021}
Nesvadba, N. P.~H., Wagner, A.~Y., Mukherjee, D., {et~al.} 2021, Astronomy and Astrophysics, 654, A8, aDS Bibcode: 2021A\&A...654A...8N

\bibitem[{Ogle {et~al.}(2007)Ogle, Antonucci, Appleton, \& Whysong}]{ogle_shocked_2007}
Ogle, P., Antonucci, R., Appleton, P.~N., \& Whysong, D. 2007, The Astrophysical Journal, 668, 699, aDS Bibcode: 2007ApJ...668..699O

\bibitem[{Ogle {et~al.}(2010)Ogle, Boulanger, Guillard, Evans, Antonucci, Appleton, Nesvadba, \& Leipski}]{ogle_jet-powered_2010}
Ogle, P., Boulanger, F., Guillard, P., {et~al.} 2010, The Astrophysical Journal, 724, 1193, aDS Bibcode: 2010ApJ...724.1193O

\bibitem[{Padoan \& Nordlund(2011)}]{padoan_star_2011}
Padoan, P. \& Nordlund, A. 2011, The Astrophysical Journal, 730, 40, aDS Bibcode: 2011ApJ...730...40P

\bibitem[{Pereira-Santaella {et~al.}(2022)Pereira-Santaella, Álvarez Márquez, García-Bernete, Labiano, Colina, Alonso-Herrero, Bellocchi, García-Burillo, Hönig, Ramos~Almeida, \& Rosario}]{pereira-santaella_low-power_2022}
Pereira-Santaella, M., Álvarez Márquez, J., García-Bernete, I., {et~al.} 2022, Astronomy and Astrophysics, 665, L11, aDS Bibcode: 2022A\&A...665L..11P

\bibitem[{{Planck Collaboration} {et~al.}(2016){Planck Collaboration}, Ade, Aghanim, Arnaud, Ashdown, Aumont, Baccigalupi, Banday, Barreiro, Bartlett, Bartolo, Battaner, Battye, Benabed, Benoît, Benoit-Lévy, Bernard, Bersanelli, Bielewicz, Bock, Bonaldi, Bonavera, Bond, Borrill, Bouchet, Boulanger, Bucher, Burigana, Butler, Calabrese, Cardoso, Catalano, Challinor, Chamballu, Chary, Chiang, Chluba, Christensen, Church, Clements, Colombi, Colombo, Combet, Coulais, Crill, Curto, Cuttaia, Danese, Davies, Davis, de~Bernardis, de~Rosa, de~Zotti, Delabrouille, Désert, Di~Valentino, Dickinson, Diego, Dolag, Dole, Donzelli, Doré, Douspis, Ducout, Dunkley, Dupac, Efstathiou, Elsner, Enßlin, Eriksen, Farhang, Fergusson, Finelli, Forni, Frailis, Fraisse, Franceschi, Frejsel, Galeotta, Galli, Ganga, Gauthier, Gerbino, Ghosh, Giard, Giraud-Héraud, Giusarma, Gjerløw, González-Nuevo, Górski, Gratton, Gregorio, Gruppuso, Gudmundsson, Hamann, Hansen, Hanson, Harrison, Helou, Henrot-Versillé, Hernández-Monteagudo,
  Herranz, Hildebrandt, Hivon, Hobson, Holmes, Hornstrup, Hovest, Huang, Huffenberger, Hurier, Jaffe, Jaffe, Jones, Juvela, Keihänen, Keskitalo, Kisner, Kneissl, Knoche, Knox, Kunz, Kurki-Suonio, Lagache, Lähteenmäki, Lamarre, Lasenby, Lattanzi, Lawrence, Leahy, Leonardi, Lesgourgues, Levrier, Lewis, Liguori, Lilje, Linden-Vørnle, López-Caniego, Lubin, Macías-Pérez, Maggio, Maino, Mandolesi, Mangilli, Marchini, Maris, Martin, Martinelli, Martínez-González, Masi, Matarrese, McGehee, Meinhold, Melchiorri, Melin, Mendes, Mennella, Migliaccio, Millea, Mitra, Miville-Deschênes, Moneti, Montier, Morgante, Mortlock, Moss, Munshi, Murphy, Naselsky, Nati, Natoli, Netterfield, Nørgaard-Nielsen, Noviello, Novikov, Novikov, Oxborrow, Paci, Pagano, Pajot, Paladini, Paoletti, Partridge, Pasian, Patanchon, Pearson, Perdereau, Perotto, Perrotta, Pettorino, Piacentini, Piat, Pierpaoli, Pietrobon, Plaszczynski, Pointecouteau, Polenta, Popa, Pratt, Prézeau, Prunet, Puget, Rachen, Reach, Rebolo, Reinecke,
  Remazeilles, Renault, Renzi, Ristorcelli, Rocha, Rosset, Rossetti, Roudier, Rouillé~d'Orfeuil, Rowan-Robinson, Rubiño-Martín, Rusholme, Said, Salvatelli, Salvati, Sandri, Santos, Savelainen, Savini, Scott, Seiffert, Serra, Shellard, Spencer, Spinelli, Stolyarov, Stompor, Sudiwala, Sunyaev, Sutton, Suur-Uski, Sygnet, Tauber, Terenzi, Toffolatti, Tomasi, Tristram, Trombetti, Tucci, Tuovinen, Türler, Umana, Valenziano, Valiviita, Van~Tent, Vielva, Villa, Wade, Wandelt, Wehus, White, White, Wilkinson, Yvon, Zacchei, \& Zonca}]{planck_collaboration_planck_2016}
{Planck Collaboration}, Ade, P. A.~R., Aghanim, N., {et~al.} 2016, Astronomy and Astrophysics, 594, A13, aDS Bibcode: 2016A\&A...594A..13P

\bibitem[{Puxley {et~al.}(1990)Puxley, Hawarden, \& Mountain}]{puxley_molecular_1990}
Puxley, P.~J., Hawarden, T.~G., \& Mountain, C.~M. 1990, The Astrophysical Journal, 364, 77, aDS Bibcode: 1990ApJ...364...77P

\bibitem[{Rawlings {et~al.}(1990)Rawlings, Saunders, Miller, Jones, \& Eales}]{rawlings_new_1990}
Rawlings, S., Saunders, R., Miller, P., Jones, M.~E., \& Eales, S.~A. 1990, Monthly Notices of the Royal Astronomical Society, 246, 21P, aDS Bibcode: 1990MNRAS.246P..21R

\bibitem[{Rodríguez-Ardila {et~al.}(2004)Rodríguez-Ardila, Pastoriza, Viegas, Sigut, \& Pradhan}]{rodriguez-ardila_molecular_2004}
Rodríguez-Ardila, A., Pastoriza, M.~G., Viegas, S., Sigut, T. A.~A., \& Pradhan, A.~K. 2004, Astronomy and Astrophysics, 425, 457, aDS Bibcode: 2004A\&A...425..457R

\bibitem[{Rémy-Ruyer {et~al.}(2014)Rémy-Ruyer, Madden, Galliano, Galametz, Takeuchi, Asano, Zhukovska, Lebouteiller, Cormier, Jones, Bocchio, Baes, Bendo, Boquien, Boselli, DeLooze, Doublier-Pritchard, Hughes, Karczewski, \& Spinoglio}]{remy-ruyer_gas--dust_2014}
Rémy-Ruyer, A., Madden, S.~C., Galliano, F., {et~al.} 2014, Astronomy and Astrophysics, 563, A31, aDS Bibcode: 2014A\&A...563A..31R

\bibitem[{Sage {et~al.}(2007)Sage, Welch, \& Young}]{sage_cool_2007}
Sage, L.~J., Welch, G.~A., \& Young, L.~M. 2007, The Astrophysical Journal, 657, 232, aDS Bibcode: 2007ApJ...657..232S

\bibitem[{Saintonge {et~al.}(2013)Saintonge, Lutz, Genzel, Magnelli, Nordon, Tacconi, Baker, Bandara, Berta, Förster~Schreiber, Poglitsch, Sturm, Wuyts, \& Wuyts}]{saintonge_validation_2013}
Saintonge, A., Lutz, D., Genzel, R., {et~al.} 2013, The Astrophysical Journal, 778, 2, aDS Bibcode: 2013ApJ...778....2S

\bibitem[{Schaye {et~al.}(2015)Schaye, Crain, Bower, Furlong, Schaller, Theuns, Dalla~Vecchia, Frenk, McCarthy, Helly, Jenkins, Rosas-Guevara, White, Baes, Booth, Camps, Navarro, Qu, Rahmati, Sawala, Thomas, \& Trayford}]{schaye_eagle_2015}
Schaye, J., Crain, R.~A., Bower, R.~G., {et~al.} 2015, Monthly Notices of the Royal Astronomical Society, 446, 521, aDS Bibcode: 2015MNRAS.446..521S

\bibitem[{Schmidt(1968)}]{schmidt_space_1968}
Schmidt, M. 1968, The Astrophysical Journal, 151, 393

\bibitem[{Scoville {et~al.}(1982)Scoville, Hall, Ridgway, \& Kleinmann}]{scoville_velocity_1982}
Scoville, N.~Z., Hall, D. N.~B., Ridgway, S.~T., \& Kleinmann, S.~G. 1982, The Astrophysical Journal, 253, 136, aDS Bibcode: 1982ApJ...253..136S

\bibitem[{Silk \& Rees(1998)}]{silk_quasars_1998}
Silk, J. \& Rees, M.~J. 1998, Astronomy and Astrophysics, 331, L1, aDS Bibcode: 1998A\&A...331L...1S

\bibitem[{Solomon {et~al.}(1997)Solomon, Downes, Radford, \& Barrett}]{solomon_molecular_1997}
Solomon, P.~M., Downes, D., Radford, S. J.~E., \& Barrett, J.~W. 1997, The Astrophysical Journal, 478, 144, aDS Bibcode: 1997ApJ...478..144S

\bibitem[{Spoon {et~al.}(2009)Spoon, Armus, Marshall, Bernard-Salas, Farrah, Charmandaris, \& Kent}]{spoon_high-velocity_2009}
Spoon, H. W.~W., Armus, L., Marshall, J.~A., {et~al.} 2009, The Astrophysical Journal, 693, 1223, aDS Bibcode: 2009ApJ...693.1223S

\bibitem[{Spoon \& Holt(2009)}]{spoon_discovery_2009}
Spoon, H. W.~W. \& Holt, J. 2009, The Astrophysical Journal, 702, L42, aDS Bibcode: 2009ApJ...702L..42S

\bibitem[{Springel {et~al.}(2005)Springel, Di~Matteo, \& Hernquist}]{springel_modelling_2005}
Springel, V., Di~Matteo, T., \& Hernquist, L. 2005, Monthly Notices of the Royal Astronomical Society, 361, 776, aDS Bibcode: 2005MNRAS.361..776S

\bibitem[{Stanley {et~al.}(2017)Stanley, Alexander, Harrison, Rosario, Wang, Aird, Bourne, Dunne, Dye, Eales, Knudsen, Michałowski, Valiante, De~Zotti, Furlanetto, Ivison, Maddox, \& Smith}]{stanley_mean_2017}
Stanley, F., Alexander, D.~M., Harrison, C.~M., {et~al.} 2017, Monthly Notices of the Royal Astronomical Society, 472, 2221, aDS Bibcode: 2017MNRAS.472.2221S

\bibitem[{Sun {et~al.}(2018)Sun, Leroy, Schruba, Rosolowsky, Hughes, Kruijssen, Meidt, Schinnerer, Blanc, Bigiel, Bolatto, Chevance, Groves, Herrera, Hygate, Pety, Querejeta, Usero, \& Utomo}]{sun_cloud-scale_2018}
Sun, J., Leroy, A.~K., Schruba, A., {et~al.} 2018, The Astrophysical Journal, 860, 172, aDS Bibcode: 2018ApJ...860..172S

\bibitem[{Veilleux {et~al.}(1994)Veilleux, Cecil, Bland-Hawthorn, Tully, Filippenko, \& Sargent}]{veilleux_nuclear_1994}
Veilleux, S., Cecil, G., Bland-Hawthorn, J., {et~al.} 1994, The Astrophysical Journal, 433, 48, aDS Bibcode: 1994ApJ...433...48V

\bibitem[{Veilleux {et~al.}(2013)Veilleux, Meléndez, Sturm, Gracia-Carpio, Fischer, González-Alfonso, Contursi, Lutz, Poglitsch, Davies, Genzel, Tacconi, de~Jong, Sternberg, Netzer, Hailey-Dunsheath, Verma, Rupke, Maiolino, Teng, \& Polisensky}]{veilleux_fast_2013}
Veilleux, S., Meléndez, M., Sturm, E., {et~al.} 2013, The Astrophysical Journal, 776, 27, aDS Bibcode: 2013ApJ...776...27V

\bibitem[{Villar~Martín {et~al.}(2023)Villar~Martín, Castro-Rodríguez, Pereira~Santaella, Lamperti, Tadhunter, Emonts, Colina, Alonso~Herrero, Cabrera-Lavers, \& Bellocchi}]{villar_martin_limited_2023}
Villar~Martín, M., Castro-Rodríguez, N., Pereira~Santaella, M., {et~al.} 2023, Astronomy and Astrophysics, 673, A25, aDS Bibcode: 2023A\&A...673A..25V

\bibitem[{Weinberger {et~al.}(2018)Weinberger, Springel, Pakmor, Nelson, Genel, Pillepich, Vogelsberger, Marinacci, Naiman, Torrey, \& Hernquist}]{weinberger_supermassive_2018}
Weinberger, R., Springel, V., Pakmor, R., {et~al.} 2018, Monthly Notices of the Royal Astronomical Society, 479, 4056, aDS Bibcode: 2018MNRAS.479.4056W

\bibitem[{Werner \& Mernier(2020)}]{werner_hot_2020}
Werner, N. \& Mernier, F. 2020, Hot {Atmospheres} of {Galaxies}, {Groups}, and {Clusters} of {Galaxies}, pages: 279-310 Publication Title: Reviews in Frontiers of Modern Astrophysics; From Space Debris to Cosmology ADS Bibcode: 2020rfma.book..279K

\bibitem[{White {et~al.}(1997)White, Becker, Helfand, \& Gregg}]{white_catalog_1997}
White, R.~L., Becker, R.~H., Helfand, D.~J., \& Gregg, M.~D. 1997, The Astrophysical Journal, 475, 479, aDS Bibcode: 1997ApJ...475..479W

\bibitem[{Willis \& Strom(1978)}]{willis_multifrequency_1978}
Willis, A.~G. \& Strom, R.~G. 1978, Astronomy and Astrophysics, 62, 375, aDS Bibcode: 1978A\&A....62..375W

\bibitem[{Winge {et~al.}(2009)Winge, Riffel, \& Storchi-Bergmann}]{winge_gemini_2009}
Winge, C., Riffel, R.~A., \& Storchi-Bergmann, T. 2009, The Astrophysical Journal Supplement Series, 185, 186, publisher: The American Astronomical Society

\bibitem[{Yoon {et~al.}(2021)Yoon, Park, Chung, \& Zhang}]{yoon_rotation_2021}
Yoon, Y., Park, C., Chung, H., \& Zhang, K. 2021, The Astrophysical Journal, 922, 249, aDS Bibcode: 2021ApJ...922..249Y

\bibitem[{Young {et~al.}(2011)Young, Bureau, Davis, Combes, McDermid, Alatalo, Blitz, Bois, Bournaud, Cappellari, Davies, de~Zeeuw, Emsellem, Khochfar, Krajnović, Kuntschner, Lablanche, Morganti, Naab, Oosterloo, Sarzi, Scott, Serra, \& Weijmans}]{young_atlas3d_2011}
Young, L.~M., Bureau, M., Davis, T.~A., {et~al.} 2011, Monthly Notices of the Royal Astronomical Society, 414, 940, aDS Bibcode: 2011MNRAS.414..940Y

\bibitem[{Zovaro {et~al.}(2019)Zovaro, Sharp, Nesvadba, Bicknell, Mukherjee, Wagner, Groves, \& Krishna}]{zovaro_jets_2019}
Zovaro, H. R.~M., Sharp, R., Nesvadba, N. P.~H., {et~al.} 2019, Monthly Notices of the Royal Astronomical Society, 484, 3393, aDS Bibcode: 2019MNRAS.484.3393Z

\bibitem[{Álvarez Márquez {et~al.}(2023)Álvarez Márquez, Labiano, Guillard, Dicken, Argyriou, Patapis, Law, Kavanagh, Larson, Gasman, Mueller, Alberts, Brandl, Colina, García-Marín, Jones, Noriega-Crespo, Shivaei, Temim, \& Wright}]{alvarez-marquez_nuclear_2023}
Álvarez Márquez, J., Labiano, A., Guillard, P., {et~al.} 2023, Astronomy and Astrophysics, 672, A108, aDS Bibcode: 2023A\&A...672A.108A

\end{thebibliography}

\begin{acknowledgements}
We would like to thank the staff at STScI, at the JVLA, and at ALMA for their fantastic assistance with preparing the data taking, and with carrying out the observations. JL, NPHN and DM acknowledge partial financial support through CEFIPRA program 6504-2. This work was supported by the French government through the National Research Agency (ANR) with funding grant ANR AGN\_MELBa (ANR-21-CE31-0011).
This work is based in large parts on observations made with the NASA/ESA/CSA James Webb Space Telescope. The data were obtained from the Mikulski Archive for Space Telescopes at the Space Telescope Science Institute, which is operated by the Association of Universities for Research in Astronomy, Inc., under NASA contract NAS 5-03127 for JWST. These observations are associated with program \#GO1-2162.
This work is partly based on data obtained with the Karl G. Jansky Very Large Array (JVLA) of the National Radio Astronomy Observatory (NRAO) through program ID 18A-285. The National Radio Astronomy Observatory is a facility of the National Science Foundation operated under cooperative agreement by Associated Universities, Inc. 
This paper makes use of the following ALMA data: 2015.1.01120.S. ALMA is a partnership of ESO (representing its member states), NSF (USA) and NINS (Japan), together with NRC (Canada), NSTC and ASIAA (Taiwan), and KASI (Republic of Korea), in cooperation with the Republic of Chile. The Joint ALMA Observatory is operated by ESO, AUI/NRAO and NAOJ.
\end{acknowledgements}

\end{document}